\documentclass[letterpaper,floatfix,reprint,aps,prx,superscriptaddress]{revtex4-2} 

\usepackage{graphicx} 
\usepackage{silence}
\usepackage{gensymb}
\usepackage{tabularx}
\usepackage{color}
\usepackage{amsmath}
\usepackage{mathtools}
\usepackage{amssymb}
\usepackage{hyperref} 
\usepackage{xcolor}

\usepackage[version=3]{mhchem}
\usepackage[pass]{geometry}

\begin{document}

\title{Gradient-based optimization of exact stochastic kinetic models}

\author{Francesco Mottes}
\thanks{fmottes@seas.harvard.edu}
\affiliation{School of Engineering and Applied Sciences, Harvard University, Cambridge MA 02138}

\author{Qian-Ze Zhu}
\affiliation{School of Engineering and Applied Sciences, Harvard University, Cambridge MA 02138}

\author{Michael P. Brenner}
\thanks{brenner@seas.harvard.edu}
\affiliation{School of Engineering and Applied Sciences, Harvard University, Cambridge MA 02138}

\date{\today}

\begin{abstract}
    Stochastic kinetic models describe systems across biology, chemistry, and physics where discrete events and small populations render deterministic approximations inadequate. Parameter inference and inverse design in these systems require optimizing over trajectories generated by the Stochastic Simulation Algorithm, but the discrete reaction events involved are inherently non-differentiable. We present an approach based on straight-through Gumbel-Softmax estimation that maintains exact stochastic simulations in the forward pass while approximating gradients through a continuous relaxation applied only in the backward pass. We demonstrate robust performance on parameter inference in stochastic gene expression, first recovering kinetic rates of telegraph promoter models from both moment statistics and full steady-state distributions across diverse and challenging synthetic parameter regimes, then inferring the kinetic parameters of a four-state promoter model from experimental single-molecule RNA timecourse measurements. We further apply the method to inverse design in stochastic thermodynamics, optimizing non-equilibrium currents in an interacting particle system under kinetic resource constraints and recovering known analytical bounds. The ability to efficiently differentiate through exact stochastic simulations provides a foundation for systematic scalable inference and rational design across the many domains governed by continuous-time Markov dynamics.
\end{abstract}

\maketitle

\section{Introduction}

Stochastic kinetic models, formalized as continuous-time Markov processes describing discrete state transitions, arise throughout the quantitative sciences. In systems biology, they are used to describe transcriptional bursting and regulatory switching underlying cell-to-cell variability in gene expression \cite{elowitz2002stochastic,raj2008nature}. Classic examples include the bistable lysis-lysogeny decision of lambda phage \cite{arkin1998stochastic} and stochastic induction of the lac operon \cite{ozbudak2004regulation}. In non-equilibrium statistical mechanics, analogous frameworks describe molecular motors, driven transport processes, and the thermodynamic costs associated with maintaining currents far from equilibrium \cite{seifert2012stochastic,jarzynski1997nonequilibrium}. In chemistry and ecology, stochastic kinetics govern reaction networks and population dynamics in regimes where small molecular counts and discrete events render deterministic approximations inadequate \cite{gillespie1992rigorous,black2012stochastic}.

The Chemical Master Equation (CME) provides the mathematical framework for describing the time evolution of probability distributions over discrete system states in stochastic kinetic models \cite{vankampen1992stochastic}. Analytical solutions, however, are available only for restricted classes of reaction networks \cite{schnakenberg1976network, shahrezaei2008analytical}. For general systems, exact sample trajectories can be generated using Gillespie's Stochastic Simulation Algorithm (SSA) \cite{gillespie1977exact}, which produces realizations of the underlying continuous-time Markov process. Parameter inference and inverse design in these systems therefore rely mostly on simulation-based approaches.

Solving inverse problems in stochastic kinetic systems is essential for extracting mechanistic insight from noisy experimental data, and substantial effort has been devoted to developing inference methods. For systems with state spaces amenable to controlled truncation, Finite State Projection and related methods enable likelihood-based inference \cite{munsky2006fsp,munsky2012science,neuert2013systematic,fang2024advanced}, but require explicit representation of the truncated state space. Moment-based methods offer computational efficiency, but only when low-order statistics are informative \cite{zechner2012moment}. Bayesian frameworks, including Approximate Bayesian Computation \cite{toni2009approximate,liepe2014abc} and particle Markov chain Monte Carlo \cite{golightly2011bayesian,andrieu2010particle}, provide rigorous uncertainty quantification but can become computationally demanding in higher-dimensional parameter spaces. Gradient-based optimization is an attractive alternative, since reverse-mode automatic differentiation provides gradients at cost comparable to a single forward evaluation \cite{baydin2018automatic}. However, the discrete reaction selections and state updates in the SSA are intrinsically non-differentiable, precluding direct pathwise differentiation. Likelihood ratio estimators yield unbiased gradients without requiring differentiable dynamics \cite{plyasunov2007efficient,gupta2013unbiased}, but exhibit variance that grows with trajectory length.
Finite difference methods provide an alternative, but their computational cost scales linearly with the number of parameters, even when using variance-reduction schemes that drastically improve convergence \cite{anderson2012efficient}.
Recent work introduced a continuous SSA relaxation enabling end-to-end differentiation \cite{rijal2023dga}, though at the cost of approximation errors in forward dynamics that accumulate over trajectory length, propagate into gradient estimates, and break permutation symmetry across reaction channels.

Here we introduce a reparameterized gradient estimator that enables efficient optimization while maintaining exact stochastic simulations. The approach preserves full fidelity of the Chemical Master Equation, retaining discrete reaction events and correct waiting time statistics in forward trajectories, while providing low-variance gradient estimates compatible with modern adaptive optimizers. We achieve this through straight-through Gumbel-Softmax estimation \cite{jang2017categorical,maddison2017concrete}, decoupling forward sampling from backward differentiation. In the forward pass, exact discrete samples are drawn according to the SSA. In the backward pass, gradients are propagated through a continuous relaxation of the categorical sampling operation. The resulting estimator exhibits substantially lower variance than likelihood ratio methods and avoids the approximation errors introduced by continuous relaxations of forward dynamics.

We demonstrate the framework on two problem classes spanning biophysics and non-equilibrium statistical mechanics. For parameter inference in stochastic gene expression, we first validate the approach on synthetic data, recovering kinetic rates of telegraph promoter models from both moment statistics and full steady-state probability distributions. We then fit a four-state promoter model to experimental smFISH timecourse data, inferring eight kinetic parameters simultaneously from time-resolved RNA count distributions. For inverse design in stochastic thermodynamics, we optimize non-equilibrium currents in an interacting particle system under kinetic resource constraints, recovering known theoretical bounds. These applications illustrate the generality of the approach: any scalar objective computable from stochastic trajectories becomes accessible to efficient gradient-based optimization.

\section{Straight-Through Gumbel-Softmax Pathwise Gradient Estimator}

\begin{figure}[t]
\includegraphics[width=.48\textwidth]{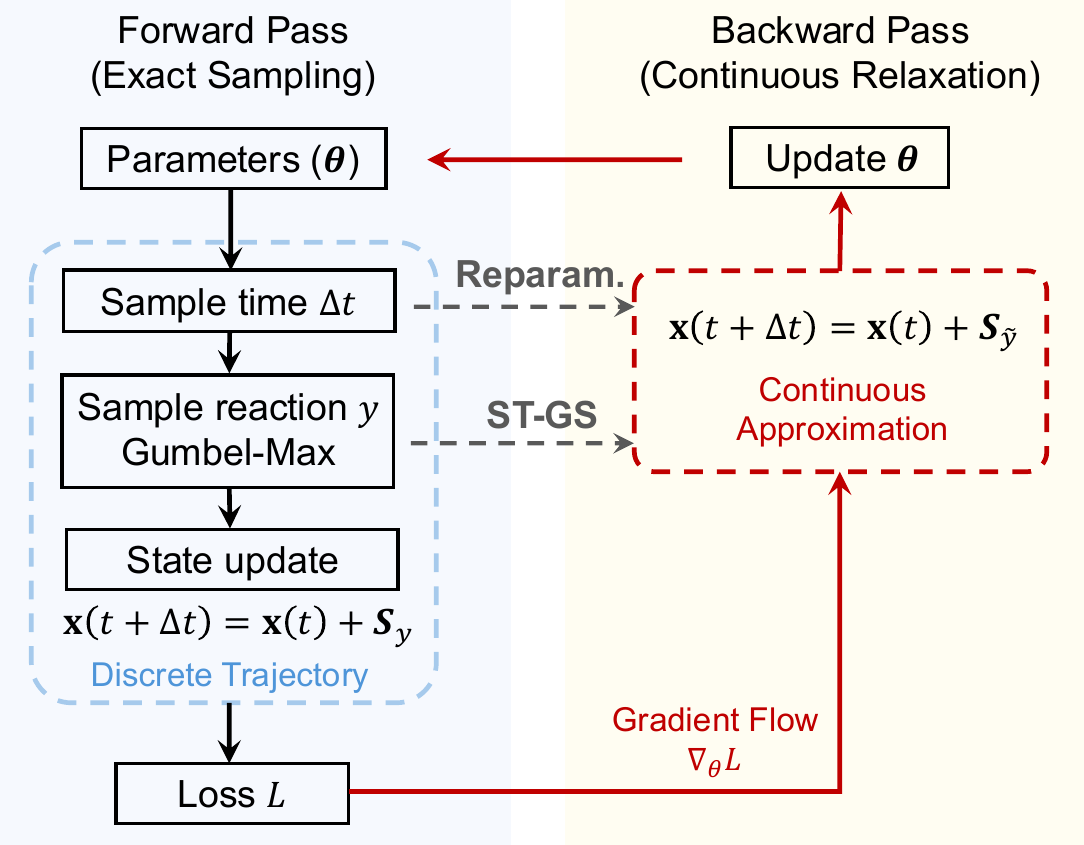}
\caption{\textbf{Modeling and Optimization Framework.} SSA trajectories are generated exactly in the forward pass by sampling waiting times and discrete reaction events (left, blue shaded box).
In the backward pass, gradients are computed by directly reparameterizing waiting times and applying a continuous Gumbel-Softmax relaxation to reaction selection (right, yellow shaded box), enabling efficient gradient-based optimization without approximating the stochastic dynamics.}
\label{fig:model}
\end{figure}

The Stochastic Simulation Algorithm involves two sampling operations at each step, selecting which reaction occurs and when it occurs (Fig.~\ref{fig:model}, left). Both can be written as deterministic functions of parameter-independent random draws, enabling gradients to flow through the simulation while preserving exact discrete dynamics (Fig.~\ref{fig:model}, right).

For waiting times, this reparameterization is standard. If $u \sim \mathcal{U}(0,1)$, then $\Delta t = -\log u / a_0(\theta)$ is exponentially distributed with rate equal to the total propensity $a_0(\theta) = \sum_j a_j(\theta)$. Since $\Delta t$ is an explicit function of the propensities, gradients flow through the waiting time automatically. The categorical sampling of reaction identity appears more problematic, since selecting reaction index $r$ with probability $\pi_r = a_r/a_0$ requires a discrete choice. However, this too admits reparameterization through the Gumbel-Max trick. If $u_k \sim \mathcal{U}(0, 1)$ are independent samples then $g_k = -\log(-\log u_k)$ follow the standard Gumbel distribution, and
$$
y = \text{one\_hot}\left(\underset{k}{\arg\max}\left(g_k + \log \pi_k\right)\right)
$$
samples from the correct categorical distribution. Expressing the selected reaction as a one-hot vector allows the state update to be written as a matrix-vector product with the stoichiometry matrix. The discrete outcome is now expressed as a deterministic function of fixed random inputs $\{g_k\}$ and the parameter-dependent reaction propensities $\{a_k(\theta)\}$.

\begin{figure*}[t]
\includegraphics[width=1.0\textwidth]{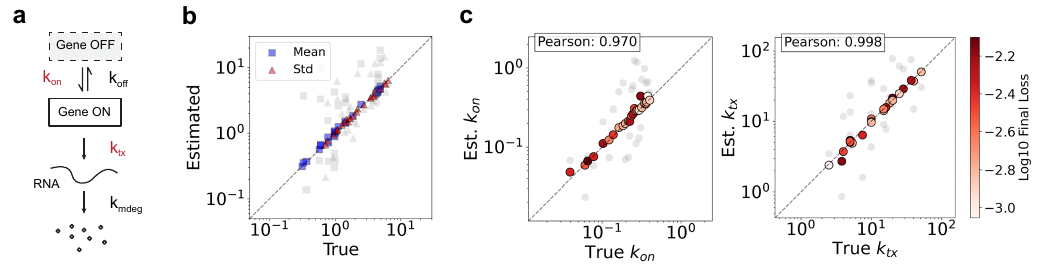}
\caption{\textbf{Moment matching inferences from synthetic data.} Inference of parameters from the mean and variance of the true underlying generative model at steady state, for 25 different parameter sets. Gray markers indicate the random initializations. (a) Two parameters of the telegraph model ($k_{\mathrm{on}}$, $k_{\mathrm{tx}}$) are simultaneously estimated. (b) Mean and standard deviation of the RNA levels at steady state for the true and optimized models. (c) Comparison between the estimated and true rates.}
\label{fig:moments}
\end{figure*}

The Gumbel-Max formulation explicitly exposes the source of non-differentiability: the $\arg\max$ operation has zero gradients almost everywhere. The \textit{straight-through Gumbel-Softmax} (ST-GS) estimator resolves this by employing different operations for forward and backward passes \cite{jang2017categorical,maddison2017concrete}. In the forward pass, we evaluate the exact $\arg\max$ to obtain a discrete one-hot vector $y$ specifying the selected reaction. In the backward pass, we substitute the softmax relaxation
$$
\tilde{y}_k = \frac{\exp\left((g_k + \log \pi_k)/\tau\right)}{\sum_{j} \exp\left((g_j + \log \pi_j)/\tau\right)}
$$
and backpropagate gradients through this continuous approximation. The temperature parameter $\tau$ controls the fidelity of the approximation. As $\tau \to 0$, the softmax concentrates toward the $\arg\max$, reducing bias but increasing gradient variance. Larger values of $\tau$ yield smoother gradients at the cost of a less accurate approximation to the true sensitivity. We find $\tau = 1.0$ provides a robust default across the applications considered in this work.

The resulting gradient estimator is biased, since derivatives computed through the softmax relaxation do not necessarily equal the true sensitivity of the discrete system. However, the bias is confined to the gradient calculation. Every forward trajectory remains an exact sample from the Chemical Master Equation, ensuring that optimized parameters are evaluated against faithful stochastic dynamics. A more thorough derivation is provided in Section~\ref{sec:stgs-derivation}.

\section{Synthetic benchmarks with a telegraph promoter model}

We first validate our inference framework on the telegraph promoter model, a canonical description of stochastic gene expression widely adopted in systems biology. First introduced by Peccoud and Ycart \cite{peccoud1995markovian}, the telegraph model has successfully explained experimental observations of transcriptional bursting across prokaryotic \cite{golding2005real,so2011general} and eukaryotic systems \cite{raj2006stochastic,suter2011mammalian}, and remains the basis for modern inference frameworks applied to single-cell RNA sequencing data \cite{gorin2022interpretable}.

The model describes a gene switching stochastically between two discrete promoter states: ON, permitting transcription, and OFF, where transcription is inhibited. The promoter transitions between states with rates $k_{\mathrm{on}}$ and $k_{\mathrm{off}}$. In the ON state, mRNA is produced at rate $k_{\mathrm{tx}}$, while mRNA degradation occurs at rate $k_{\mathrm{mdeg}}$. The system is represented by three reactions:
\begin{align*}
    \text{OFF} &\xrightleftharpoons[k_{\mathrm{off}}]{k_{\mathrm{on}}} \text{ON} \\
    \text{ON} &\xrightarrow{k_{\mathrm{tx}}} \text{ON} + \text{mRNA} \\
    \text{mRNA} &\xrightarrow{k_{\mathrm{mdeg}}} \emptyset
\end{align*}

We generate a diverse synthetic dataset by sampling 25 parameter sets from ranges consistent with bacterial transcriptional dynamics (\textit{Methods}). For each parameter set, we simulate the stochastic process until steady state and record the resulting single-cell RNA count distribution. These distributions serve as input for inference, allowing direct comparison between inferred and ground-truth parameters. We restrict our analysis to parameter regimes where the model is structurally identifiable in theory, isolating the performance of the inference method from fundamental limitations of the model itself.

\begin{figure*}[t]
\includegraphics[width=.9\textwidth]{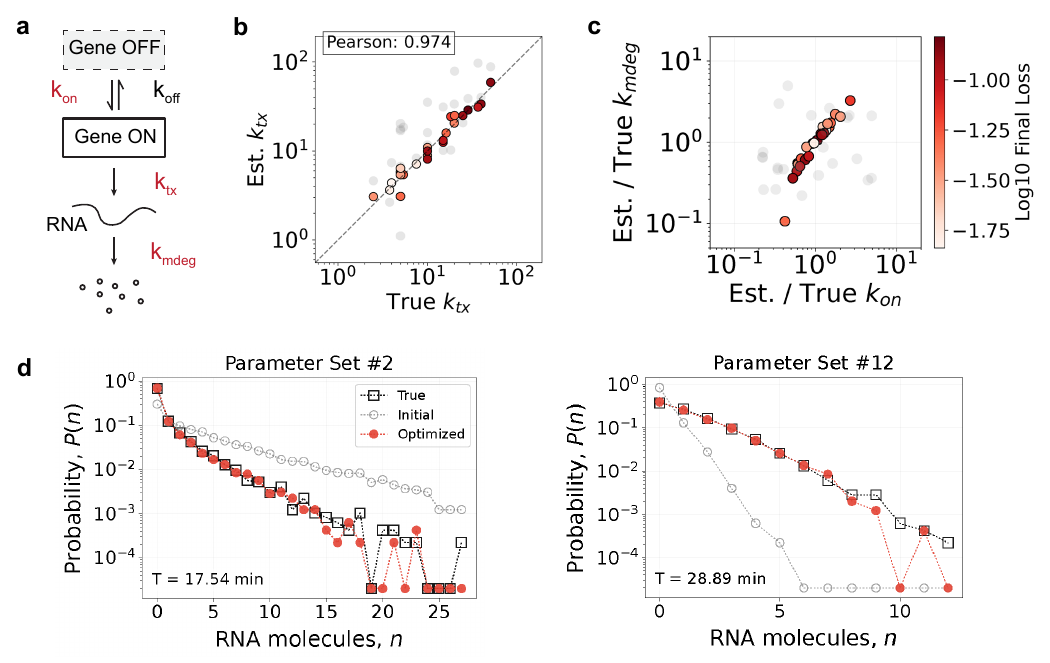}
\caption{\textbf{Distribution matching inferences from synthetic data.} Inference of parameters from the RNA histogram of the true underlying generative model at steady state, for 25 different parameter sets. Gray markers indicate the random initializations. (a) Three parameters of the telegraph model ($k_{\mathrm{on}}$, $k_{\mathrm{tx}}$, $k_{\mathrm{mdeg}}$) are simultaneously estimated. (b) Comparison between the estimated and true transcription rate. (c) Relative errors in the estimation of the degradation rate $k_{\mathrm{mdeg}}$ and the on-rate $k_{\mathrm{on}}$ illustrate sloppy parameter directions. (d) Two representative examples comparing the target data distribution (black) and the distribution generated by the optimized model (red), showing agreement over several orders of magnitude.}
\label{fig:synth-hist}
\end{figure*}

\subsection{Moment Matching Loss}

We first consider inferring two parameters of the telegraph model ($k_{\mathrm{on}}$ and $k_{\mathrm{tx}}$) while fixing the others to their ground-truth values (Fig.~\ref{fig:moments}a). We initialize the free parameters randomly within a log-uniform range spanning $[0.2\times, 5\times]$ their true values. With only two free parameters, matching the mean and variance of the stationary distribution uniquely determines the solution.

We minimize the following loss function:
$$
\mathcal{L} = \left[\log_{10}\left(\frac{\mu_{\text{true}}}{\mu_{\text{sim}}}\right)\right]^2 + \left[\log_{10}\left(\frac{\sigma^2_{\text{true}}}{\sigma^2_{\text{sim}}}\right)\right]^2
$$
where $\mu_{\text{true}}, \sigma^2_{\text{true}}$ and $\mu_{\text{sim}}, \sigma^2_{\text{sim}}$ are the mean and variance of the RNA copy number at steady state for the ground truth and simulated model, respectively. The logarithmic ratio formulation ensures that relative errors are penalized equally regardless of the absolute scale of each statistic. Rate constants are optimized in logarithmic space, enforcing positivity and normalizing sensitivity across parameters of different magnitudes.

The optimized models accurately reproduce target moments (Fig.~\ref{fig:moments}b), and recover the true values of $k_{\mathrm{on}}$ and $k_{\mathrm{tx}}$ across all 25 parameter sets (Fig.~\ref{fig:moments}c). Full results are reported in Fig.~\ref{fig:moment-matching-results}. We note that a previous study using automatic differentiation (but without the current methodology) on this class of problems achieved Pearson correlations of 0.68-0.74 using less challenging parameter regimes \cite{rijal2023dga}. 

Despite its apparent simplicity, this problem presents non-trivial optimization challenges. Although it admits a unique and analytically derivable minimum, the loss landscape is severely ill-conditioned around the global optimum, due to flat ridges corresponding to directions of low sensitivity~\cite{gutenkunst2007universally}. This poses significant challenges to first-order optimizers, even when using adaptive methods such as Adam~\cite{kingma2014adam}. Achieving full convergence required deviating from standard hyperparameter settings, as detailed in Methods. The moment matching problem can be reformulated into a better-conditioned problem by reparameterizing to fit burst frequency $k_{\mathrm{on}} k_{\mathrm{off}}/(k_{\mathrm{on}} + k_{\mathrm{off}})$ and mean burst size $k_{\mathrm{tx}}/k_{\mathrm{off}}$, effectively preconditioning the landscape. However, identifying such reparameterizations is generally non-trivial, and we demonstrate here that our method is sufficiently robust to solve the original ill-conditioned problem directly.

\begin{figure*}[t]
    \includegraphics[width=.95\textwidth]{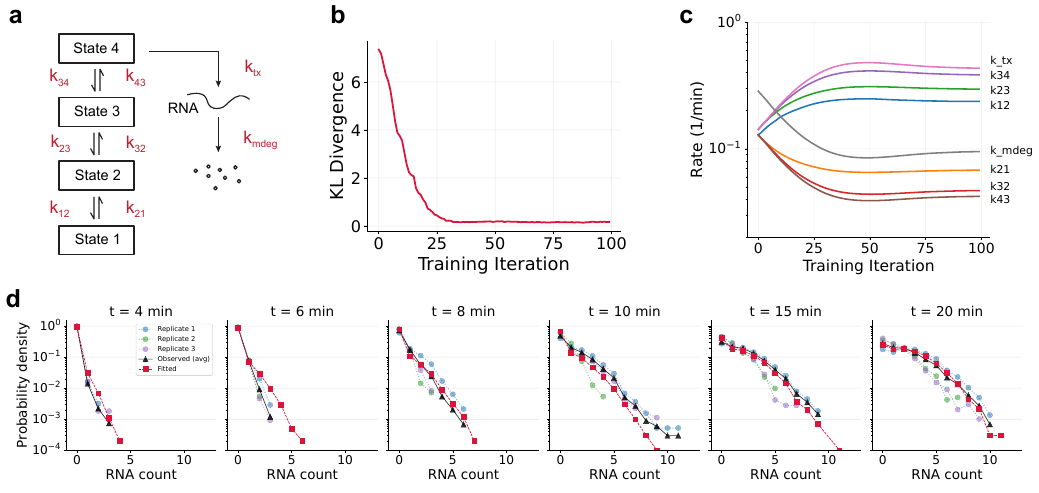}
    \caption{\textbf{Parameter inference from experimental single-molecule FISH timecourse.} Inference of kinetic parameters for a four-state promoter model from single-molecule FISH measurements of STL1 osmotic stress response in yeast. (a) Schematic of the four-state promoter model: the promoter transitions between states 1--4 with rates $k_{ij}$, transcription occurs from the final state $S_4$ at rate $k_{\mathrm{tx}}$, and mRNA gets depleted at rate $k_{\mathrm{mdeg}}$. (b) Kullback-Leibler divergence between observed and simulated RNA distributions during optimization. (c) Convergence of kinetic rates over training iterations. (d) Comparison between observed RNA count distributions (black triangles: replicate average; colored circles: individual replicates) and the fitted model (red squares) at six of the eight fitted time points (selected for clarity) following osmotic stress, on a logarithmic probability scale.}
    \label{fig:hist-data}
\end{figure*}

\subsection{Distribution Matching Loss}

We next consider inference from the full steady-state RNA copy-number distribution. The complete distribution contains substantially more information than moments alone and in principle can uniquely determine additional kinetic parameters. With three free parameters ($k_{\mathrm{on}}$, $k_{\mathrm{tx}}$, $k_{\mathrm{mdeg}}$) and the remaining rate fixed to its ground-truth value (Fig.~\ref{fig:synth-hist}a), the optimization problem admits a unique minimum. In practice, however, the loss landscape exhibits pronounced near-degeneracies: coupled changes in promoter switching and mRNA degradation can produce nearly indistinguishable steady-state distributions, making the global optimum difficult to locate numerically.

We quantify distributional mismatch using the 1-Wasserstein (Earth Mover's) distance, which for one-dimensional distributions can be calculated as the $L_1$ distance between two cumulative distribution functions~\cite{peyre2019computational}:
$$\mathcal{L} = \sum_{n} \left|F_{\text{true}}(n) - F_{\text{sim}}(n)\right|,
$$
where $F_{\text{true}}$ and $F_{\text{sim}}$ denote the empirical CDFs of the target and simulated RNA counts, respectively.

Optimizing this loss requires differentiating through the simulated RNA histogram. For this we employ a triangular kernel density estimator with bandwidth $h = 1.01$. For integer-valued data, this closely approximates standard hard binning while providing well-defined gradients (\textit{Methods}). Accurately resolving the distribution, particularly in the tails, requires a large number of simulations, but backpropagating through all of them would exceed even very loose memory constraints. We therefore compute distributional statistics from a large pool of forward-only simulations combined with a smaller set of gradient-tracked trajectories (\textit{Methods}). This decouples the sample size needed for accurate distribution estimates from the memory cost of gradient computation, with the benefit of reducing gradient variance. Given the stiffness of this inverse problem, we adopt a multi-start protocol, running three independent optimizations from random initializations spanning $[0.2\times, 5\times]$ the ground-truth values and reporting the solution with the lowest loss.

Across the 25 synthetic parameter sets, this procedure recovers $k_{\mathrm{tx}}$ accurately (Fig.~\ref{fig:synth-hist}b). The remaining parameters exhibit correlated deviations (Fig.~\ref{fig:synth-hist}c): errors in $k_{\mathrm{on}}$ and $k_{\mathrm{mdeg}}$ are anticorrelated, reflecting a near-degenerate direction in the loss landscape where changes in switching kinetics can be partially compensated by changes in degradation. This pattern is consistent with the ``sloppy parameter'' structure characteristic of kinetic models \cite{gutenkunst2007universally}. Importantly, despite the difficulty of pinpointing the unique optimum along this degenerate direction, the optimized models faithfully reproduce the full target distribution over several orders of magnitude in probability (Fig.~\ref{fig:synth-hist}d; full results in Figs.~\ref{fig:hist-match-all}--\ref{fig:hist-match-all-nolog}). This demonstrates that ST-GS gradients provide sufficient signal to solve challenging distributional inverse problems while maintaining exact SSA dynamics throughout.

\section{Parameter Inference from experimental gene expression timecourse distributions}

Inferring kinetic parameters of stochastic gene expression models from single-cell measurements is a central challenge in quantitative biology, with growing interest in fitting such models at scale \cite{gorin2025monod}. Because fitting the full RNA distribution shape---rather than summary statistics---is essential for producing mechanistic predictive models \cite{munsky2018distribution}, inference methods must be both computationally efficient and expressive enough to handle complex dynamics. Differentiable stochastic simulation addresses these requirements naturally, enabling direct gradient-based optimization over full RNA count distributions without moment closures, state-space truncations, or likelihood-free approximations.

We demonstrate this approach by fitting a four-state promoter model to single-molecule fluorescence \textit{in situ} hybridization (smFISH) data from the osmotic stress response of \textit{S.~cerevisiae}, using time-resolved measurements of STL1 nuclear mRNA following exposure to 0.4M NaCl \cite{li2019multiplex}. This extends the two-state telegraph model to capture the multi-step activation process underlying transcriptional induction (Fig.~\ref{fig:hist-data}a): the promoter transitions through a linear chain of four states with forward and reverse rates $k_{ij}$, accommodating effects such as chromatin remodeling and transcription factor binding. Transcription occurs exclusively from the final state $S_4$ at rate $k_{\mathrm{tx}}$, and nuclear mRNA is depleted at rate $k_{\mathrm{mdeg}}$ through degradation and nuclear export. This four-state topology is a simplified instance of the richer architecture originally developed to explain these data \cite{neuert2013systematic,munsky2018distribution}, and has been independently supported by Bayesian model selection \cite{kilic2023gene}.

We restrict the analysis to the rising phase of the transcriptional response (0--20 min after stress), during which the upstream Hog1 kinase signal is active and rates can be approximated as constant. The data comprise smFISH measurements of nuclear STL1 mRNA counts at eight time points ($t = 0, 2, 4, 6, 8, 10, 15, 20$ min), pooled across three biological replicates. All cells are initialized in state $S_1$ and with zero mRNA molecules, reflecting the basal transcriptionally silent condition prior to osmotic stress. To match experimental count distributions we minimize the Kullback--Leibler divergence, a commonly used loss for distribution fitting, between observed and simulated RNA histograms simultaneously across all eight time points. Simulated distributions are constructed from a combined pool of gradient-tracked and baseline trajectories, as described in \textit{Distribution Matching}. All eight kinetic parameters---the six inter-state transition rates $k_{ij}$, the transcription rate $k_{\mathrm{tx}}$, and the mRNA degradation rate $k_{\mathrm{mdeg}}$---are optimized simultaneously in log-space using the Adam optimizer with default hyperparameters, reflecting the standard choices we expect most practitioners to adopt.

The optimization converges within approximately 100 iterations (Fig.~\ref{fig:hist-data}b), requiring less than five minutes of wall-clock time on a single NVIDIA A100 GPU. Kinetic rate constants separate into physically meaningful values (Fig.~\ref{fig:hist-data}c), and the fitted model captures the time evolution of the full mRNA count distributions, reproducing both the dominant zero-count peak at early times and the gradual emergence of higher copy numbers as the transcriptional response develops (Fig.~\ref{fig:hist-data}d; Figs.~\ref{fig:smfish-fitted}--\ref{fig:smfish-replicates-linear}).

The fitted values (Table~\ref{tab:smfish-rates}) are physically compatible with an induced yeast stress gene: the nuclear mRNA decay rate $k_{\mathrm{mdeg}} = 0.095~\mathrm{min^{-1}}$ implies a $\sim 7.3$~min half-life, and the active-state production rate $k_{\mathrm{tx}} = 0.434~\mathrm{min^{-1}}$ corresponds to one transcript every $\sim 2.3$~min while transcription-competent. These values are of the same order of magnitude as previous estimates for this system, though direct comparison is complicated by differences in model structure \cite{munsky2018distribution}. The promoter progression is strongly biased toward activation, with forward/reverse ratios increasing along the chain ($k_{12}/k_{21} \approx 3.50$, $k_{23}/k_{32} \approx 6.33$, $k_{34}/k_{43} \approx 9.13$), consistent with increasingly committed steps over the induction window.

\begin{figure*}[t]
    \includegraphics[width=\textwidth]{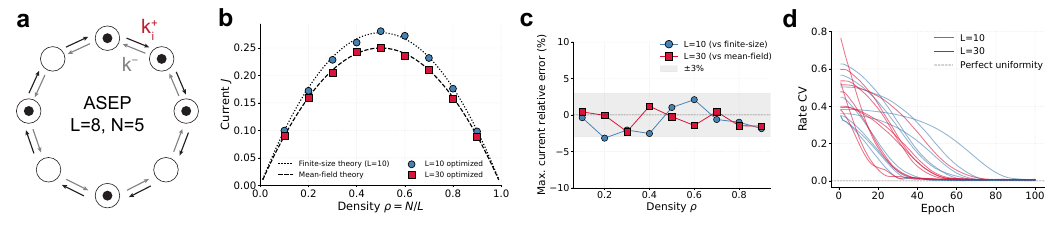}
    \caption{\textbf{Fundamental diagram of the Asymmetric Simple Exclusion Process.} Optimization of particle current in the ASEP on a periodic lattice. (a) Schematic of the ASEP with $L$ sites and $N$ particles, each particle hops forward with rate $k_i^+$ and backward with rate $k^-$. (b) Current $J$ versus particle density $\rho = N/L$ for $L=10$ (blue circles) and $L=30$ (red squares), compared to finite-size theory (dotted line) and mean-field theory (dashed line). (c) Relative error in optimized current versus density, shaded band indicates $\pm 3\%$. (d) Coefficient of variation (CV) of optimized forward rates during training, showing convergence toward uniform rate allocation.}
    \label{fig:asep}
\end{figure*}

\section{Inverse design in nonequilibrium stochastic thermodynamics}

We next apply our framework to an inverse design problem in stochastic thermodynamics \cite{seifert2012stochastic}: optimizing steady-state current in an asymmetric simple exclusion process (ASEP) \cite{spitzer1970Interaction, derrida1998exactly} on a periodic lattice under a kinetic resource constraint. The ASEP couples exclusion interactions to nonequilibrium driving on a minimal many-body model with an analytically characterized optimum, providing a rigorous benchmark for gradient-based optimization of interacting particle systems.

We study a reversible ASEP ring with $L$ sites and hard-core exclusion, occupied by $N$ particles at density $\rho = N/L$ (Fig.~\ref{fig:asep}a). Each bond $(i,i{+}1)$ supports both forward and backward hopping events:
\begin{align*}
    P_i + H_{i+1} &\xrightleftharpoons[k^-]{k_i^+} H_i + P_{i+1},
\end{align*}
where $P_i$ and $H_i$ denote a particle and a hole at site $i$, respectively. The forward rates $\{k_i^+\}_{i=1}^L$ are the design variables, while the backward rate $k^-$ is fixed and homogeneous across bonds, establishing a reference scale that removes the degeneracy between absolute rates and their ratios. Breaking detailed balance induces a net stationary particle current $J$ around the ring, with corresponding entropy production rate $\sigma = J\,\mathcal{A}$, where $\mathcal{A} = \sum_{i=1}^L \log(k_i^+/k^-)$ is the cycle affinity.

The inverse design objective is to maximize $J$ subject to a fixed average forward hopping budget $\bar{k}^+ = 1/L\sum_i k_i^+$. This prevents the trivial strategy of uniformly increasing all rates and isolates the question of how to optimally distribute limited kinetic resources across bonds.

The exact cycle current (net reaction counts divided by time) is not differentiable with respect to rate parameters. We instead estimate it from simulated trajectories as
$$
\hat{J} = \frac{1}{T}\sum_n \left[\frac{1}{L} \sum_{i=1}^L \left(a_i^+(t_n) - a_i^-(t_n)\right)\right] \Delta t_n,
$$
where $a_i^\pm(t_n)$ are the forward and backward propensities at bond $i$ and $\Delta t_n$ is the waiting time at step $n$. This time-weighted propensity average is differentiable and provides gradients with respect to all rate parameters, with the entropy production estimate following as $\hat{\sigma} = \hat{J}\,\mathcal{A}$. We optimize a maximum entropy production objective with a logarithmic penalty enforcing the forward-rate budget:
$$
\mathcal{L} = -\hat{\sigma} + \lambda_K\left[\log\frac{1/L \sum_i k_i^+}{\bar{k}^+}\right]^2.
$$

All optimizations start from the same random initial rates and are run to stable convergence using the modified Adam optimizer described in \textit{Methods}.

For a ring with exclusion interactions and fixed mean forward rate $\bar{k}^+$, the steady-state current is maximized when the forward rates are uniform, $k_i^+ \equiv \bar{k}^+$, yielding the fundamental diagram
$$
J^*(\rho) = (\bar{k}^+ - k^-)\rho(1-\rho),
$$
with a finite-size correction of $L/(L-1)$ for a ring of $L$ sites. The parabolic density dependence reflects the competition between particle availability ($\rho$) and vacant sites for hopping ($1-\rho$), with maximum current at half-filling.

The optimized currents closely track the theoretical predictions across the full range of densities (Fig.~\ref{fig:asep}b), following the finite-size prediction for $L=10$ and approaching the mean-field limit for $L=30$. Relative errors in optimized current remain within $\pm 3\%$ across all densities for both system sizes (Fig.~\ref{fig:asep}c), demonstrating quantitative agreement with the analytical predictions.
The optimization recovers both the optimal current magnitude and the optimal rate allocation strategy. Starting from random initializations with CV $\approx 0.3$--$0.7$, the coefficient of variation of the optimized forward rates decreases monotonically during training and approaches zero within 100 updates, confirming convergence to the theoretically optimal uniform allocation (Fig.~\ref{fig:asep}d). This holds reliably across both system sizes.

It is worth noting the scale of this optimization problem. For $L=30$ at half-filling, the underlying state space comprises $\binom{30}{15} \approx 10^8$ configurations, making master-equation-based optimization prohibitively expensive in general. Differentiable simulation sidesteps this problem by operating directly on sampled trajectories.

A complementary benchmark---characterizing the Pareto-optimal current-dissipation tradeoff in a three-state ring under simultaneous entropy production and kinetic budget constraints---is presented in Section~\ref{sec:three-state-ring}, where the optimized solutions closely track the analytically derived bound across two orders of magnitude in dissipation.

\section{Discussion}

The ST-GS estimator introduces a biased approximation to gradient computation, controlled by a single temperature hyperparameter $\tau$. Crucially, this bias affects only the backward pass: forward trajectories remain exact samples from the Chemical Master Equation throughout optimization. While a complete theoretical characterization of the bias remains an open question, the default value $\tau=1$ empirically yields sufficiently accurate gradients for effective parameter recovery across diverse kinetic regimes and close agreement with analytical predictions, without requiring problem-specific tuning.

With differentiable but exact SSA, any differentiable objective computable from stochastic trajectories becomes in principle accessible to efficient gradient-based optimization, including complex objectives defined over full probability distributions rather than summary statistics alone. The approach enables all of the optimizations demonstrated here to converge in under 15 minutes on a single GPU --- often in a matter of minutes --- with the potential of substantially accelerating iterative cycles of model refinement and hypothesis testing. Because computational cost scales with the number of sampled trajectories rather than with state-space size, the method extends naturally to systems with large configuration spaces that may prove challenging for alternative approaches.

These capabilities complement existing inference strategies, each of which involves different tradeoffs: explicit state-space representation for Finite State Projection~\cite{munsky2006fsp}, growing variance with trajectory length for likelihood ratio estimators~\cite{plyasunov2007efficient} and poor scaling to high-dimensional parameter spaces for Approximate Bayesian Computation~\cite{toni2009approximate}. The ST-GS estimator trades unbiased gradients for exact forward dynamics, and through reverse-mode automatic differentiation provides gradient cost independent of parameter dimensionality --- a distinct and practically useful point in this space of methods.

Our method also presents specific practical limitations. Backpropagating through long trajectories requires storing all intermediate states, imposing memory constraints that necessitate strategies such as the gradient-tracked and baseline trajectory split used here. Objectives that are not naturally differentiable --- such as histograms and cycle currents --- require carefully designed differentiable surrogates, and identifying such surrogates may not be straightforward for new problem classes. The ST-GS gradient bias, while empirically benign at $\tau=1$ across all applications considered here, lacks theoretical convergence guarantees, and no principled criterion exists for selecting the temperature in new settings. More broadly, differentiability does not eliminate the intrinsic difficulty of inverse problems in stochastic kinetic systems, whose loss landscapes typically exhibit flat ridges and sloppy parameter directions \cite{gutenkunst2007universally}.

Several extensions are natural. Reverse-mode automatic differentiation scales independently of parameter dimensionality, enabling in principle inference in larger mechanistic models. Concurrent independent work applied the same estimator to optimization benchmarks in high-dimensional parameter spaces, confirming its scalability~\cite{vilar2026exact}. Time-dependent rate parameters would extend the framework to non-stationary processes, and integration with automatic differentiation variational inference techniques~\cite{kucukelbir2017automatic} could enable full posterior estimation. More broadly, stochastic kinetic models arise throughout epidemiology, ecology, neuroscience, and the social sciences. The ability to efficiently differentiate through exact stochastic simulations provides a foundation for scalable inference and rational design across these domains.

\section*{Methods}

End-to-end differentiable, hardware-accelerated simulations were coded using the JAX Python library \cite{jax2018github} and other libraries in the JAX ecosystem, in particular Equinox \cite{kidger2021equinox} as the simulation backbone. JAX provides automatic differentiation, just-in-time compilation and automatic vectorization that greatly improve computational performance of Monte Carlo gradient estimation. We provide a succinct overview of the main components of our model. A full mathematical account of simulation and optimization details is given in the \textit{Supplementary Information}.

\subsection*{Synthetic Data}

We generate 25 parameter sets $\{k_{\mathrm{on}}, k_{\mathrm{off}}, k_{\mathrm{tx}}, k_{\mathrm{mdeg}}\}$ chosen to span a range of biologically plausible kinetic regimes characteristic of bacterial gene expression. The off-rate is fixed at $k_{\mathrm{off}} = 5.0~\text{min}^{-1}$, corresponding to a mean ON-state duration ($1/k_{\mathrm{off}}$) of approximately 12 seconds. This implies short transcriptional bursts, consistent with rapid transcription factor binding and unbinding dynamics. The mRNA degradation rate $k_{\mathrm{mdeg}}$ ranges from 0.12 to 0.7~min$^{-1}$, corresponding to half-lives of approximately 1--6 minutes. This range encompasses typical \textit{E.~coli} mRNA lifetimes, where the average half-life is 2--5 minutes. The transcription rate $k_{\mathrm{tx}}$ spans 2.5--50~min$^{-1}$. The lower end represents standard constitutive promoters, while rates approaching 50~min$^{-1}$ (nearly one initiation per second) correspond to highly active promoters such as those driving ribosomal RNA synthesis or fully induced viral genes. These rates produce mean burst sizes ranging from 0.5 to 10 transcripts per burst, consistent with experimental measures in prokaryotic systems. The resulting mean mRNA copy numbers at steady state range from approximately 0.3 to 5 transcripts per cell, typical of low-to-moderate expression bacterial genes where the majority of transcripts are present at fewer than 10 copies per cell. This bursty, low-copy number regime represents a challenging test case for inference due to the inherent sparsity and stochasticity of the data.

\subsection*{Differentiable Histograms}

\noindent \textbf{Triangular KDE.} To compare simulated and target distributions, we construct differentiable histograms using triangular kernel density estimation (KDE). For a batch of $N_\text{tot}$ simulation outputs $\{S_i\}_{i=1}^{N_\text{tot}}$ and $B$ histogram bins with centers $\{c_j\}_{j=1}^B$, we first compute raw kernel affinities with bandwidth $h$:
$$
k_{ij} = \max\left(0, 1 - \frac{|S_i - c_j|}{h}\right)
$$
Soft bin assignments are obtained by normalizing per sample, $w_{ij} = k_{ij} / \sum_{l} k_{il}$. This normalization ensures conservation of probability mass for samples near the grid boundaries, without biasing the estimation in the bulk. We employ a bandwidth slightly larger than the bin width, corresponding to $h=1.01$ for integer data. This preserves small but nonzero kernel overlaps at neighboring bin centers, ensuring that gradients remain well-defined everywhere while introducing negligible smoothing bias. The resulting estimator closely recovers the exact probability mass function for integer data, in contrast e.g. to the more common Gaussian kernel which introduces broader smoothing artifacts.

\noindent \textbf{Dirichlet smoothing.} The histogram count for bin $j$ is $h_j = \sum_{i=1}^{N_\text{tot}} w_{ij}$. To ensure full support and numerical stability, we apply Dirichlet smoothing \cite{bishop2006pattern} with per-bin pseudocount $\alpha$ and total number of bins $B$:
$$
\hat{p}_j = \frac{h_j + \alpha}{\sum_{k=1}^B h_k + \alpha B}
$$
This regularization prevents zero-probability singularities (e.g., in KL divergence) with minimal distortion to the distribution, avoiding the significant bias introduced by naive background mass addition.

\noindent \textbf{Variance reduction.} Accurately resolving the distribution, particularly in the tails, requires a large number of simulations, but backpropagating through all of them would exceed memory constraints. We therefore compute distributional statistics from a combined pool of $N$ gradient-tracked simulations and $M$ forward-only baseline simulations (detached from the computation graph), with $N_\text{tot} = N+M$. This decouples the sample size needed for accurate distribution estimates from the memory cost of gradient computation. As an additional benefit, this yields lower-variance gradient estimates at the cost of a controlled bias that is effectively absorbed by adaptive learning rate methods. A full account of the bias-variance trade-off is provided in Section~\ref{sec:bias-variance}.

\subsection*{Optimization}

The gradient estimates from the ST-GS estimator are used to optimize the simulation parameters via gradient-based methods. We employ the Adam optimizer \cite{kingma2014adam}, a modern adaptive variant of stochastic gradient descent (SGD). Adam maintains exponential moving averages of gradients and their second moments to adapt learning rates on a per-parameter basis.

We found that the specific geometry of the loss landscapes in kinetic model inference requires careful optimizer tuning. These landscapes typically feature deep, narrow, and flat ridges (manifolds of sloppy parameters \cite{gutenkunst2007universally}) surrounded by steep walls. When initialization places parameters far from the optimum, initial gradients are extremely large, causing rapid accumulation of momentum and second-moment estimates in adaptive optimizers like Adam. This leads to two critical issues: (i) high accumulated momentum causes severe oscillations upon reaching the low-gradient valley, and (ii) the second-moment estimator ``remembers'' the large initial gradients, excessively dampening the effective learning rate once in the flat region where gradients are small. To mitigate these effects, we employ Adam with modified hyperparameters: $\beta_1=0.8$ (reduced momentum) to minimize oscillations, and $\beta_2=0.9$ (short-term memory for dampening) to allow the variance estimate to adapt rapidly to the drastic change in gradient magnitude between the steep walls and the flat valley. 

\acknowledgments{We thank Aidan Zentner and all members of the Brenner Group for valuable feedback and discussions. This work was supported by NSF AI Institute of Dynamic Systems 2112085, the Harvard MRSEC (NSF DMR-2011754) and the Office of Naval Research through grant number ONR N00014-23-1-2654.}

\section*{Data Availability}
The Python library written to conduct this study is available at \href{https://github.com/fmottes/stochastix/}{https://github.com/fmottes/stochastix/}. Code and data for reproducing the experiments in this paper are available at \href{https://github.com/fmottes/stochastix-paper/}{https://github.com/fmottes/stochastix-paper/}.

\bibliography{Refs/refs}


\clearpage
\onecolumngrid
\newgeometry{left=1.25in, right=1.25in, top=1in, bottom=1in}

\setcounter{equation}{0}
\setcounter{figure}{0}
\setcounter{table}{0}
\setcounter{section}{0}

\renewcommand{\theequation}{S\arabic{equation}}
\renewcommand{\thefigure}{S\arabic{figure}}
\renewcommand{\thetable}{S\arabic{table}}
\renewcommand{\thesection}{S\arabic{section}}

\begin{center}
\textbf{\Large SUPPLEMENTARY INFORMATION}\\[0.7em]
\textbf{\large Gradient-based optimization of exact stochastic kinetic models }\\[1em]
Francesco Mottes, Qian-Ze Zhu, and Michael P. Brenner\\[0.5em]
\textit{School of Engineering and Applied Sciences, Harvard University, Cambridge MA 02138}
\end{center}
\vspace{1em}

\section{Details of Straight-Through Gumbel-Softmax Gradient Estimation}
\label{sec:stgs-derivation}

Stochastic trajectories are generated using the Gillespie Direct Method \cite{gillespie1977exact}, an exact algorithm for sampling from the probability distribution defined by the Chemical Master Equation. For a system in state $\mathbf{x}$ at time $t$, the probability density of reaction $j$ occurring at time $t+\tau$ is
$$
p(j, \tau \mid \mathbf{x}, t) = a_j(\mathbf{x}; \theta) \exp\left(-a_0(\mathbf{x}; \theta)\tau\right),
$$
where $a_j(\mathbf{x}; \theta)$ is the propensity of reaction $j$ in state $\mathbf{x}$ with parameter set $\theta$, and $a_0(\mathbf{x}; \theta) = \sum_{j} a_j(\mathbf{x}; \theta)$ is the total propensity. The Direct Method involves two sampling operations at each step. First, the waiting time $\Delta t$ is drawn from an exponential distribution with rate $a_0$ via inverse transform sampling,
$$
\Delta t = -\frac{\ln u}{a_0},
$$
with $u \sim \mathcal{U}(0, 1)$. Second, the reaction index $r$ is drawn from a categorical distribution with probabilities $\pi_k = a_k / a_0$. The system state is then updated according to the selected reaction,
$$
\mathbf{x}(t + \Delta t) = \mathbf{x}(t) + \mathbf{S}_{:,r},
$$
where $\mathbf{S}_{:,r}$ denotes the $r$-th column of the stoichiometry matrix $\mathbf{S}$, representing the net change in species counts due to reaction $r$.

The waiting time $\Delta t$ is already expressed as a differentiable function of the parameters through $a_0(\theta)$, with the stochastic component $u$ held fixed. The categorical sampling of reaction identity requires additional treatment. The Gumbel-Max trick expresses the discrete sample as a deterministic function of parameter-independent random draws. If $u_k \sim \mathcal{U}(0, 1)$ are independent samples and $g_k = -\log(-\log u_k)$ are the corresponding standard Gumbel variates, then
$$
y = \text{one\_hot}\left(\underset{k}{\arg\max}\left(g_k + \log \pi_k\right)\right)
$$
samples from the categorical distribution with probabilities $\pi_k = a_k/a_0$. The one-hot vector $y$ satisfies $y_j = \delta_{j,r}$, where $\delta_{j,r}$ is the Kronecker delta and $r$ is the selected reaction index. The state update can then be written as
$$
\mathbf{x}(t + \Delta t) = \mathbf{x}(t) + \mathbf{S} y,
$$
expressing the update as a matrix-vector product rather than an indexing operation.

The $\arg\max$ operation has zero gradients almost everywhere. The Straight-Through Gumbel-Softmax (ST-GS) estimator approximates gradients using the continuous Gumbel-Softmax relaxation,
$$
\tilde{y}_k = \frac{\exp\left((g_k + \log \pi_k)/\tau\right)}{\sum_{j} \exp\left((g_j + \log \pi_j)/\tau\right)},
$$
where the temperature parameter $\tau > 0$ controls the smoothness of the approximation. As $\tau \to 0$, the softmax output concentrates toward the $\arg\max$. The straight-through estimator uses the discrete sample $y$ in the forward pass while backpropagating gradients through the continuous approximation $\tilde{y}$. This is implemented as
$$
y_{\text{ST}} = \tilde{y} + \text{stop\_gradient}(y - \tilde{y}),
$$
where $\text{stop\_gradient}(\cdot)$ denotes an operation that evaluates to its argument in the forward pass but has zero gradient in the backward pass. In the forward pass, $y_{\text{ST}} = y$ since the stop-gradient term evaluates to $y - \tilde{y}$. In the backward pass, the gradient flows only through $\tilde{y}$.

Using the normalized probabilities $\pi_k = a_k/a_0$ rather than raw propensities $a_k$ in the softmax relaxation ensures that gradients correctly capture the constraint $\sum_k \pi_k = 1$. Specifically, increasing one propensity $a_j$ affects all probabilities through the normalization,
$$
\frac{\partial \pi_k}{\partial a_j} = \frac{\delta_{k,j}}{a_0} - \frac{a_k}{a_0^2} = \frac{1}{a_0}\left(\delta_{k,j} - \pi_k\right),
$$
and this coupling is correctly propagated when gradients are computed through $\log \pi_k$.

\clearpage
\section{Variance-Reduced Gradient Estimation with Baseline Simulations}
\label{sec:bias-variance}

We optimize model parameters $\theta$ by minimizing a loss function $L(\theta)$ that quantifies the discrepancy between target statistics and distributions generated by stochastic simulations $S(\theta)$. The loss involves expectations over simulation outputs, $L(\theta) = f(\mathbb{E}[g_1(S)], \mathbb{E}[g_2(S)], \dots)$, which we estimate via Monte Carlo sampling to yield $L_{MC}(\theta)$. Standard gradient descent on $\nabla_\theta L_{MC}(\theta)$ often exhibits high variance, impeding stable convergence.

We implement a variance reduction technique that introduces a controlled bias in exchange for substantially lower gradient variance. At each optimization step, we generate two distinct simulation sets: $N$ gradient-tracked simulations $\{S_i(\theta)\}_{i=1}^N$ with retained computation graphs, and $M$ baseline simulations $\{B_j(\theta)\}_{j=1}^M$ detached from the gradient computation (e.g., via \texttt{jax.lax.stop\_gradient} in JAX).

We compute all empirical estimators using the combined pool of $N+M$ simulations. For a function $g(S)$, the estimator is:
$$
\hat{\mathbb{E}}[g(S)] = \frac{1}{N+M} \left( \sum_{i=1}^N g(S_i(\theta)) + \sum_{j=1}^M g(B_j(\theta)) \right)
$$
This yields low-variance estimates for the summary statistics, with Monte Carlo error scaling as $(N+M)^{-1/2}$. When computing $\nabla_\theta L_{MC}(\theta)$, the baseline simulations act as constants:
$$
\nabla_\theta \hat{\mathbb{E}}[g(S)] = \frac{1}{N+M} \sum_{i=1}^N \nabla_\theta g(S_i(\theta))
$$

This approach decouples the precision of the loss evaluation (using $N+M$ samples) from the cost of gradient computation (using $N$ samples). 

\noindent \textbf{Bias characterization.} The resulting gradient estimator is biased. To derive the bias, we first note that the reparameterization underlying the ST-GS estimator allows interchange of differentiation and expectation (the pathwise gradient identity):
$$
\nabla_\theta \mathbb{E}[g(S)] = \mathbb{E}[\nabla_\theta g(S)].
$$
The standard Monte Carlo estimator using $N$ samples yields an unbiased estimate:
$$
\frac{1}{N} \sum_{i=1}^N \nabla_\theta g(S_i)
\quad \Rightarrow \quad
\mathbb{E}\left[\frac{1}{N} \sum_{i=1}^N \nabla_\theta g(S_i)\right]
= \mathbb{E}[\nabla_\theta g(S)]
= \nabla_\theta \mathbb{E}[g(S)].
$$
In contrast, our estimator uses only $N$ gradient-tracked samples but normalizes by $N+M$. Denoting this estimator by $\hat{\Sigma}_{\text{biased}}$:
$$
\hat{\Sigma}_{\text{biased}} = \frac{1}{N+M} \sum_{i=1}^N \nabla_\theta g(S_i).
$$
Taking expectations over the simulation randomness gives
$$
\mathbb{E}\left[\hat{\Sigma}_{\text{biased}}\right]
= \frac{N}{N+M} \, \mathbb{E}[\nabla_\theta g(S)]
= \frac{N}{N+M} \, \nabla_\theta \mathbb{E}[g(S)].
$$
Thus the estimator is shrunk by a factor $N/(N+M)$. 

This characterization holds exactly for estimating sensitivities of the form $\nabla_\theta \mathbb{E}[g(S)]$. In the full loss $L = f(\mathbb{E}[g(S)])$, additional bias can arise from the nonlinearity of $f$. However, in the large-$M$ limit, the forward estimate of $\mathbb{E}[g(S)]$ becomes effectively deterministic, and the multiplicative shrinkage derived above becomes the dominant source of bias.

\noindent \textbf{Variance reduction.} 
To analyze the variance reduction, we decompose the gradient using the chain rule. For simplicity, consider a loss depending on a single expectation, $L = f(\mathbb{E}[g(S)])$. The gradient is:
$$
\nabla_\theta L = \underbrace{\frac{\partial f}{\partial \mathbb{E}[g(S)]}}_{\text{Error Signal } E} \cdot \underbrace{\nabla_\theta \mathbb{E}[g(S)]}_{\text{Sensitivity } \Sigma}
$$

Both terms benefit from using the larger sample pool: the first through reduced error signal variance $\mathcal{O}(1/(N+M))$, the second through reduced sensitivity variance $\mathcal{O}(N/(N+M)^2)$.

\paragraph{Error signal variance.} The error signal $E$ depends on the estimated expectation $\hat{\mathbb{E}}[g(S)]$. Using all $N+M$ samples:
$$
\text{Var}(\hat{\mathbb{E}}[g(S)]) = \frac{\text{Var}(g(S))}{N+M}
$$
By the delta method, the variance of $E = f'(\hat{\mathbb{E}}[g(S)])$ scales as:
$$
\text{Var}(E) \approx \left(f''(\mathbb{E}[g(S)])\right)^2 \frac{\text{Var}(g(S))}{N+M} = \mathcal{O}\left(\frac{1}{N+M}\right)
$$
compared to $\mathcal{O}(1/N)$ when using only the $N$ gradient-tracked samples.

\paragraph{Sensitivity variance.} For the unbiased (bias-corrected) estimator, the sensitivity estimate is:
$$
\hat{\Sigma}_{\text{unbiased}} = \frac{1}{N} \sum_{i=1}^N \nabla_\theta g(S_i(\theta))
$$
with variance $\text{Var}(\hat{\Sigma}_{\text{unbiased}}) = \text{Var}(\nabla_\theta g(S))/N = \mathcal{O}(1/N)$.

For the biased estimator:
$$
\hat{\Sigma}_{\text{biased}} = \frac{1}{N+M} \sum_{i=1}^N \nabla_\theta g(S_i(\theta))
$$
The variance is:
$$
\text{Var}(\hat{\Sigma}_{\text{biased}}) = \frac{1}{(N+M)^2} \cdot N \cdot \text{Var}(\nabla_\theta g(S)) = \frac{N}{(N+M)^2} \text{Var}(\nabla_\theta g(S)) = \mathcal{O}\left(\frac{N}{(N+M)^2}\right)
$$

Note that if we rescale $\hat{\Sigma}_{\text{biased}}$ by $(N+M)/N$ to remove the shrinkage bias, its variance returns to $\mathcal{O}(1/N)$; thus the primary variance reduction comes from stabilizing the error signal $E$ via the larger $N+M$ sample pool.

\noindent \textbf{Practical considerations.} During optimization, the multiplicative shrinkage factor $N/(N+M)$ in the sensitivity estimate is effectively absorbed by adaptive optimizers such as Adam: if the gradient is scaled by an approximately constant factor, Adam's moment normalization renders the update nearly scale-invariant. Empirically, we find that optimization proceeds well without explicit corrections.

While this technique can stabilize the estimation of all summary statistics, it is particularly useful for differentiable histogram-based objectives, which require large sample sizes to accurately populate bins. Memory constraints often limit the number of concurrent gradient-tracked simulations, but forward-only simulations are much cheaper in both memory and computation time. By augmenting a small set of gradient-tracked trajectories with a large pool of forward-only simulations, we can greatly reduce Monte Carlo noise in the histogram---and hence in the error signal $E$---without exceeding hardware limits.

\clearpage
\section{Moment Matching Loss -- Additional Details}
\label{sec:supp-mom-match}

\textbf{Experiment Details.} We inferred two parameters of the telegraph promoter model ($k_\text{on}$ and $k_\text{tx}$) from the mean and variance of steady-state RNA copy numbers, with $k_\text{off}$ and $k_\text{mdeg}$ fixed to their ground-truth values. Each optimization used 512 gradient-tracked simulations combined with 1000 baseline (forward-only) simulations per epoch for variance reduction. Parameters were initialized randomly within $[0.2\times, 5\times]$ the true values. Rate constants were parameterized as $k = e^\phi$ and optimization was performed over the log-transformed variables $\phi$. Training ran for 100 epochs using Adam with learning rate $\text{lr} = 0.1$ and modified hyperparameters ($\beta_1 = 0.8$, $\beta_2 = 0.9$).

\begin{figure}[h!]
\centering
\begin{minipage}[t]{0.33\textwidth}
\centering
\includegraphics[width=\textwidth]{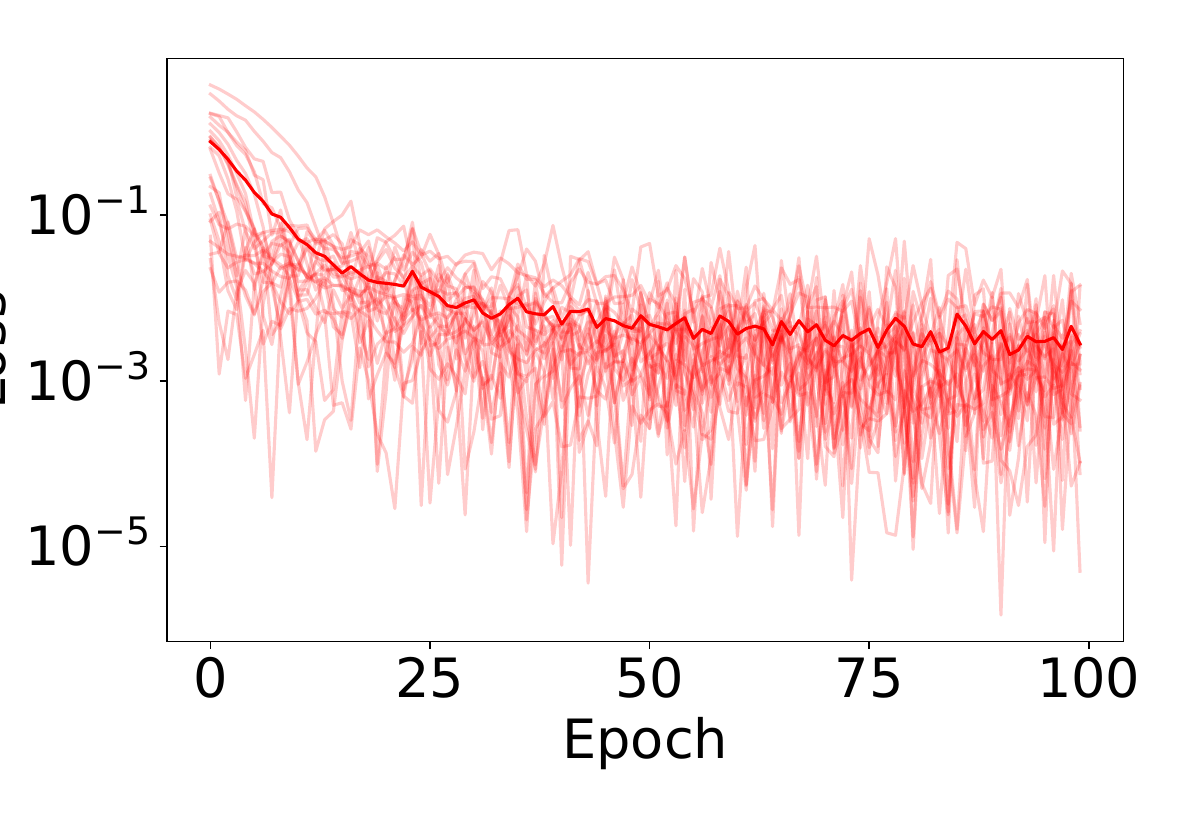}
\end{minipage}%
\hfill
\begin{minipage}[t]{0.33\textwidth}
\centering
\includegraphics[width=\textwidth]{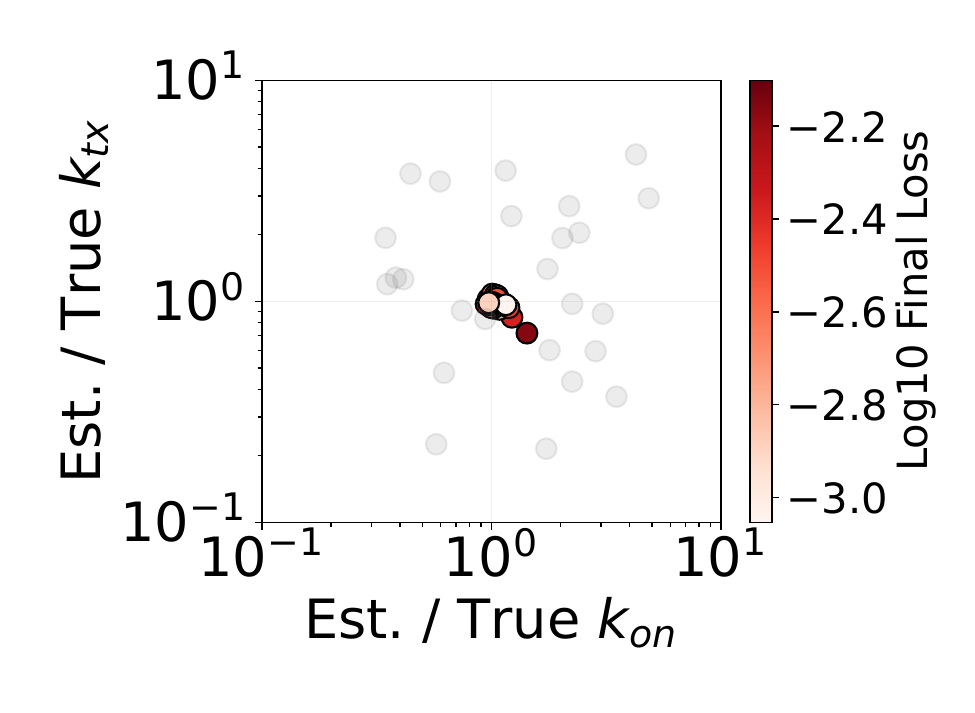}
\end{minipage}%
\hfill
\begin{minipage}[t]{0.33\textwidth}
\centering
\includegraphics[width=\textwidth]{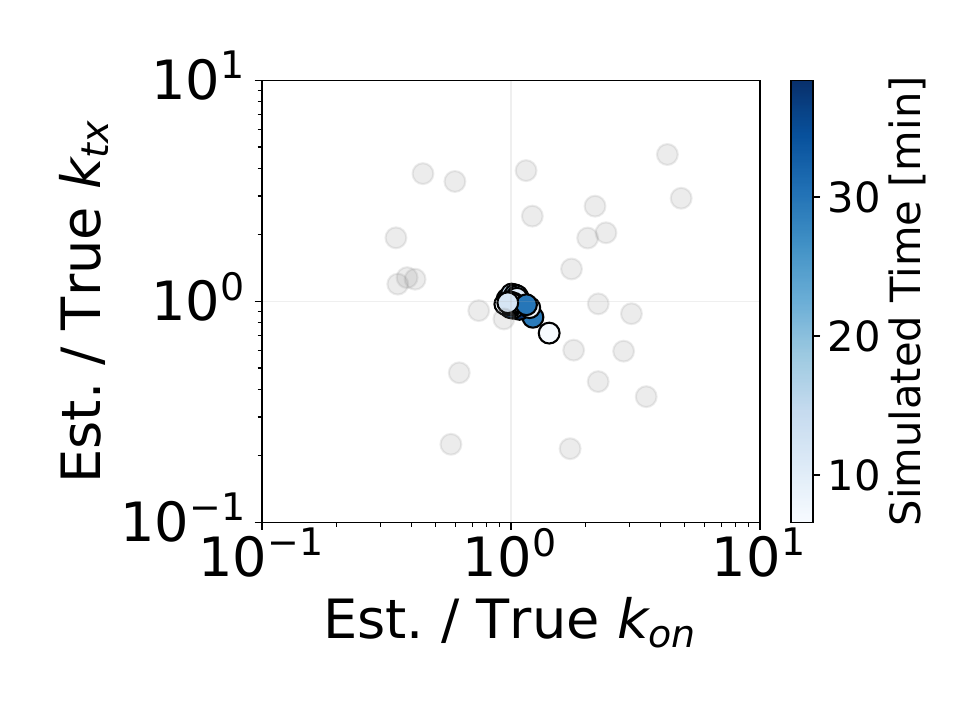}
\end{minipage}
\caption{\textbf{Moment matching optimization results.} \textbf{(Left)} Loss trajectories for all 25 parameter sets of the telegraph promoter model. Light curves show individual runs; dark curve shows the mean. \textbf{(Center)} Parameter recovery accuracy: ratio of estimated to true parameter values for $k_\text{on}$ and $k_\text{tx}$. Gray points show initial guesses; colored points show optimized values, with color indicating final loss. Optimized parameters cluster near the true values (intersection of gray lines at ratio 1). \textbf{(Right)} Same data as center panel, with color indicating the simulated trajectory duration.}
\label{fig:moment-matching-results}
\end{figure}

\textbf{Additional Results.} Fig.~\ref{fig:moment-matching-results} shows loss trajectories and parameter recovery for all 25 parameter sets. Both $k_\text{on}$ and $k_\text{tx}$ are recovered accurately, with optimized values clustering near the true parameters regardless of trajectory duration.

\clearpage
\section{Distribution Matching Loss -- Additional Details}
\label{sec:supp-hist-match}

\textbf{Experiment Details.} We inferred three parameters of the telegraph promoter model ($k_\text{on}$, $k_\text{tx}$, $k_\text{mdeg}$) from steady-state RNA copy number distributions, with $k_\text{off}$ fixed to its ground-truth value. Target distributions were generated by simulating 5000 trajectories to steady state for each of the 25 parameter sets. We minimized the 1-Wasserstein distance between target and simulated distribution functions. Differentiable histograms were constructed using triangular kernel density estimation with unit bandwidth and Dirichlet smoothing ($\alpha = 0.1$). Each optimization used 512 gradient-tracked simulations combined with 4488 baseline (forward-only) simulations per epoch. Parameters were initialized randomly within $[0.2\times, 5\times]$ the true values. Rate constants were parameterized as $k = e^\phi$ and optimization was performed over the log-transformed variables $\phi$. Training ran for 100 epochs using Adam with learning rate $\text{lr} = 0.1$ and modified hyperparameters ($\beta_1 = 0.8$, $\beta_2 = 0.9$).

\begin{figure}[h!]
\centering
\begin{minipage}[t]{0.5\textwidth}
\centering
\includegraphics[width=\textwidth]{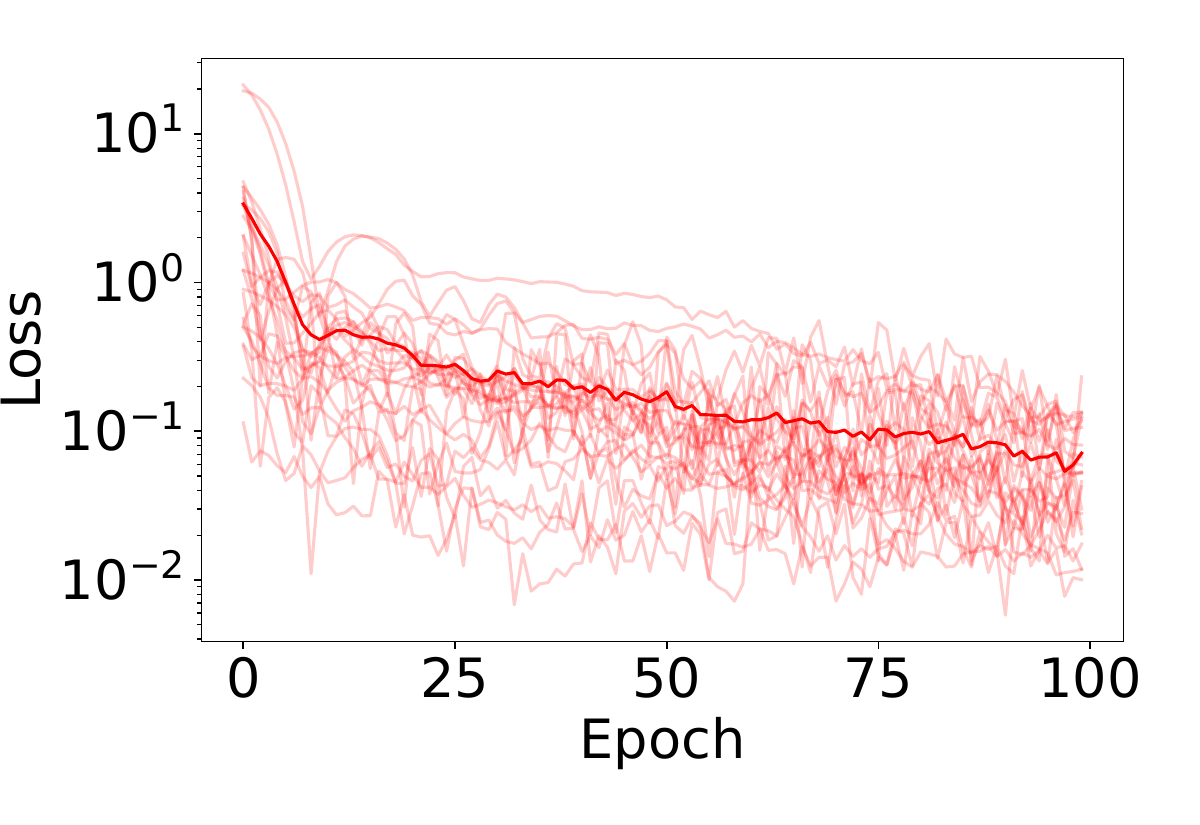}
\end{minipage}%
\hfill
\begin{minipage}[t]{0.5\textwidth}
\centering
\includegraphics[width=\textwidth]{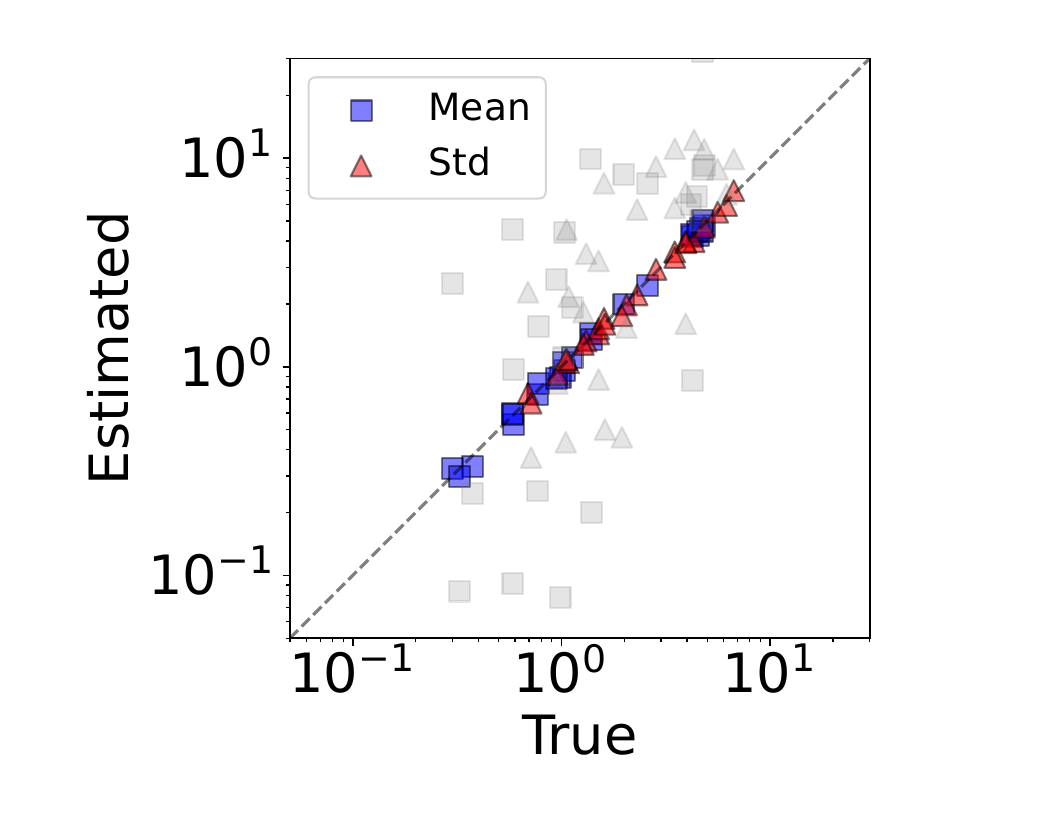}
\end{minipage}
\caption{\textbf{Distribution matching optimization summary.} \textbf{(Left)} Loss trajectories for all 25 parameter sets. Light curves show individual runs; dark curve shows the mean. \textbf{(Right)} Recovery of target distribution statistics. Estimated vs.\ true mean (blue squares) and standard deviation (red triangles) of the RNA copy number distribution. Gray markers show initial guesses; colored markers show optimized values.}
\label{fig:hist-loss-moments}
\end{figure}

\begin{figure}[h!]
\centering
\begin{minipage}[t]{0.33\textwidth}
\centering
\includegraphics[width=\textwidth]{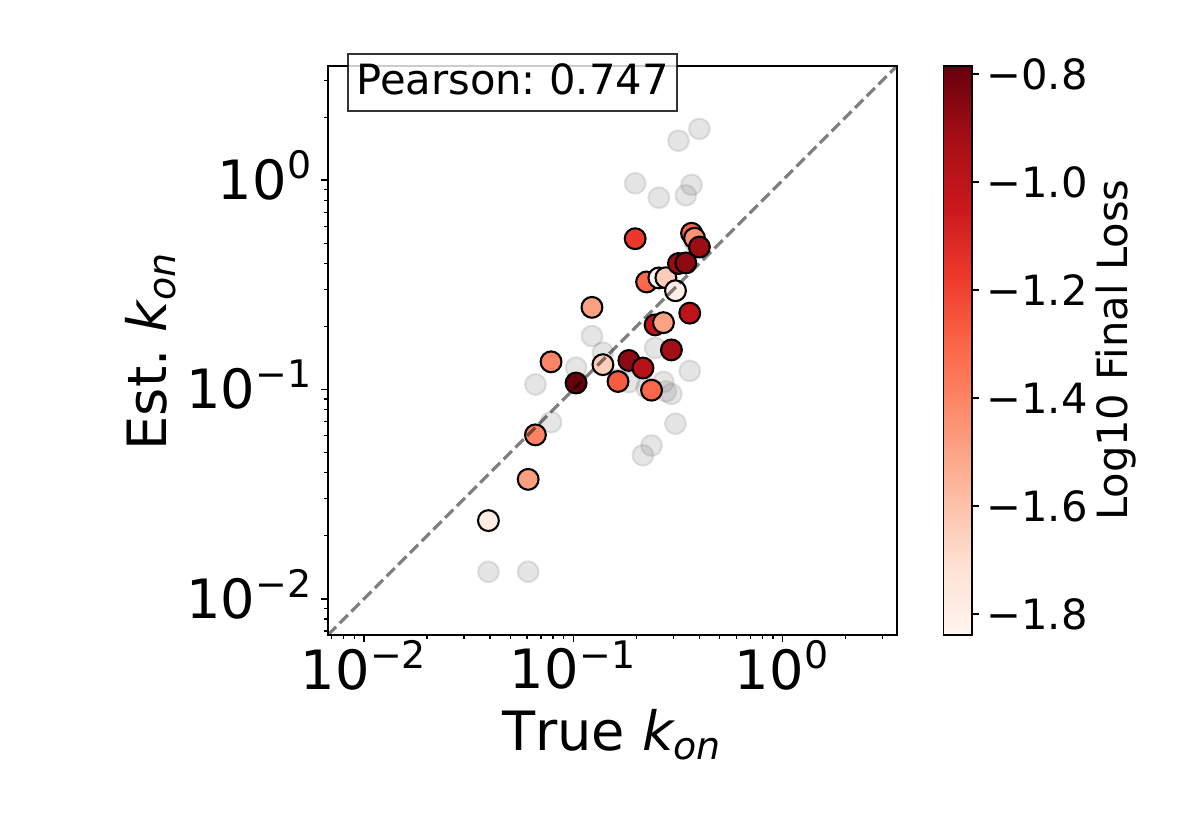}
\end{minipage}%
\hfill
\begin{minipage}[t]{0.33\textwidth}
\centering
\includegraphics[width=\textwidth]{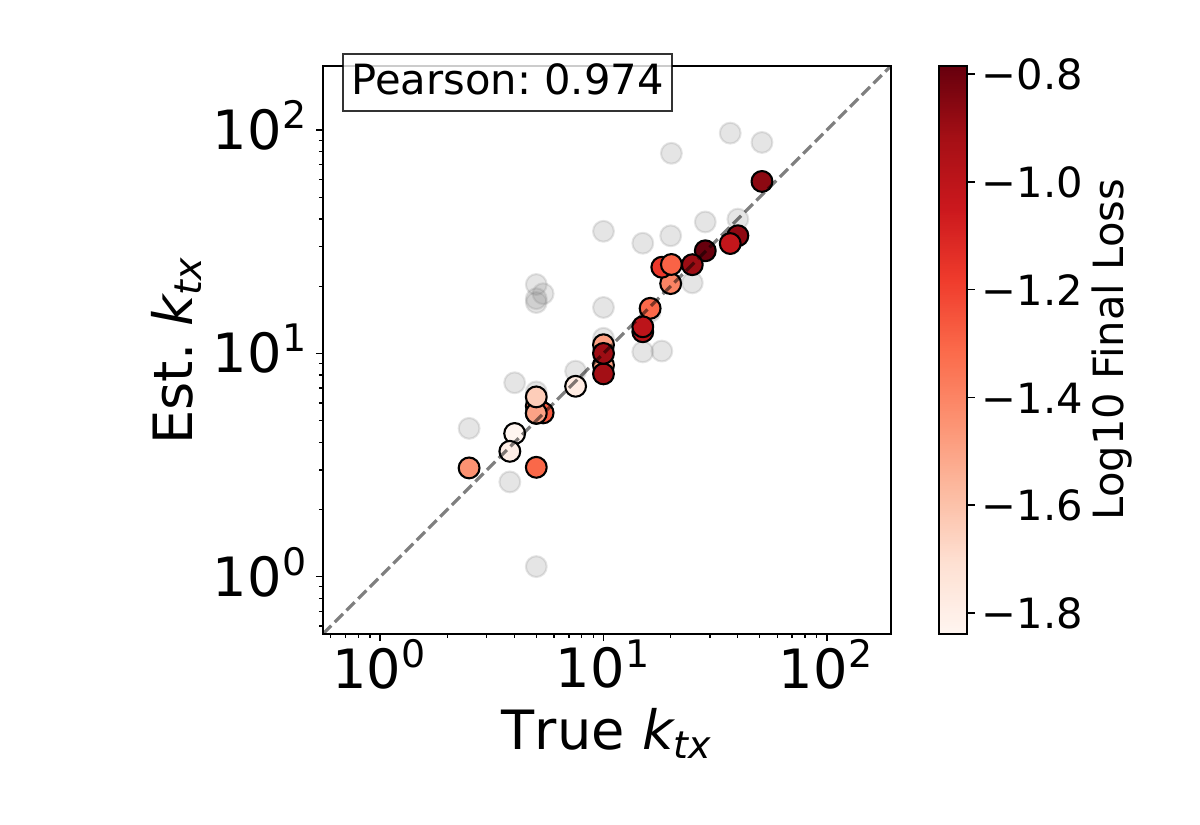}
\end{minipage}%
\hfill
\begin{minipage}[t]{0.33\textwidth}
\centering
\includegraphics[width=\textwidth]{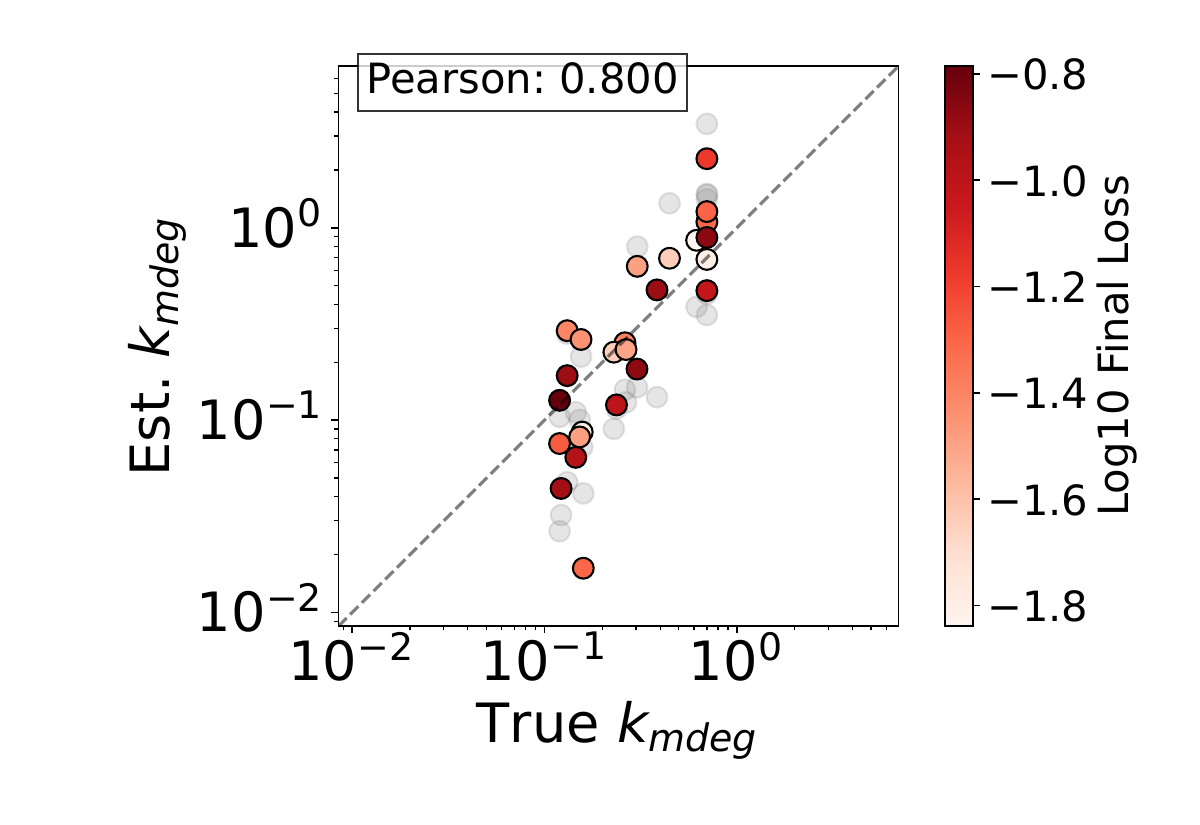}
\end{minipage}
\caption{\textbf{Individual parameter recovery for distribution matching.} Estimated vs.\ true values for \textbf{(Left)} $k_\text{on}$, \textbf{(Center)} $k_\text{tx}$, and \textbf{(Right)} $k_\text{mdeg}$. Gray points show initial guesses; colored points show optimized values, with color indicating final loss. Pearson correlation coefficients are shown for each parameter. The transcription rate $k_\text{tx}$ is recovered most accurately ($r = 0.974$), while $k_\text{on}$ and $k_\text{mdeg}$ show moderate correlations ($r = 0.747$ and $r = 0.800$, respectively).}
\label{fig:hist-param-recovery}
\end{figure}

\begin{figure}[h!]
\centering
\begin{minipage}[t]{0.33\textwidth}
\centering
\includegraphics[width=\textwidth]{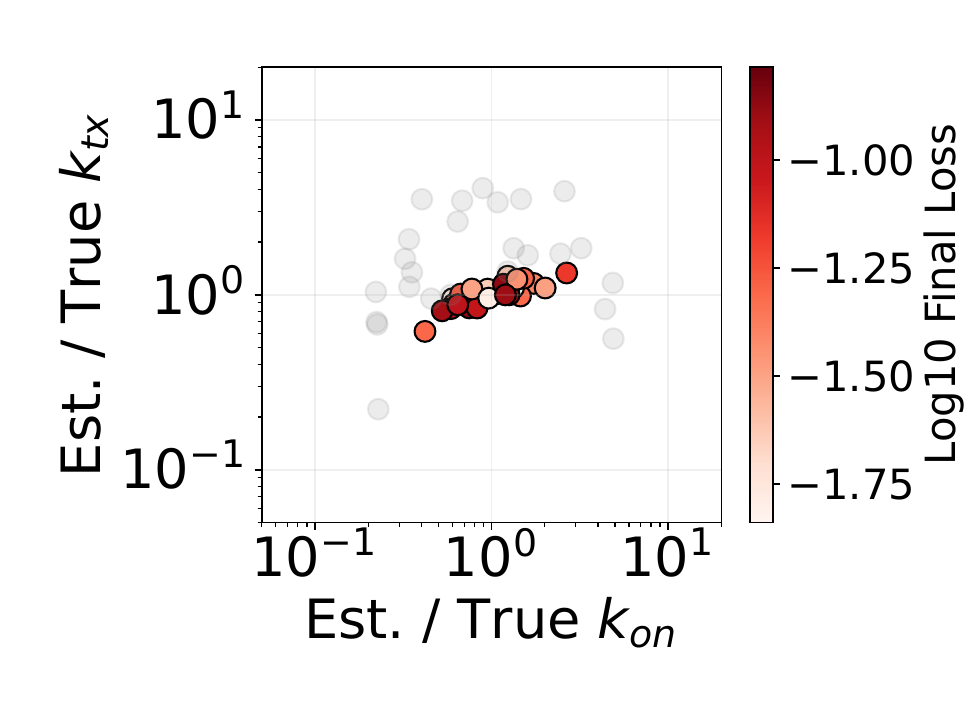}
\end{minipage}%
\hfill
\begin{minipage}[t]{0.33\textwidth}
\centering
\includegraphics[width=\textwidth]{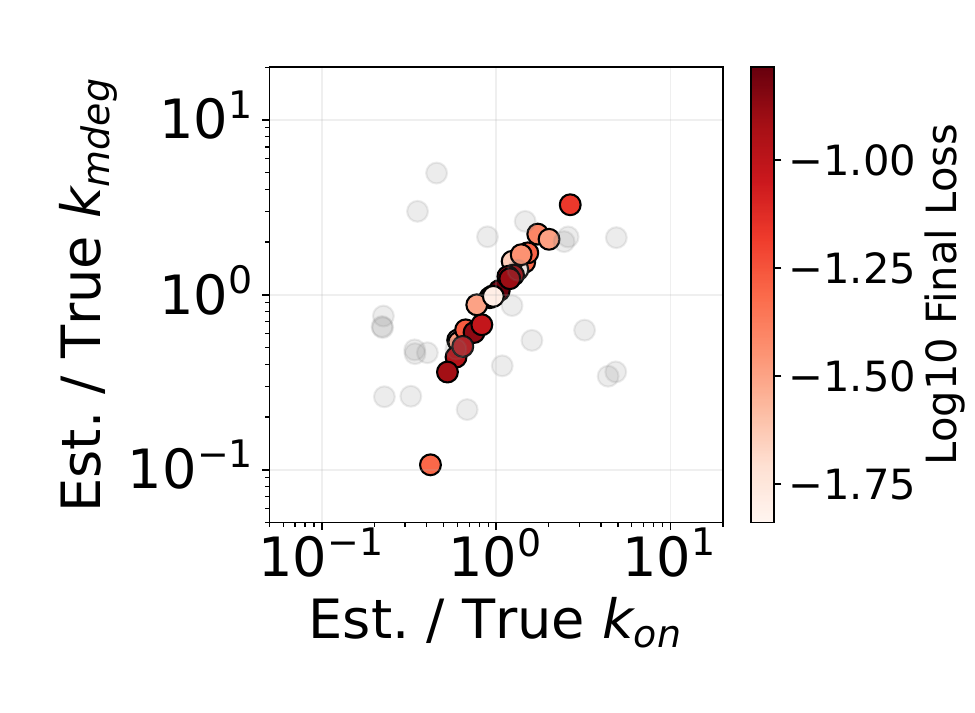}
\end{minipage}%
\hfill
\begin{minipage}[t]{0.33\textwidth}
\centering
\includegraphics[width=\textwidth]{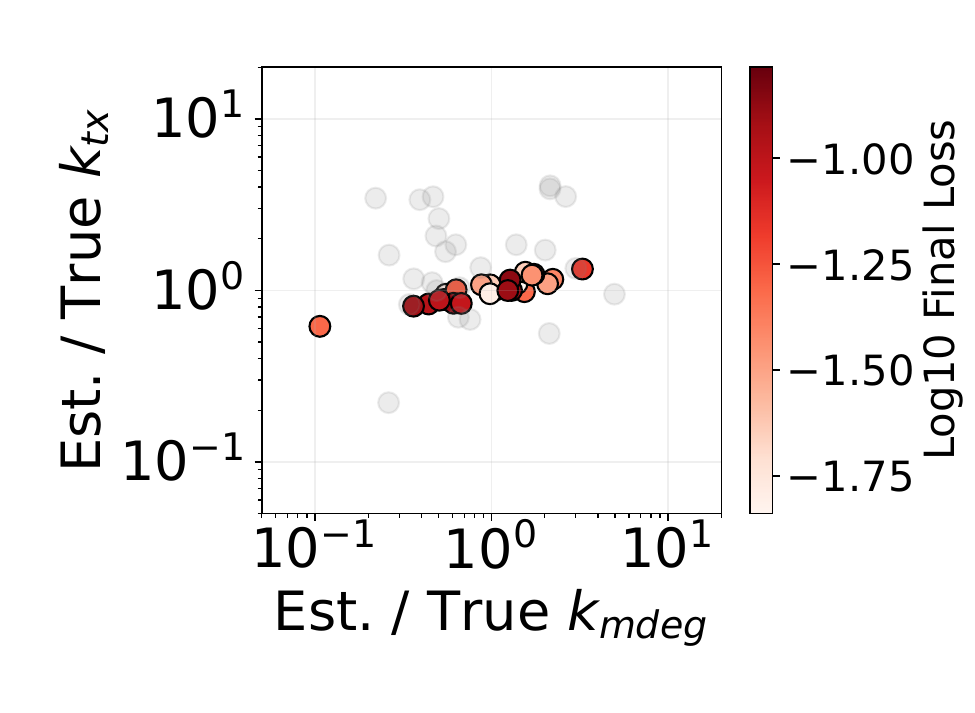}
\end{minipage}
\caption{\textbf{Pairwise parameter recovery for distribution matching.} Ratio of estimated to true parameter values shown in two-dimensional projections: \textbf{(Left)} $k_\text{on}$ vs.\ $k_\text{tx}$, \textbf{(Center)} $k_\text{on}$ vs.\ $k_\text{mdeg}$, and \textbf{(Right)} $k_\text{mdeg}$ vs.\ $k_\text{tx}$. Gray points show initial guesses; colored points show optimized values, with color indicating final loss.}
\label{fig:hist-param-pairs}
\end{figure}

\begin{figure*}[h!]
\includegraphics[width=\textwidth]{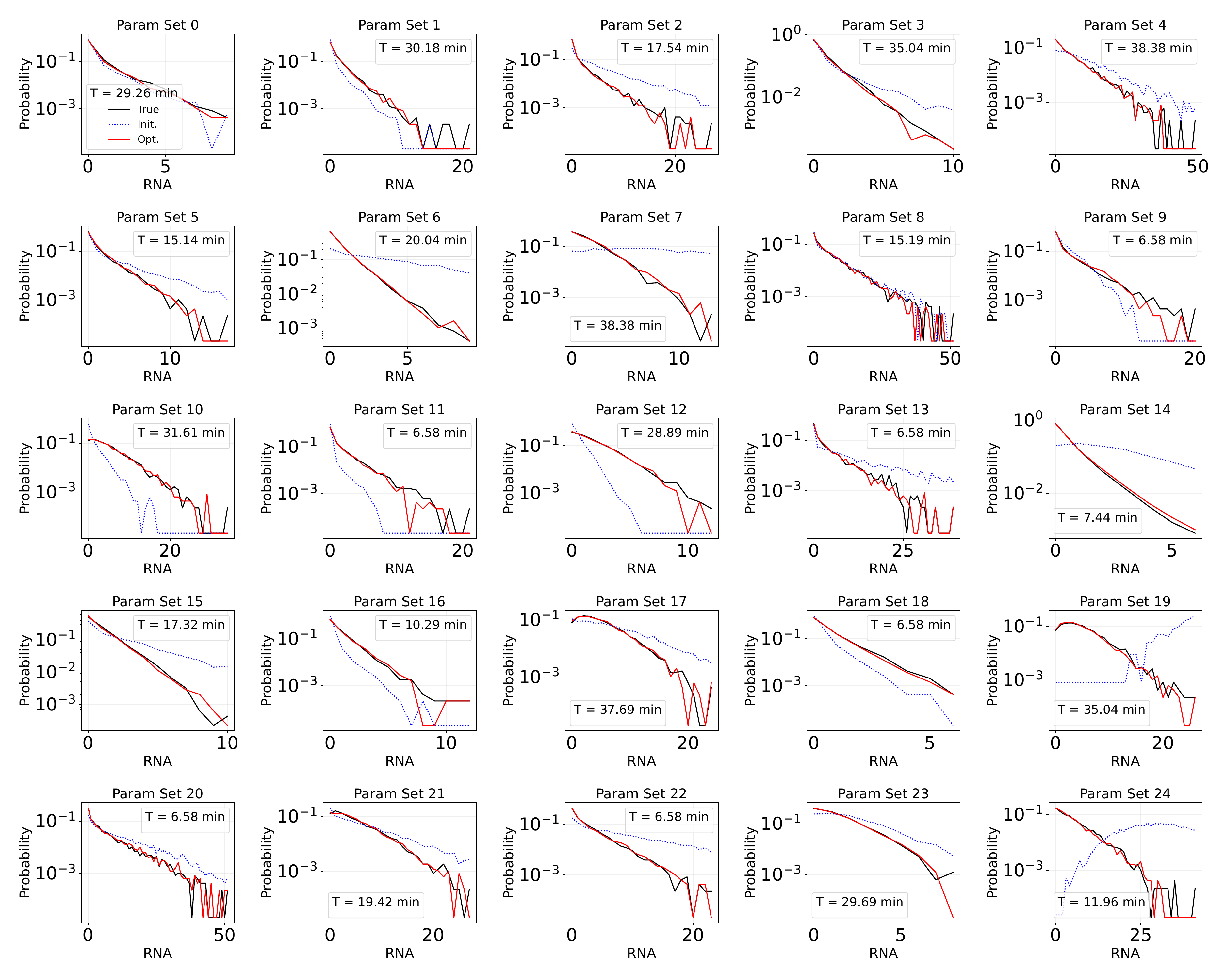}
\caption{\textbf{Distribution matching results across 25 parameter sets (log scale).} Comparison of steady-state RNA copy number distributions for the telegraph promoter model. Black: true distribution; blue dotted: initial guess; red: optimized parameters. The $y$-axis shows probability on a logarithmic scale, highlighting the fit quality in the distribution tails. Optimization time $T$ is indicated for each parameter set.}
\label{fig:hist-match-all}
\end{figure*}

\begin{figure*}[h!]
\includegraphics[width=\textwidth]{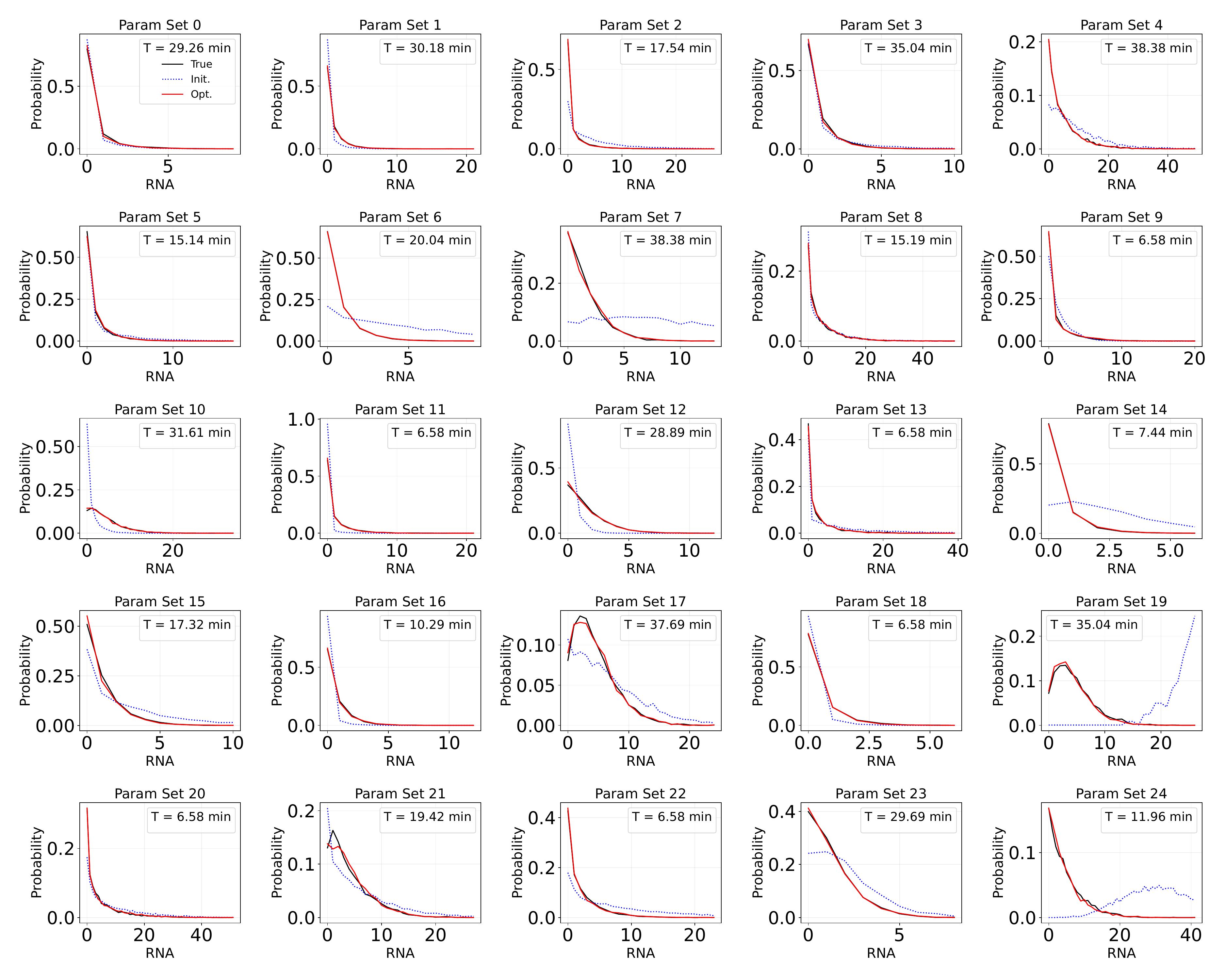}
\caption{\textbf{Distribution matching results across 25 parameter sets (linear scale).} Same data as Fig.~\ref{fig:hist-match-all} plotted with linear probability axes.}\label{fig:hist-match-all-nolog}
\end{figure*}

\textbf{Additional Results.} Figs.~\ref{fig:hist-loss-moments}--\ref{fig:hist-match-all-nolog} show optimization results. The transcription rate $k_\text{tx}$ is recovered most accurately ($r = 0.974$), while $k_\text{on}$ and $k_\text{mdeg}$ show moderate correlations ($r = 0.747$ and $r = 0.800$). The correlated scatter in the $k_\text{on}$--$k_\text{mdeg}$ plane reflects the known degeneracy where these parameters jointly control burst frequency.

\clearpage
\section{Experimental Timecourse Data Fitting -- Additional Details}
\label{sec:supp-smfish}

\textbf{Experiment Details.} We infer all eight kinetic parameters of the four-state promoter model ($k_{12}$, $k_{21}$, $k_{23}$, $k_{32}$, $k_{34}$, $k_{43}$, $k_\text{tx}$, $k_\text{mdeg}$) from nuclear STL1 mRNA count distributions measured by smFISH at eight time points ($t = 0, 2, 4, 6, 8, 10, 15, 20$~min) during the rising phase of the osmotic stress response, using data from Li \& Neuert~\cite{li2019multiplex}. Counts from three biological replicates were pooled at each time point. All trajectories were initialized in state $S_1$ with zero mRNA. We minimized the sum of per-time-point Kullback--Leibler divergences between observed and simulated RNA count distributions. Differentiable histograms were constructed using triangular kernel density estimation with bandwidth $h = 1.01$ on a fixed grid of 11 points over $[0, 10]$. Each optimization used 1024 gradient-tracked simulations combined with a variable number of baseline (forward-only) simulations per epoch. The baseline count was adjusted per time point so that the total pool matched the number of observed cells at that time point. Rate constants were parameterized as $k = e^\phi$ and optimization was performed over the log-transformed variables $\phi$, initialized uniformly at $\phi_0 = -2$. Training ran for 100 epochs using standard Adam ($\beta_1 = 0.9$, $\beta_2 = 0.999$) and learning rate $\text{lr} = 0.05$.

\begin{figure*}[h!]
\centering
\includegraphics[width=0.8\textwidth]{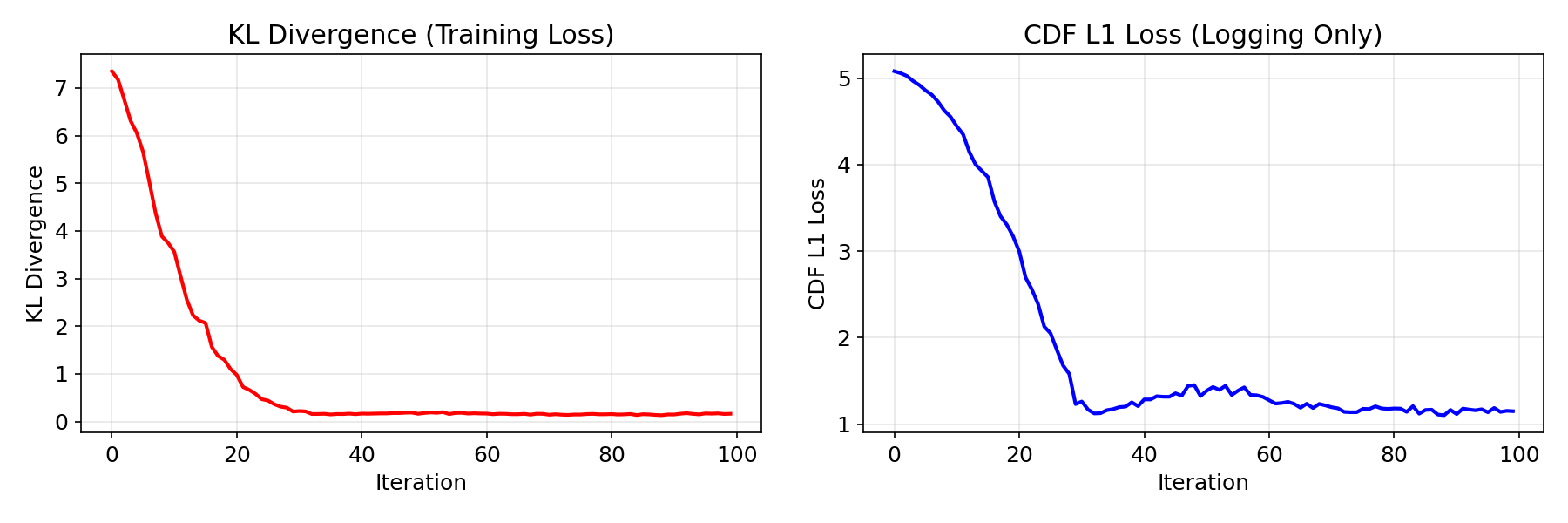}
\caption{\textbf{smFISH fitting loss trajectories.} KL divergence (training objective, left) and CDF $L_1$ distance (logged for reference, right) over 100 training epochs.}
\label{fig:smfish-training}
\end{figure*}

\begin{figure*}[h!]
\includegraphics[width=0.8\textwidth]{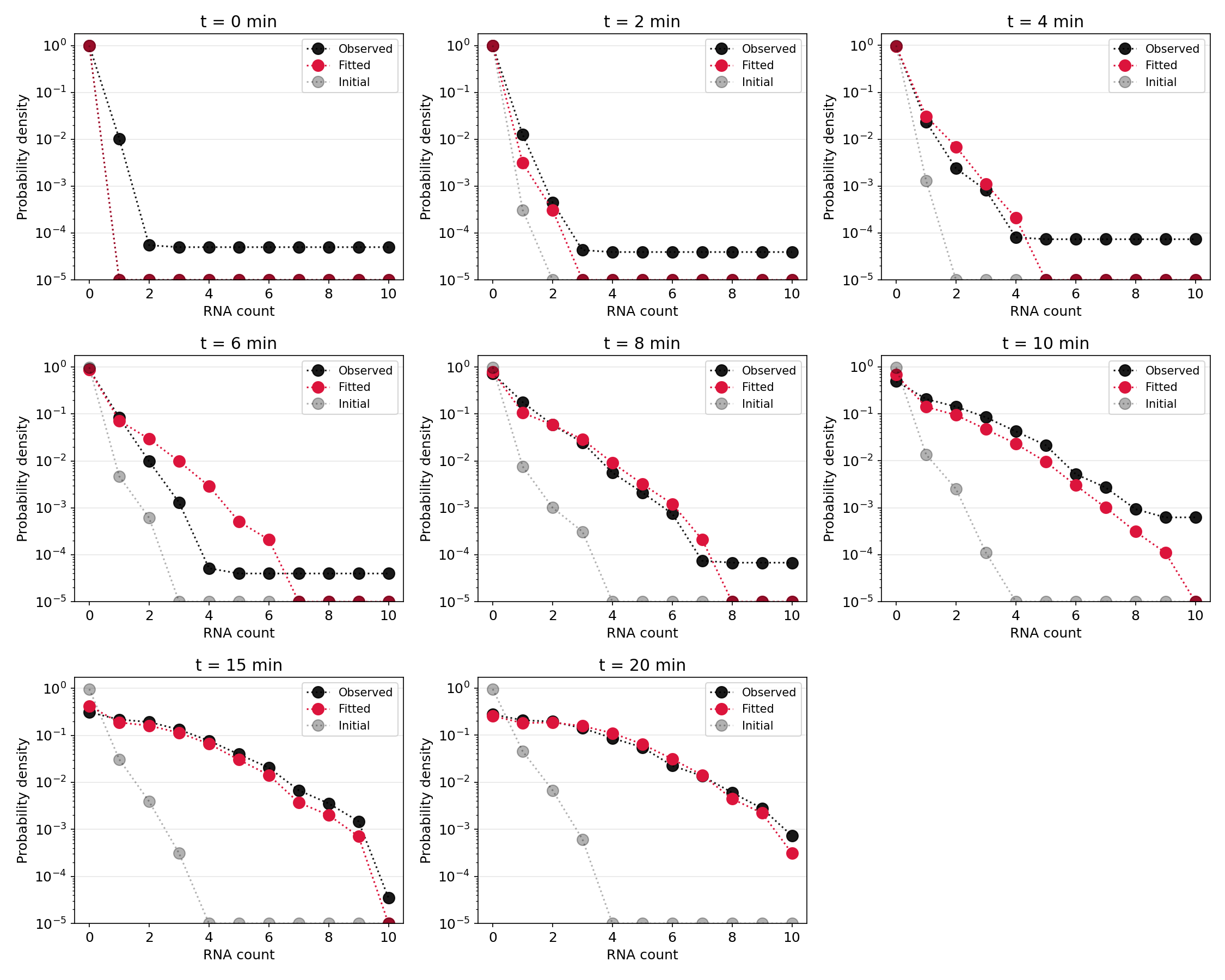}
\caption{\textbf{smFISH distribution fitting results (log scale).} Comparison of RNA count distributions at all eight time points. Black: target distribution; gray dotted: initial model (pre-optimization); red: fitted model. The $y$-axis shows probability density on a logarithmic scale.}
\label{fig:smfish-fitted}
\end{figure*}

\begin{figure*}[h!]
\includegraphics[width=0.8\textwidth]{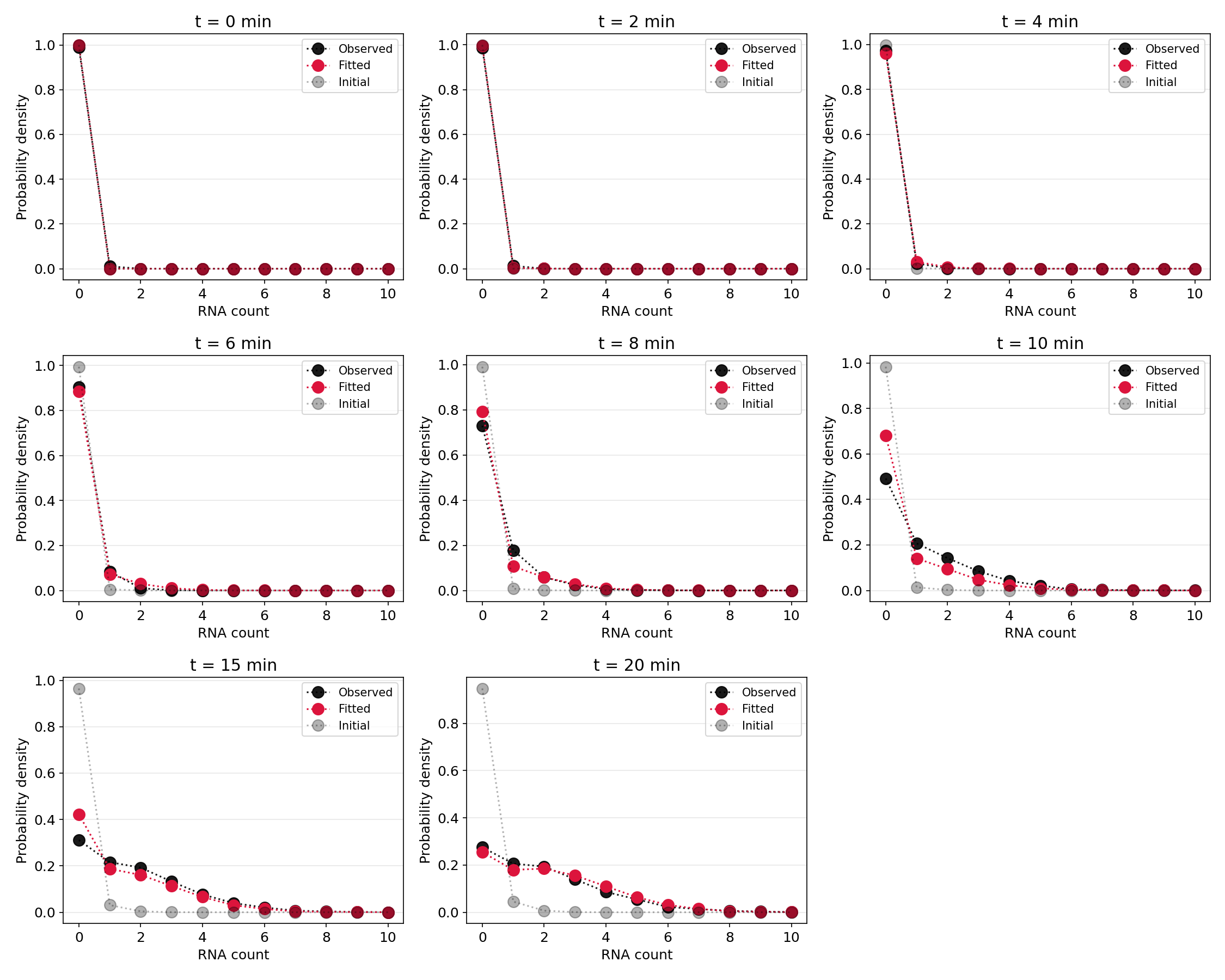}
\caption{\textbf{smFISH distribution fitting results (linear scale).} Same data as Fig.~\ref{fig:smfish-fitted} plotted with linear probability axes.}
\label{fig:smfish-fitted-linear}
\end{figure*}

\begin{figure*}[h!]
\includegraphics[width=0.8\textwidth]{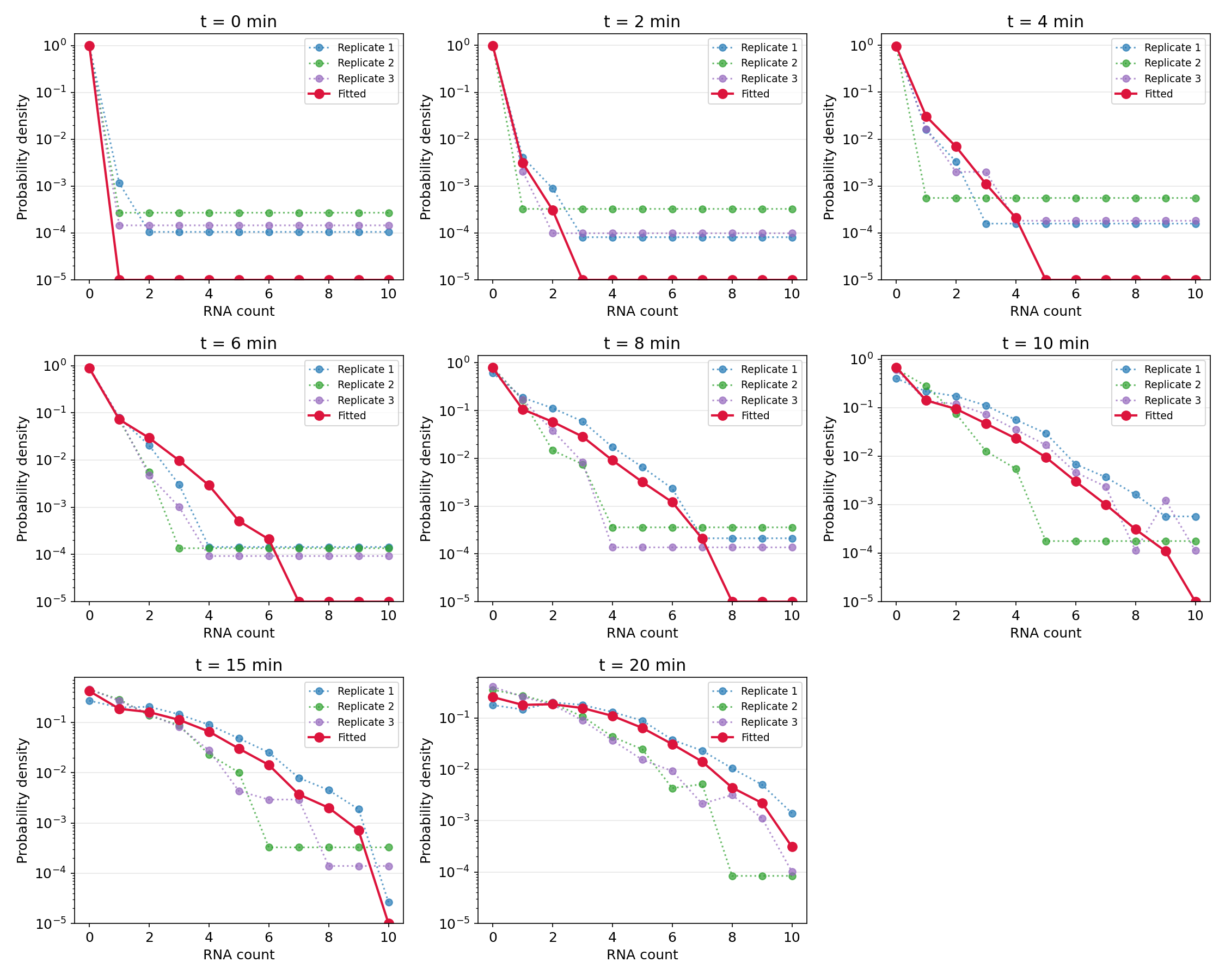}
\caption{\textbf{Fitted model vs.\ individual biological replicates (log scale).} Comparison of RNA count distributions at all eight fitted time points. Colored circles: individual replicates (1--3); red circles: fitted model. The $y$-axis shows probability density on a logarithmic scale.}
\label{fig:smfish-replicates}
\end{figure*}

\begin{figure*}[h!]
\includegraphics[width=0.8\textwidth]{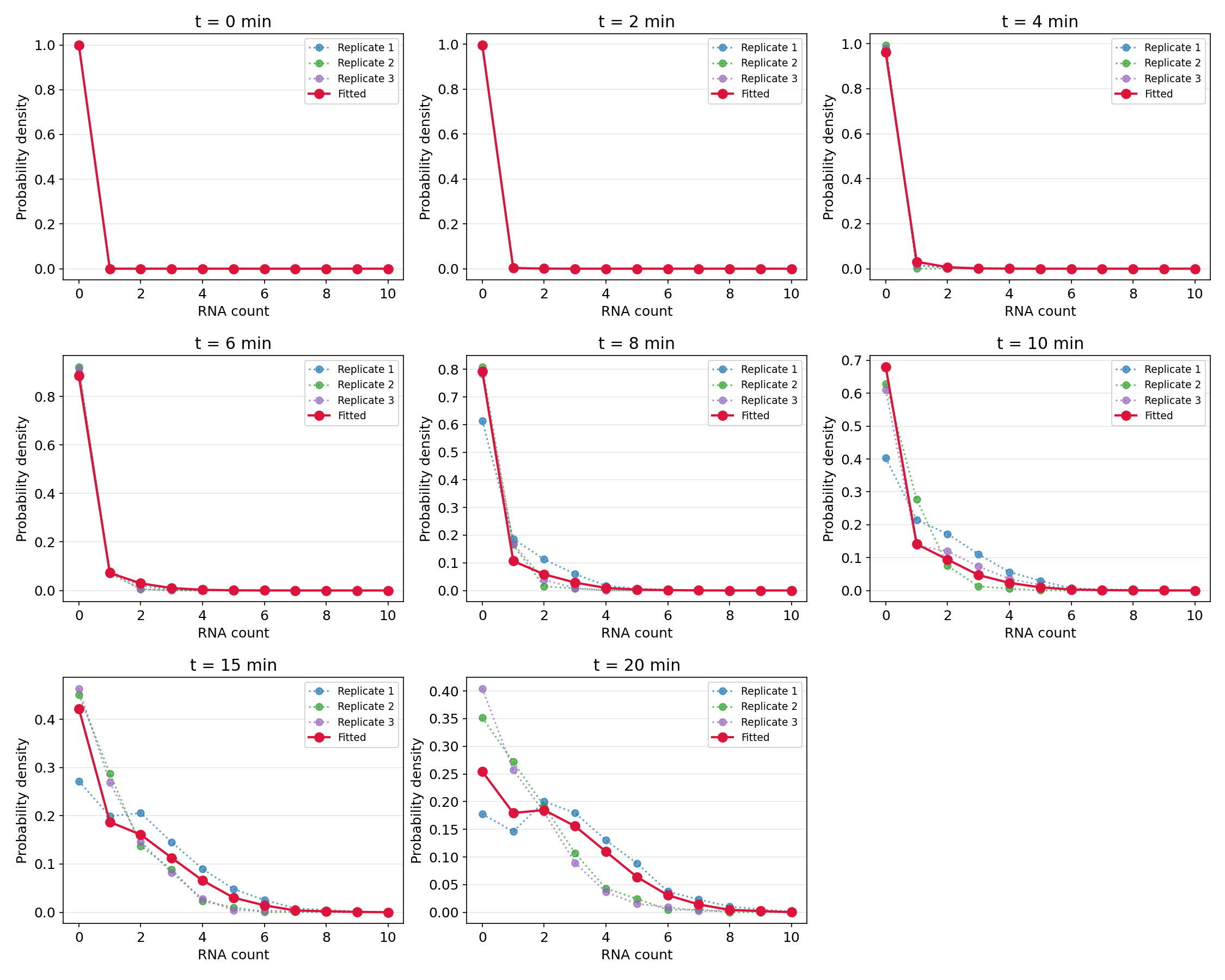}
\caption{\textbf{Fitted model vs.\ individual biological replicates (linear scale).} Same data as Fig.~\ref{fig:smfish-replicates} plotted with linear probability axes.}
\label{fig:smfish-replicates-linear}
\end{figure*}

\textbf{Additional Results.} Table~\ref{tab:smfish-rates} reports the optimized rate constants. Fig.~\ref{fig:smfish-training} shows the loss trajectories. Figs.~\ref{fig:smfish-fitted}--\ref{fig:smfish-fitted-linear} show the fitted distributions compared to the pooled target data alongside the initial (pre-optimization) model predictions. Figs.~\ref{fig:smfish-replicates}--\ref{fig:smfish-replicates-linear} compare the fitted model distributions against individual biological replicates at all eight time points, complementing the comparison shown in the main text (Fig.~\ref{fig:hist-data}d).

\vspace{1cm}
\begin{table*}[h!]
\centering
\normalsize
\begin{tabular}{l l l l}
\hline
Parameter & Reaction & Rate (min$^{-1}$) & $\log k$ \\
\hline
$k_{12}$ & $S_1 \to S_2$ & 0.237 & $-1.438$ \\
$k_{21}$ & $S_2 \to S_1$ & 0.068 & $-2.691$ \\
$k_{23}$ & $S_2 \to S_3$ & 0.296 & $-1.217$ \\
$k_{32}$ & $S_3 \to S_2$ & 0.047 & $-3.062$ \\
$k_{34}$ & $S_3 \to S_4$ & 0.384 & $-0.957$ \\
$k_{43}$ & $S_4 \to S_3$ & 0.042 & $-3.169$ \\
$k_\text{tx}$   & $S_4 \to S_4 + \text{RNA}$ & 0.434 & $-0.835$ \\
$k_\text{mdeg}$ & $\text{RNA} \to \varnothing$ & 0.095 & $-2.354$ \\
\hline
\end{tabular}
\caption{\textbf{Optimized rate constants for the four-state promoter model.}}
\label{tab:smfish-rates}
\end{table*}

\vspace{1cm}

\clearpage
\section{ASEP Current Maximization -- Additional Details}
\label{sec:supp-asep}

\textbf{Experiment Details.} We optimized the $L$ site-dependent forward hopping rates $\{k_i^+\}_{i=1}^L$ of a reversible ASEP ring to maximize steady-state current under a fixed mean forward rate budget $\bar{k}^+ = \sum_i k_i^+/L = 1$. The backward rate was fixed homogeneously at $k^- = 0.01$ for all bonds. We swept nine particle counts for each system size ($L = 10$: $N = 1, \ldots, 9$; $L = 30$: $N = 3, 6, \ldots, 27$), covering densities $\rho = 0.1, 0.2, \ldots, 0.9$. Rather than directly maximizing current, we equivalently maximized entropy production $\hat{\sigma} = \hat{J}\,\mathcal{A}$, where $\hat{J}$ is the time-weighted propensity current estimate and $\mathcal{A} = \sum_i \log(k_i^+/k^-)$ is the cycle affinity. The mean forward rate constraint was enforced via a logarithmic penalty with weight $\lambda = 100$. Forward rates were parameterized as $k_i^+ = e^{\phi_i}$ and optimized in log-space, initialized with log-normal disorder ($\phi_i \sim \mathcal{N}(0, 0.5^2)$, corresponding to CV $\approx 0.5$). Each optimization used batches of 8 trajectories ($L=10$) or 16 trajectories ($L=30$), each simulated for $T_\text{sim} = 100$ time units. Training ran for 100 epochs using Adam with modified hyperparameters ($\beta_1 = 0.8$, $\beta_2 = 0.9$) and learning rate $\text{lr} = 0.05$.

\begin{figure*}[h!]
\centering
\includegraphics[width=\textwidth]{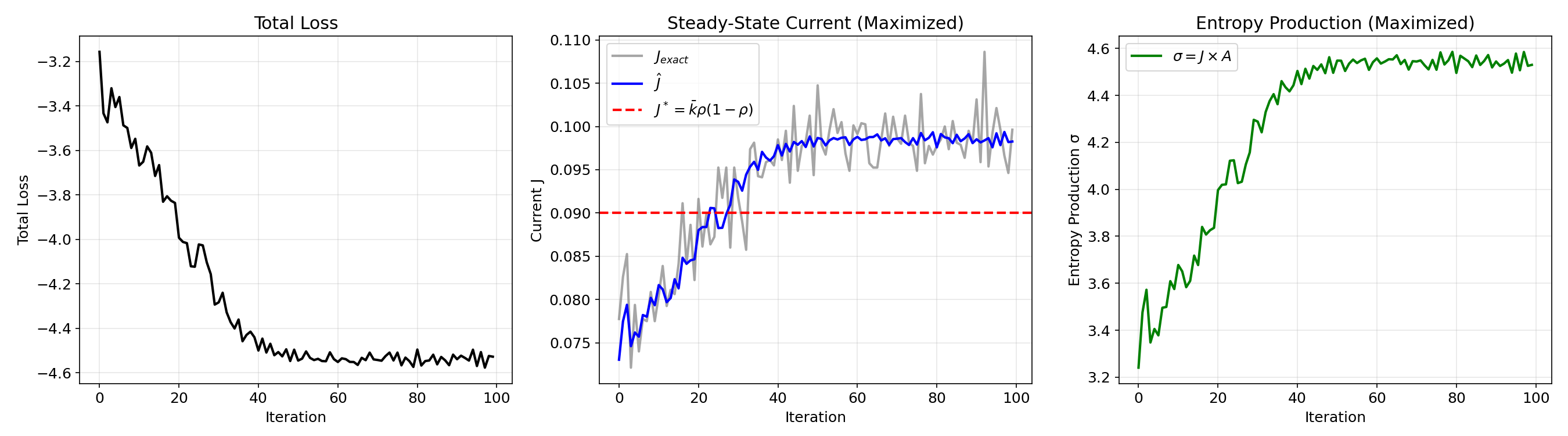}\\[0.3em]
\includegraphics[width=\textwidth]{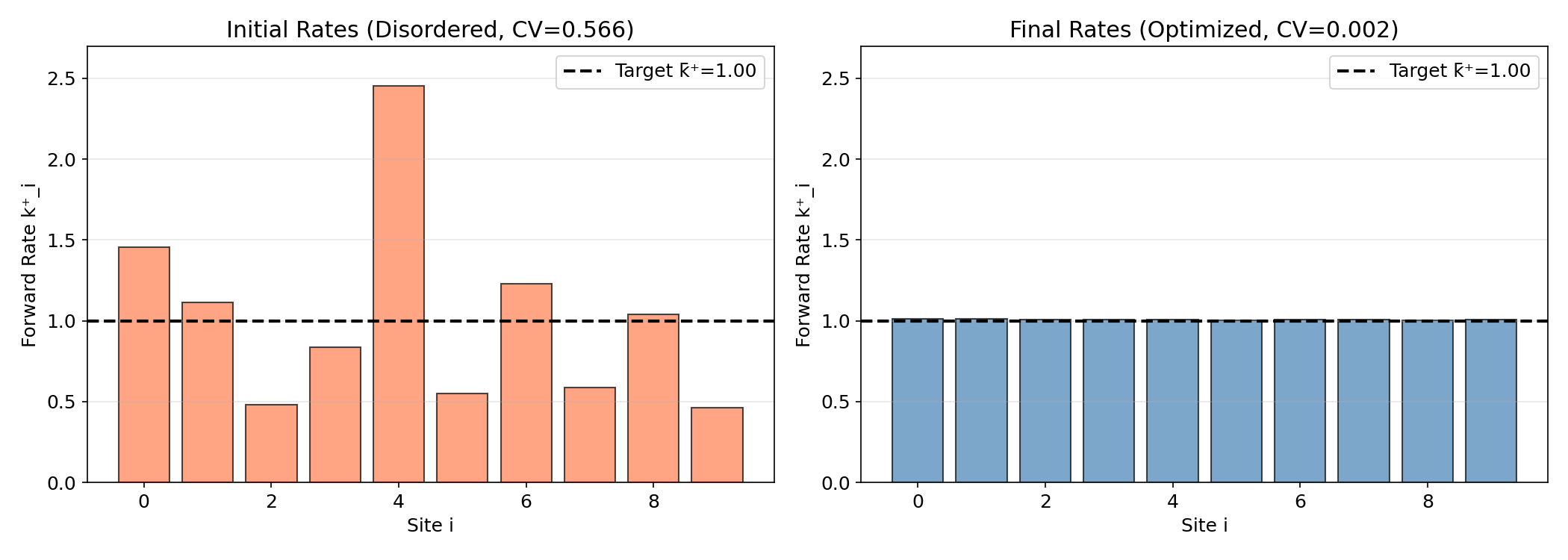}
\caption{\textbf{ASEP optimization details, $L=10$, $\rho=0.1$ ($N=1$).} \textbf{(Top)} Training dynamics: loss, optimized approximate current $J$ (blue), exact discrete current (gray), and entropy production rate (green) over training iterations. The red dashed line indicates the mean-field current $J_\text{MF}$ \textbf{(Bottom)} Forward rate profiles across bonds at initialization (orange) and at convergence (blue); dashed line indicates the target mean rate $\bar{k}^+$.}
\label{fig:asep-L10-d01}
\end{figure*}

\begin{figure*}
\centering
\includegraphics[width=\textwidth]{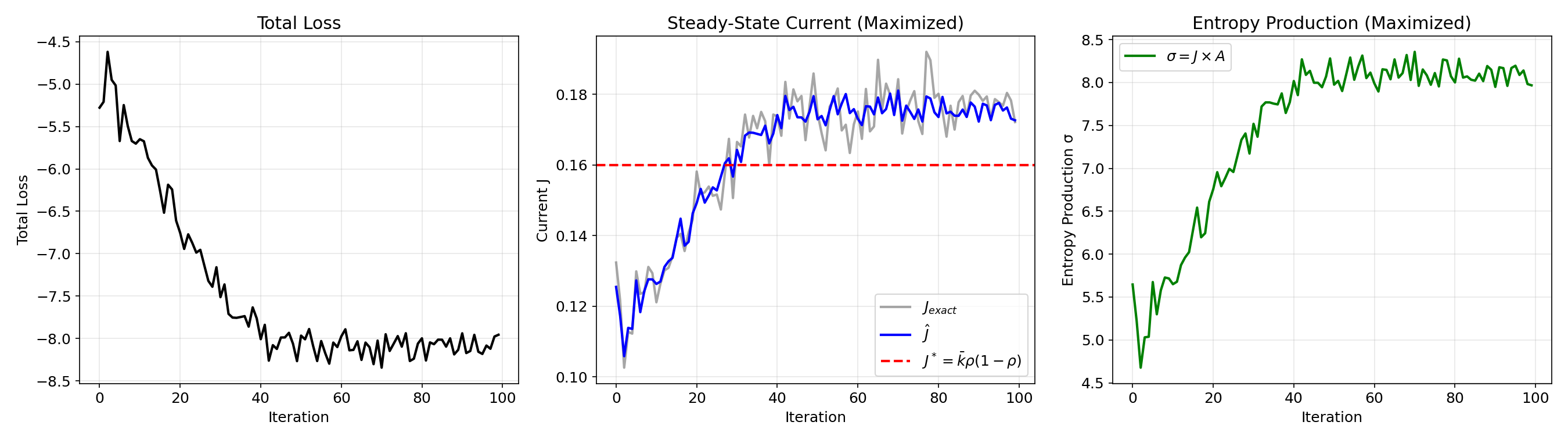}\\[0.3em]
\includegraphics[width=\textwidth]{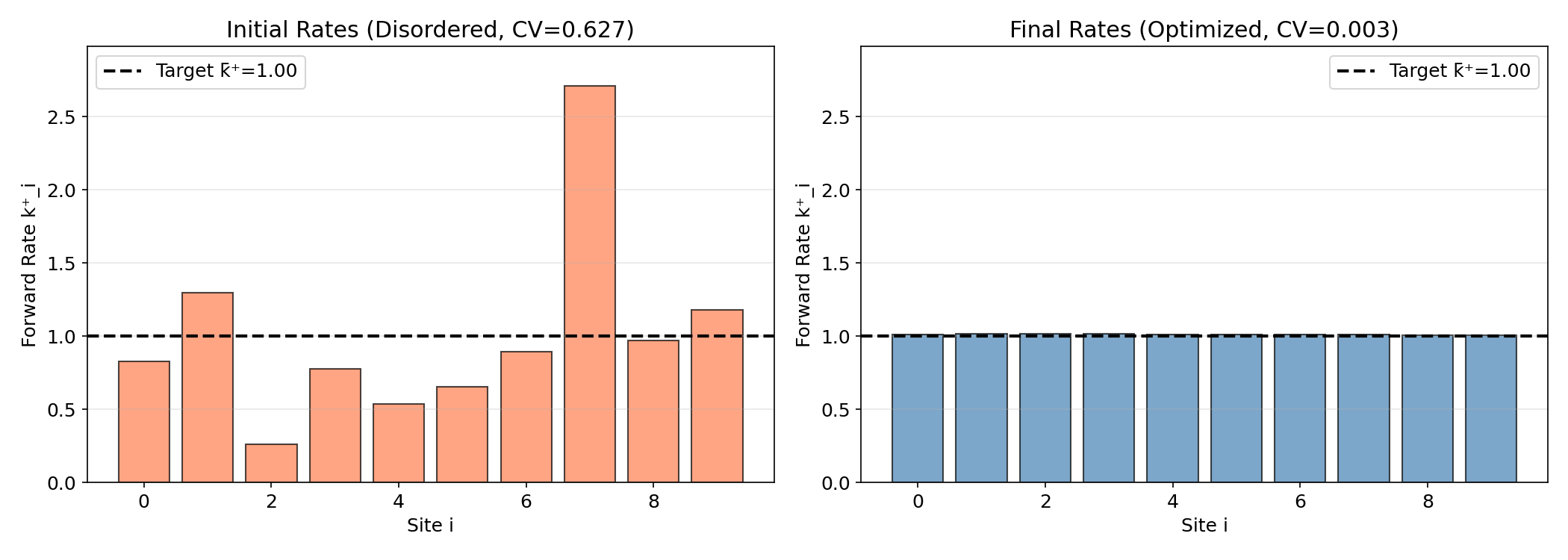}
\caption{\textbf{ASEP optimization details, $L=10$, $\rho=0.2$ ($N=2$).} Same format as Fig.~\ref{fig:asep-L10-d01}.}
\label{fig:asep-L10-d02}
\end{figure*}

\begin{figure*}[p]
\centering
\includegraphics[width=\textwidth]{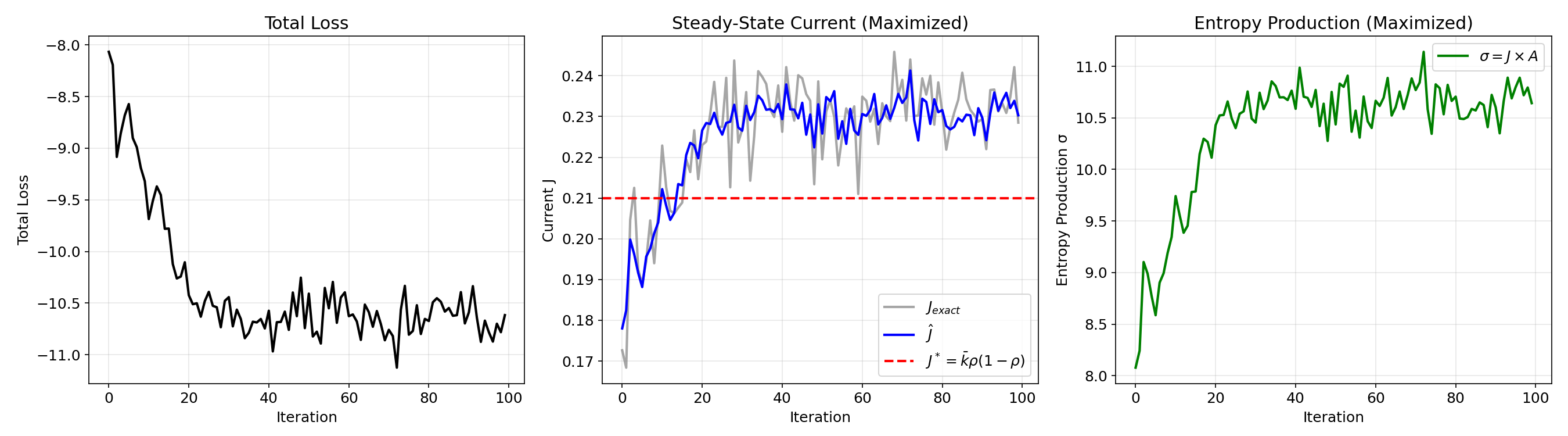}\\[0.3em]
\includegraphics[width=\textwidth]{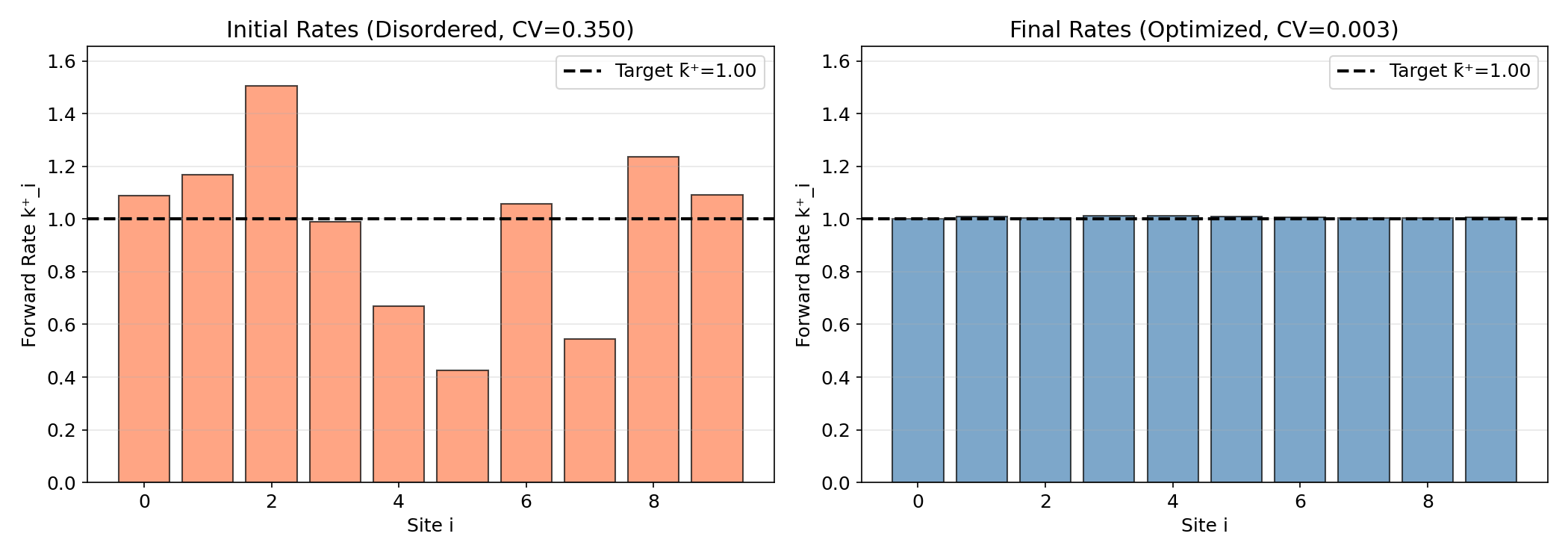}
\caption{\textbf{ASEP optimization details, $L=10$, $\rho=0.3$ ($N=3$).} Same format as Fig.~\ref{fig:asep-L10-d01}.}
\label{fig:asep-L10-d03}
\end{figure*}

\begin{figure*}[p]
\centering
\includegraphics[width=\textwidth]{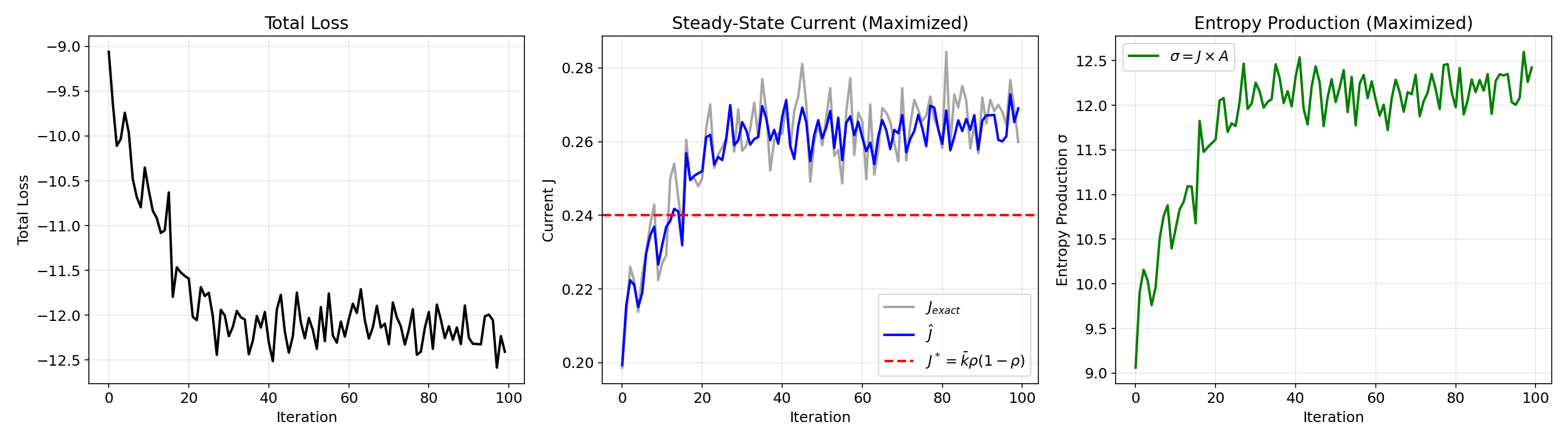}\\[0.3em]
\includegraphics[width=\textwidth]{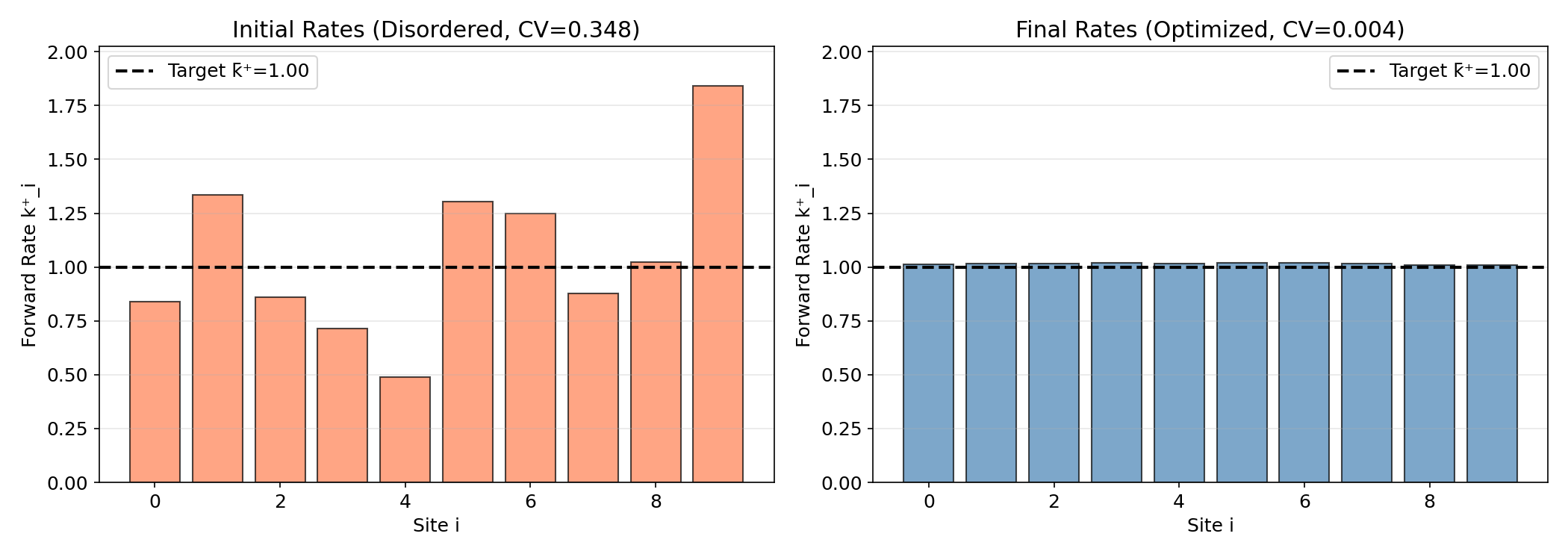}
\caption{\textbf{ASEP optimization details, $L=10$, $\rho=0.4$ ($N=4$).} Same format as Fig.~\ref{fig:asep-L10-d01}.}
\label{fig:asep-L10-d04}
\end{figure*}

\begin{figure*}[p]
\centering
\includegraphics[width=\textwidth]{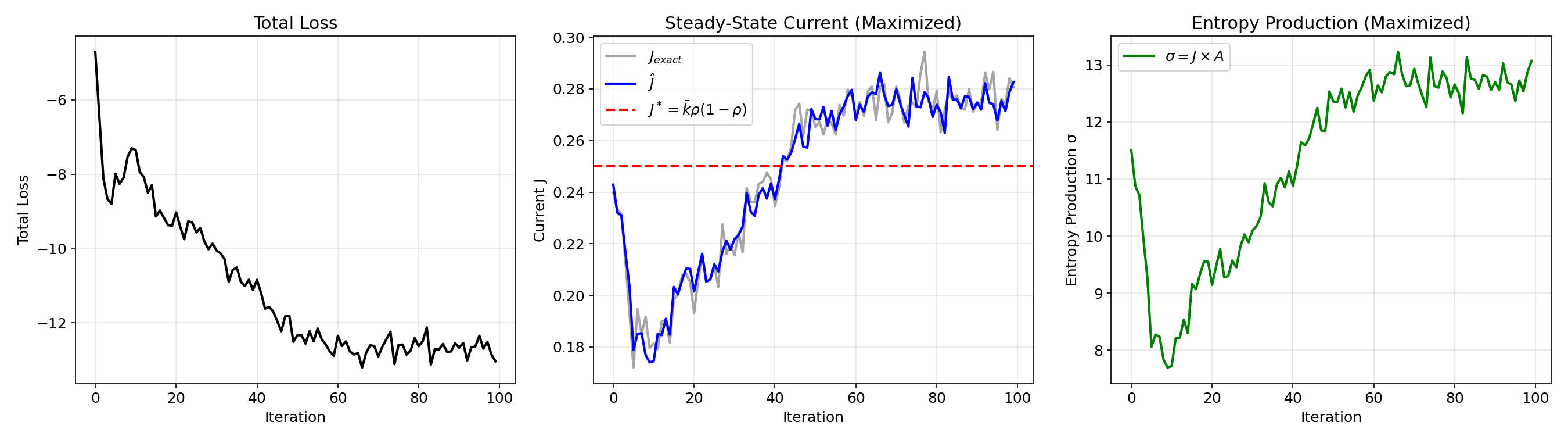}\\[0.3em]
\includegraphics[width=\textwidth]{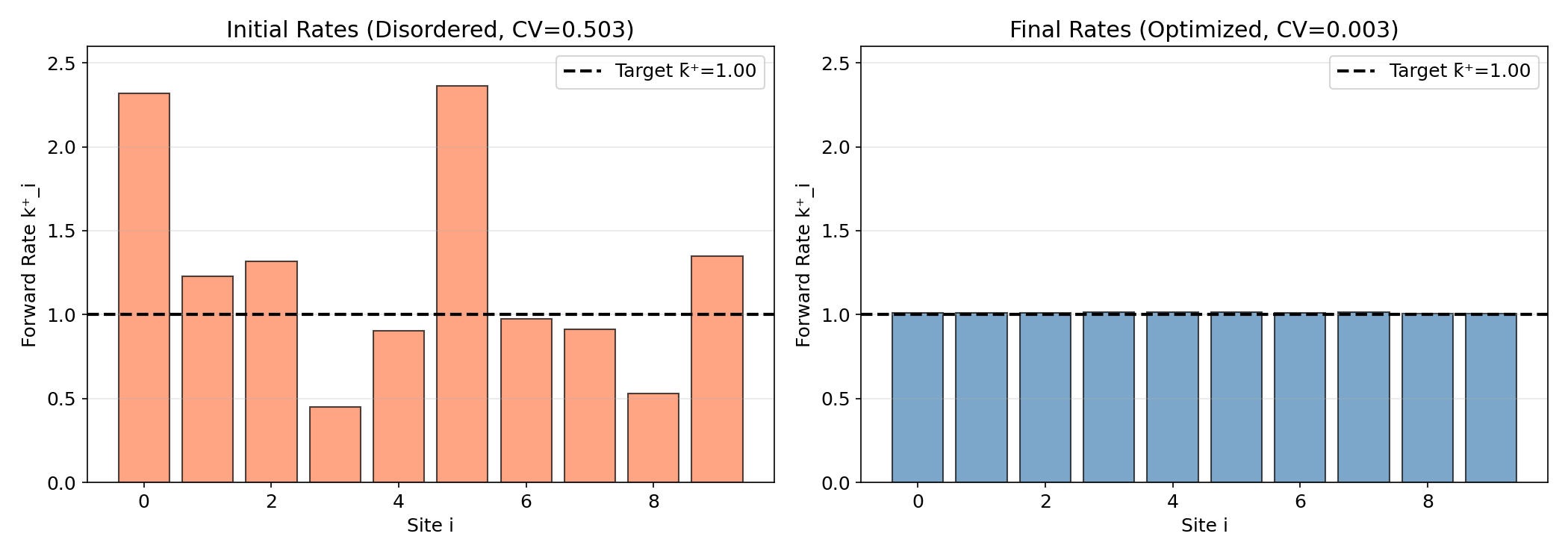}
\caption{\textbf{ASEP optimization details, $L=10$, $\rho=0.5$ ($N=5$).} Same format as Fig.~\ref{fig:asep-L10-d01}.}
\label{fig:asep-L10-d05}
\end{figure*}

\begin{figure*}[p]
\centering
\includegraphics[width=\textwidth]{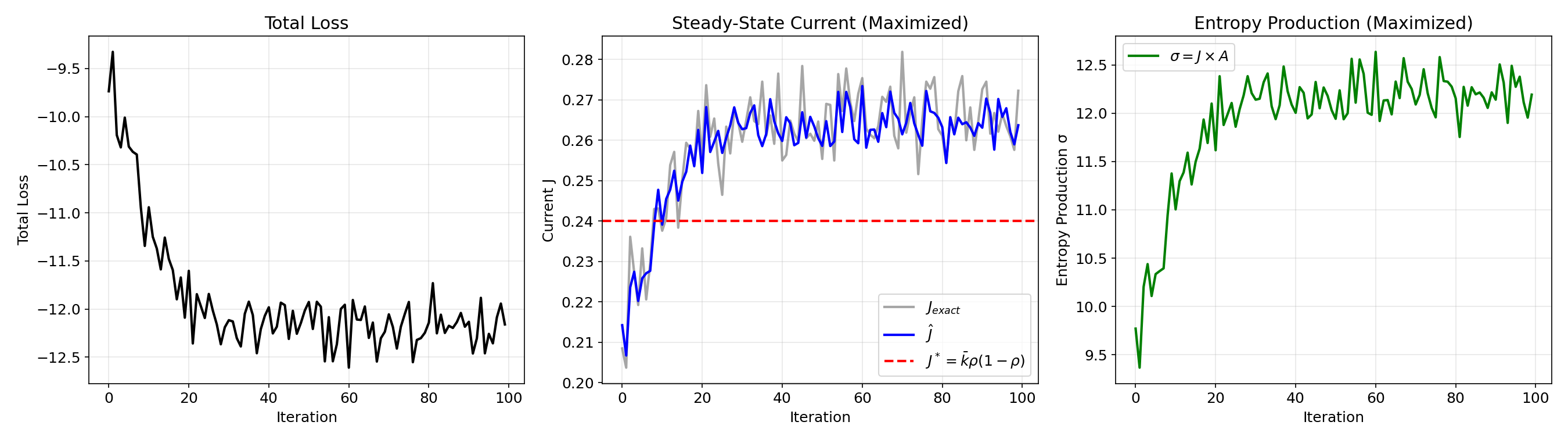}\\[0.3em]
\includegraphics[width=\textwidth]{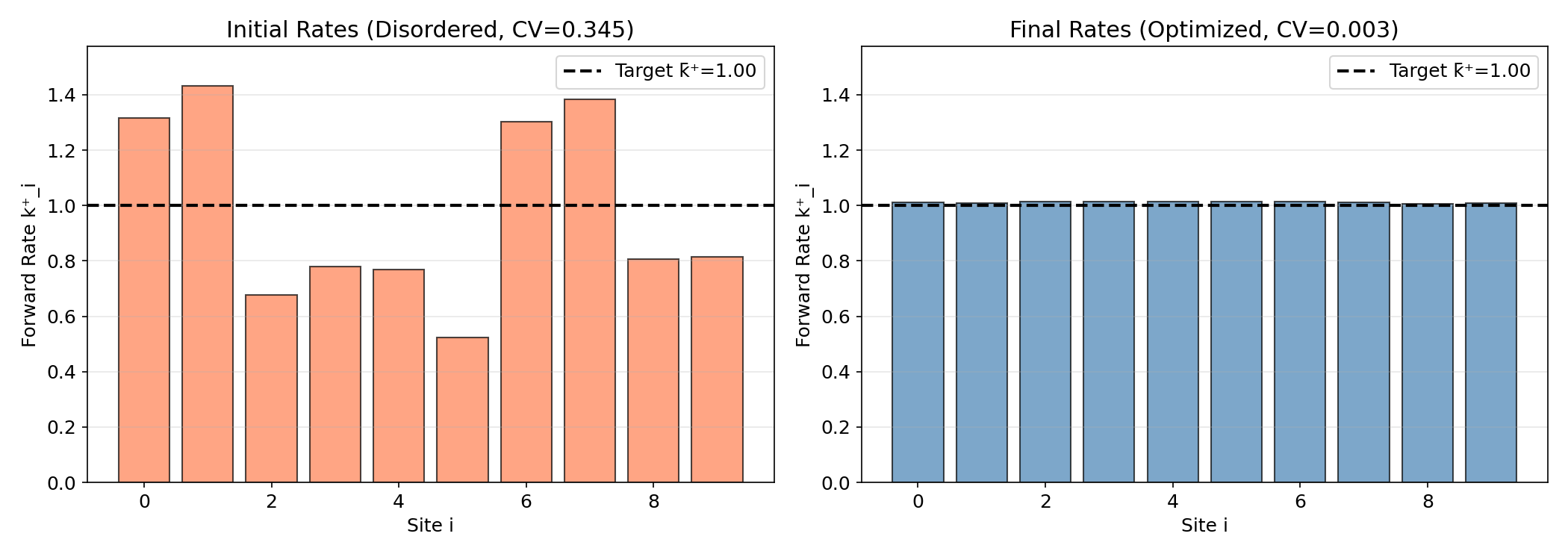}
\caption{\textbf{ASEP optimization details, $L=10$, $\rho=0.6$ ($N=6$).} Same format as Fig.~\ref{fig:asep-L10-d01}.}
\label{fig:asep-L10-d06}
\end{figure*}

\begin{figure*}[p]
\centering
\includegraphics[width=\textwidth]{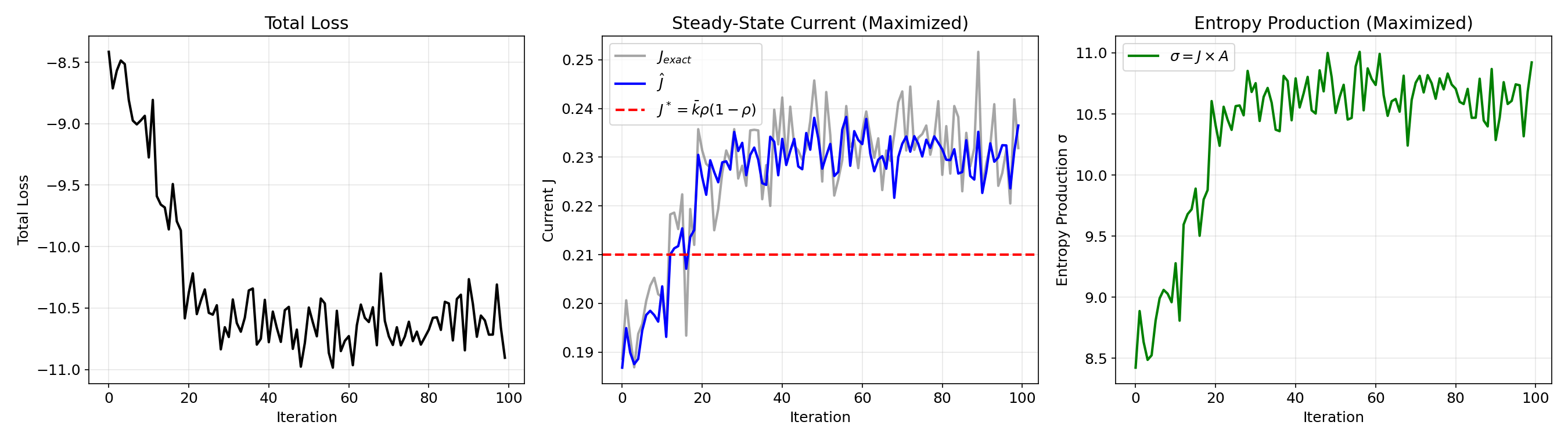}\\[0.3em]
\includegraphics[width=\textwidth]{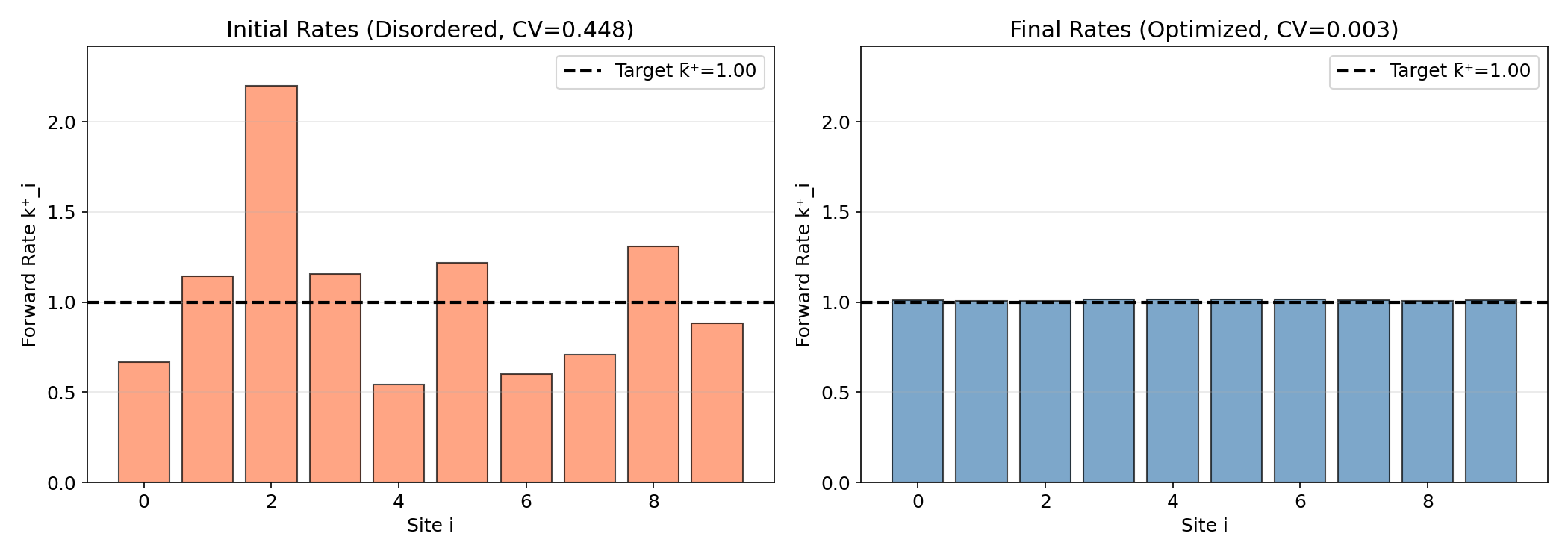}
\caption{\textbf{ASEP optimization details, $L=10$, $\rho=0.7$ ($N=7$).} Same format as Fig.~\ref{fig:asep-L10-d01}.}
\label{fig:asep-L10-d07}
\end{figure*}

\begin{figure*}[p]
\centering
\includegraphics[width=\textwidth]{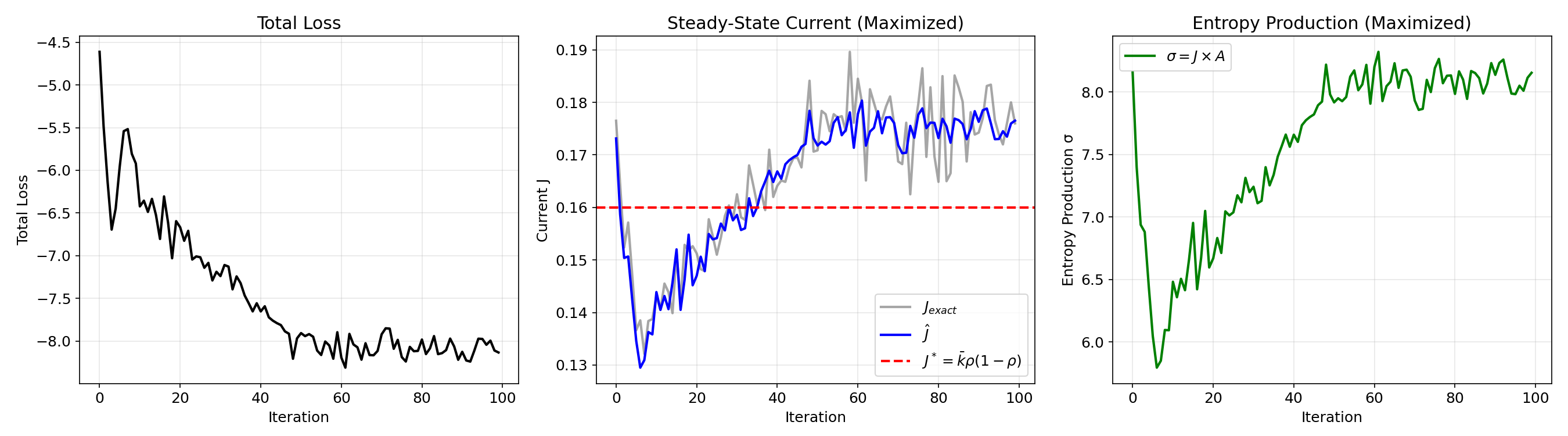}\\[0.3em]
\includegraphics[width=\textwidth]{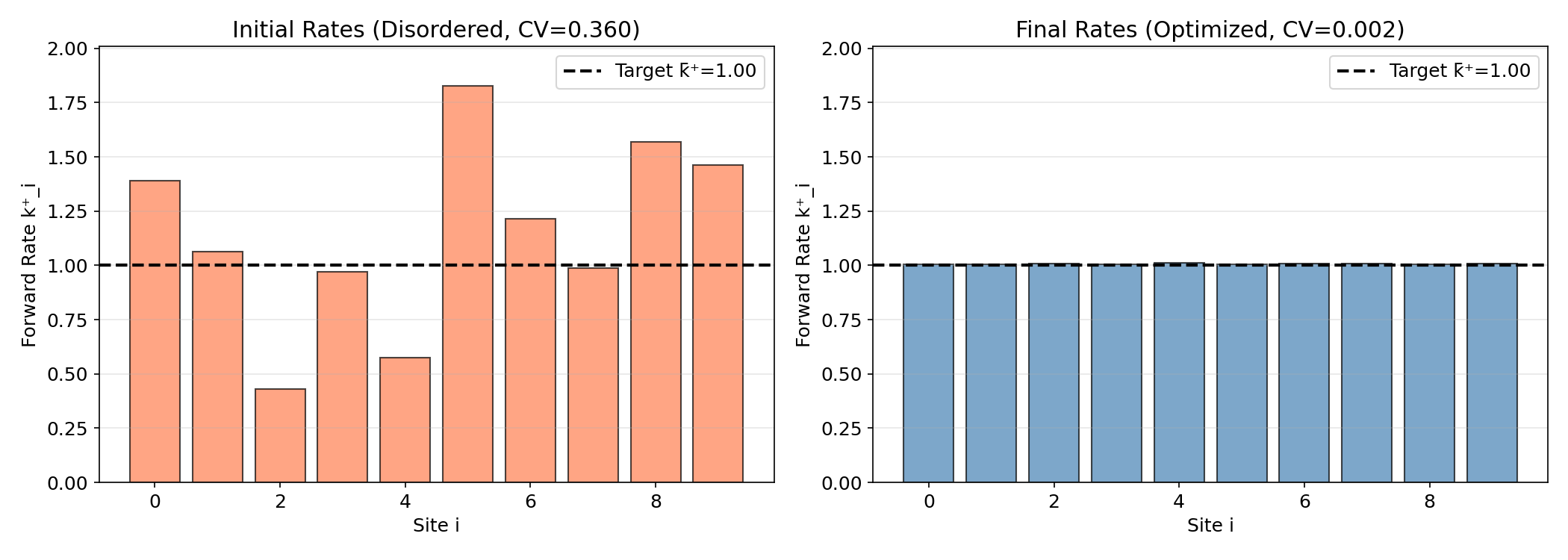}
\caption{\textbf{ASEP optimization details, $L=10$, $\rho=0.8$ ($N=8$).} Same format as Fig.~\ref{fig:asep-L10-d01}.}
\label{fig:asep-L10-d08}
\end{figure*}

\begin{figure*}[p]
\centering
\includegraphics[width=\textwidth]{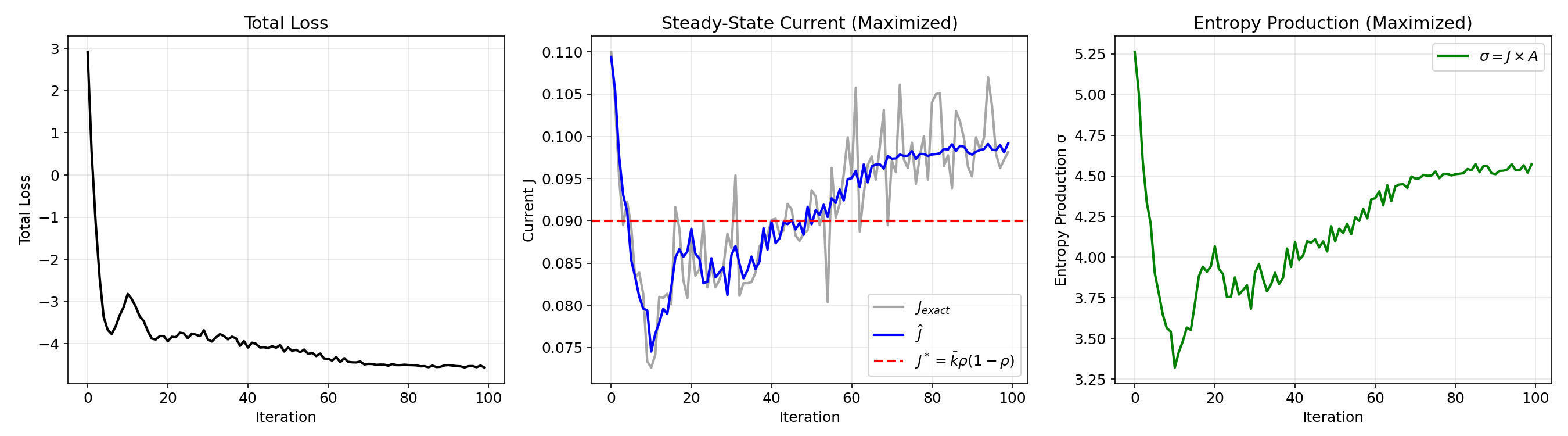}\\[0.3em]
\includegraphics[width=\textwidth]{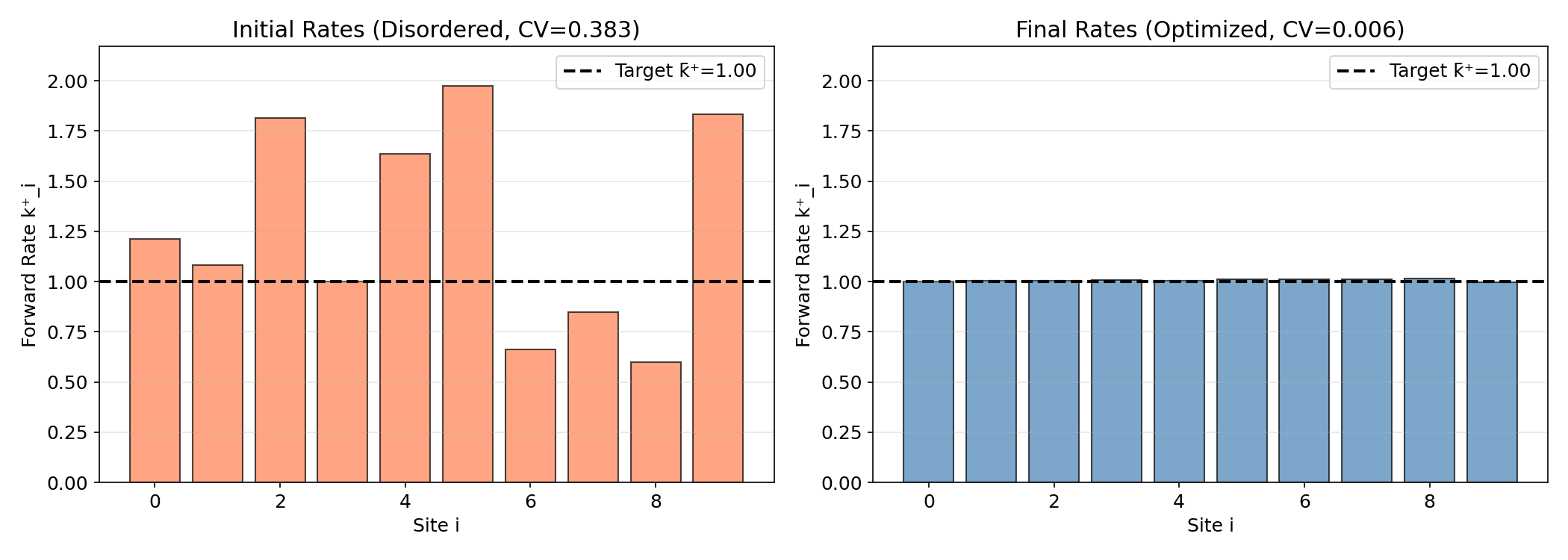}
\caption{\textbf{ASEP optimization details, $L=10$, $\rho=0.9$ ($N=9$).} Same format as Fig.~\ref{fig:asep-L10-d01}.}
\label{fig:asep-L10-d09}
\end{figure*}

\begin{figure*}[h!]
\centering
\includegraphics[width=\textwidth]{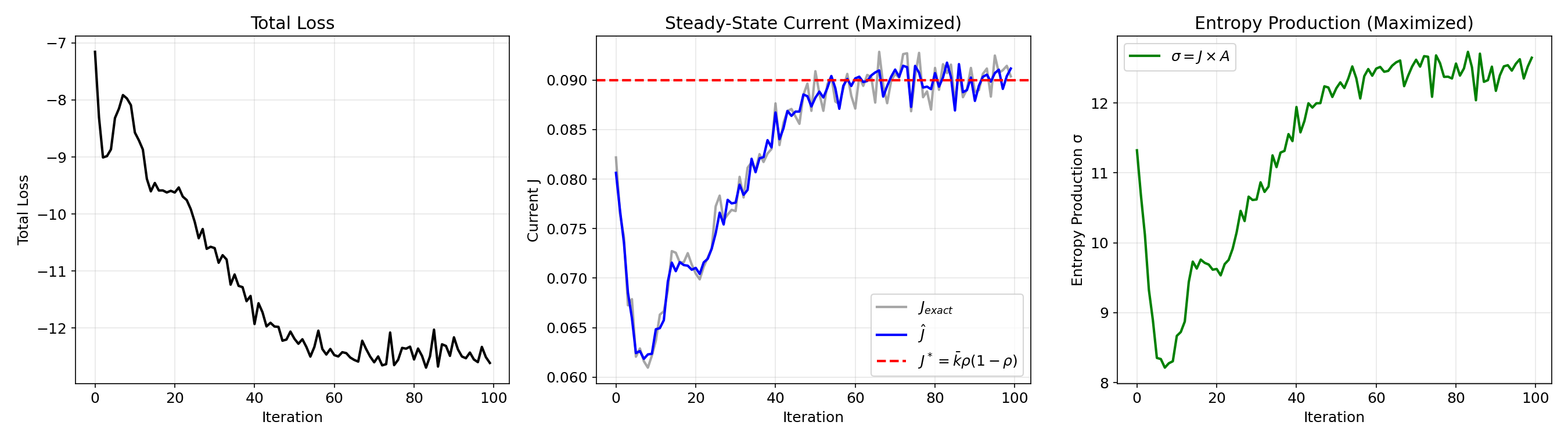}\\[0.3em]
\includegraphics[width=\textwidth]{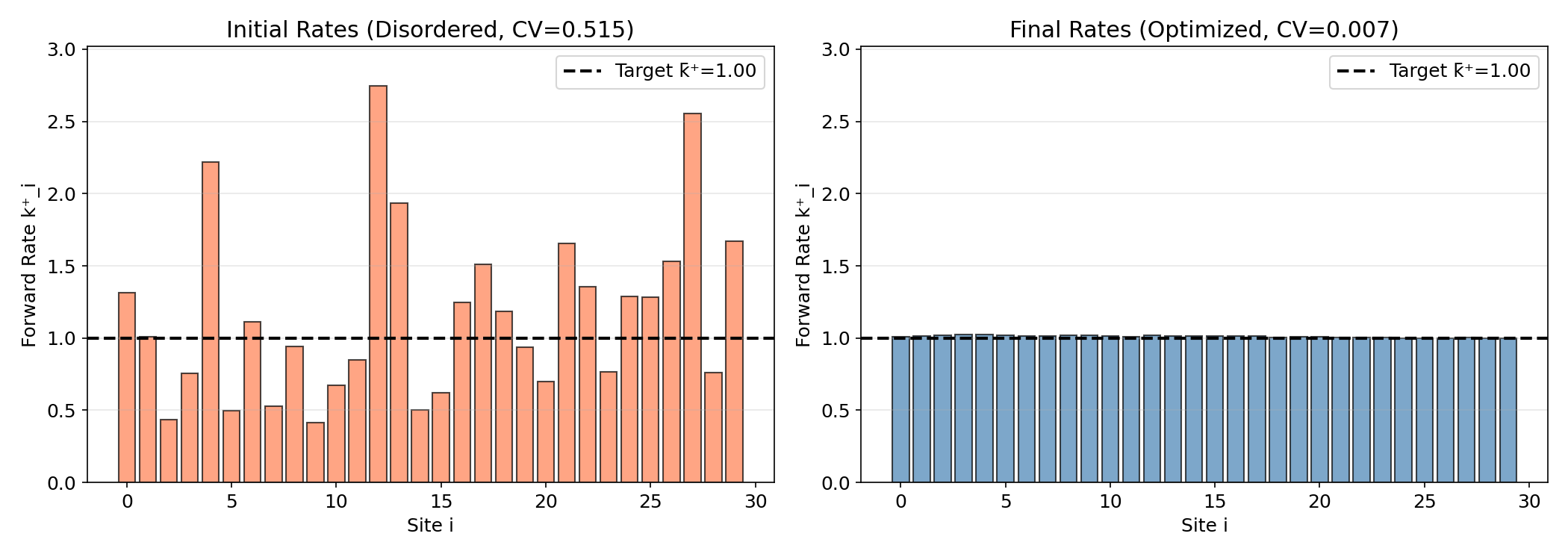}
\caption{\textbf{ASEP optimization details, $L=30$, $\rho=0.1$ ($N=3$).} Same format as Fig.~\ref{fig:asep-L10-d01}.}
\label{fig:asep-L30-d01}
\end{figure*}

\begin{figure*}[h]
\centering
\includegraphics[width=\textwidth]{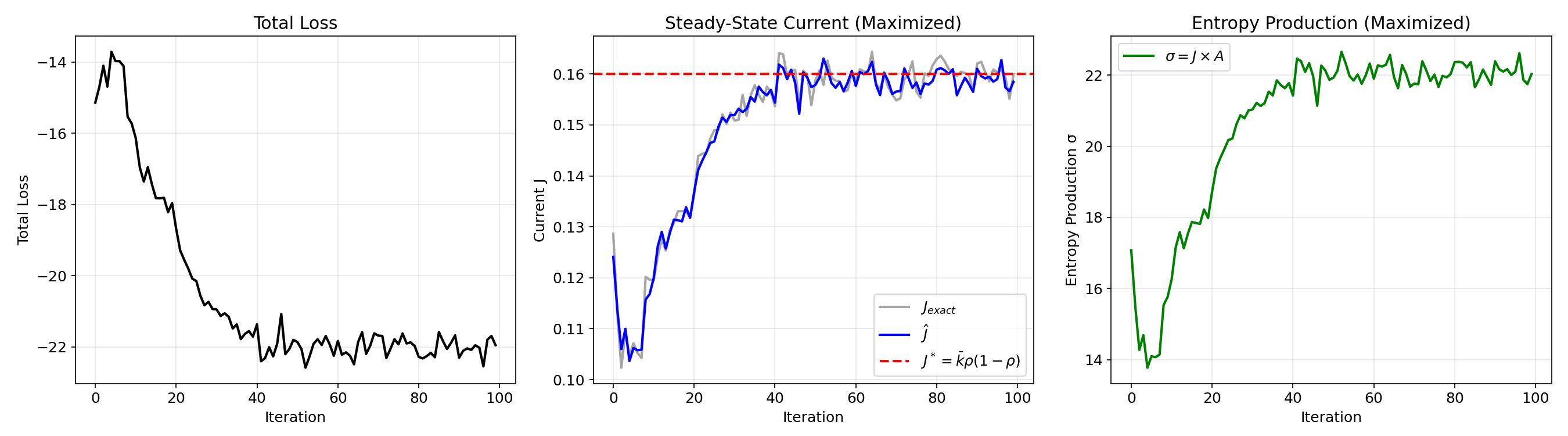}\\[0.3em]
\includegraphics[width=\textwidth]{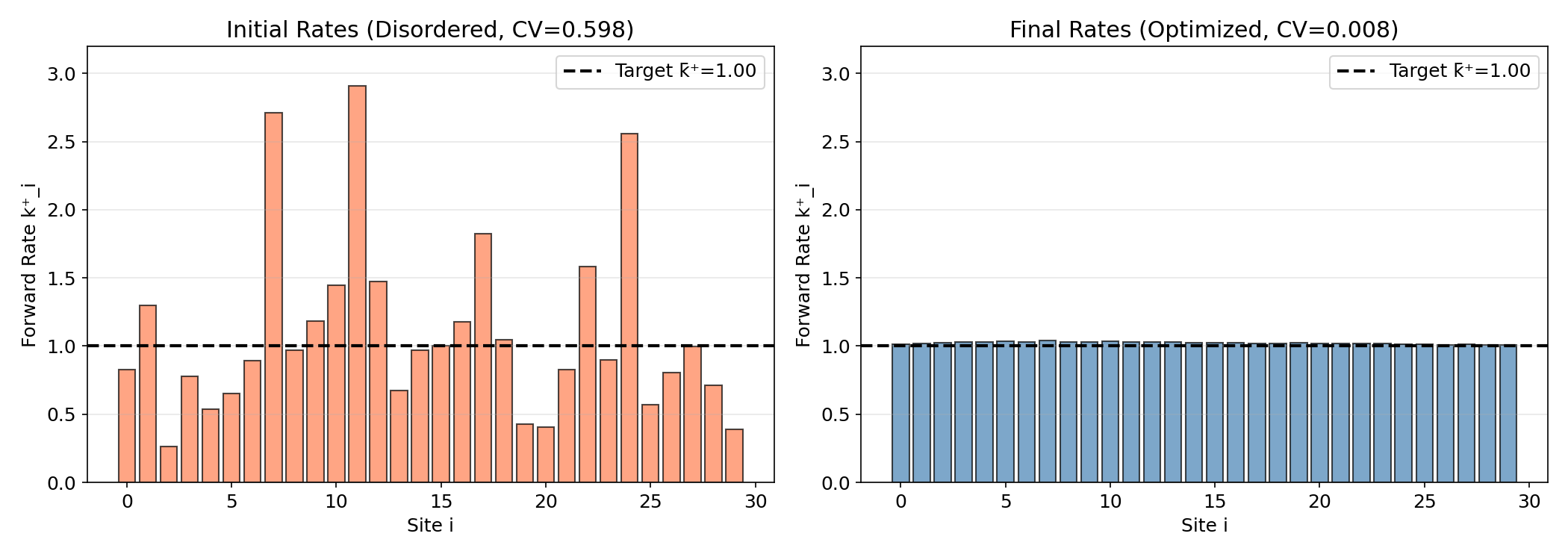}
\caption{\textbf{ASEP optimization details, $L=30$, $\rho=0.2$ ($N=6$).} Same format as Fig.~\ref{fig:asep-L10-d01}.}
\label{fig:asep-L30-d02}
\end{figure*}

\begin{figure*}[p]
\centering
\includegraphics[width=\textwidth]{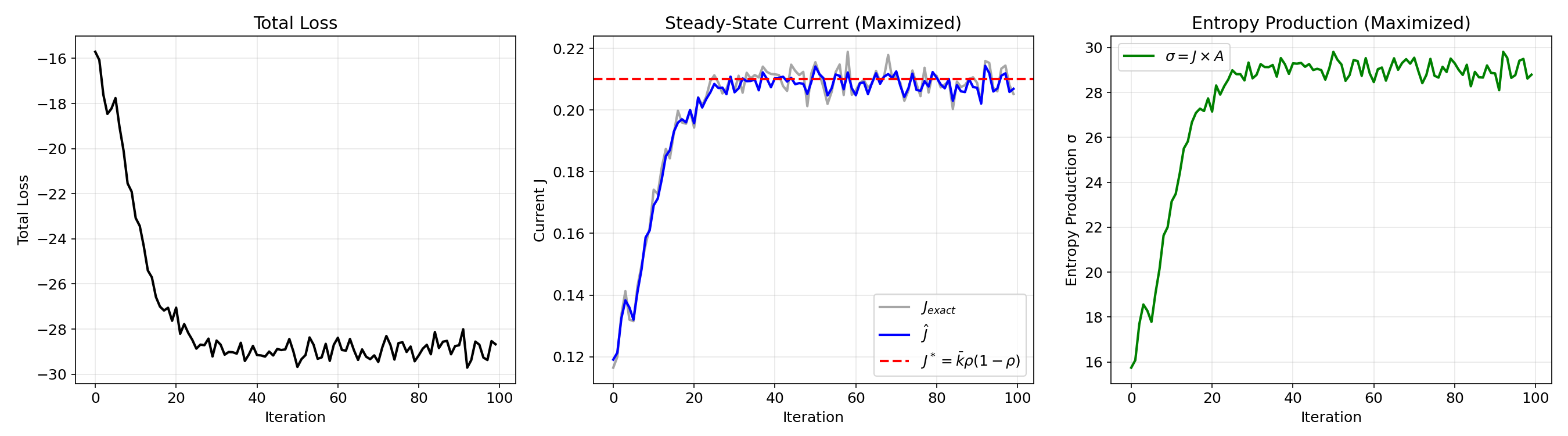}\\[0.3em]
\includegraphics[width=\textwidth]{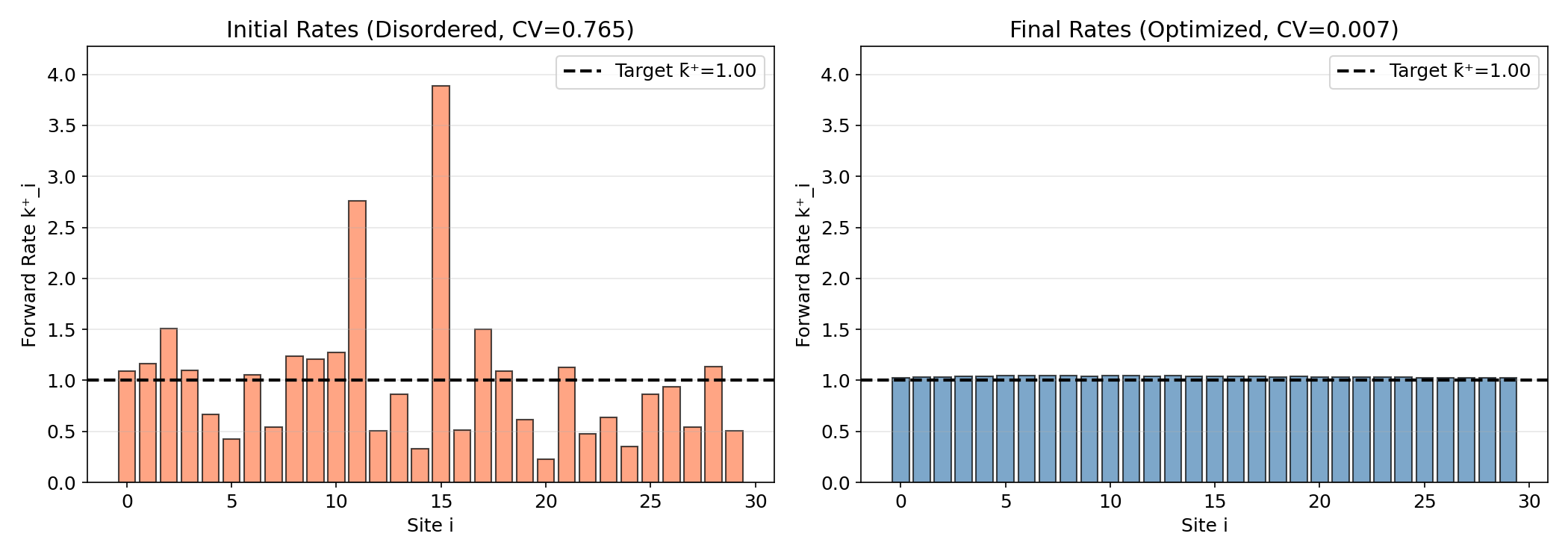}
\caption{\textbf{ASEP optimization details, $L=30$, $\rho=0.3$ ($N=9$).} Same format as Fig.~\ref{fig:asep-L10-d01}.}
\label{fig:asep-L30-d03}
\end{figure*}

\begin{figure*}[p]
\centering
\includegraphics[width=\textwidth]{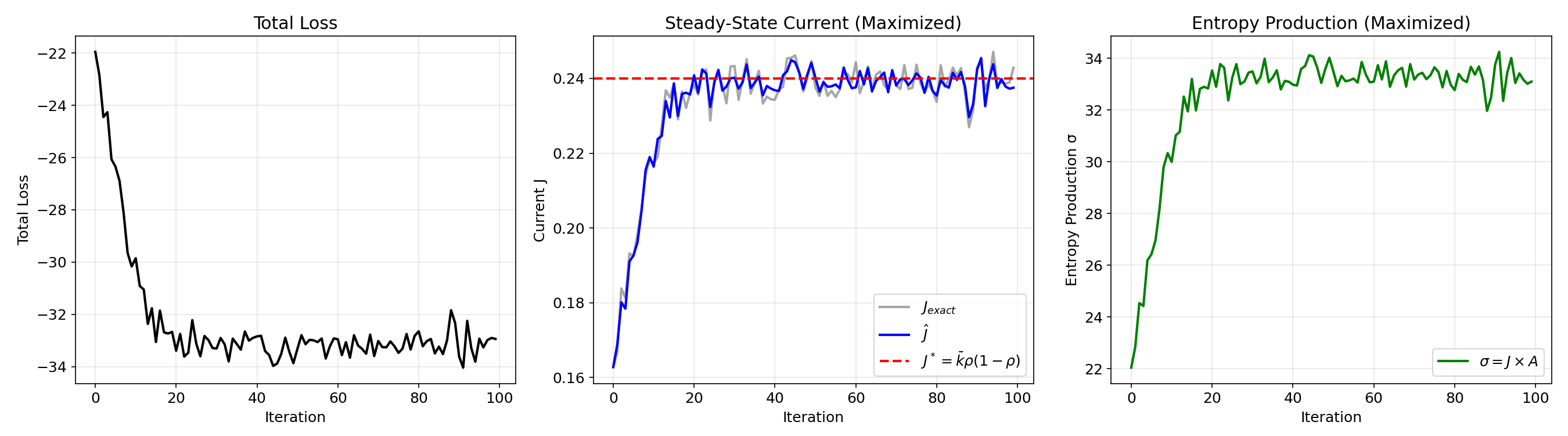}\\[0.3em]
\includegraphics[width=\textwidth]{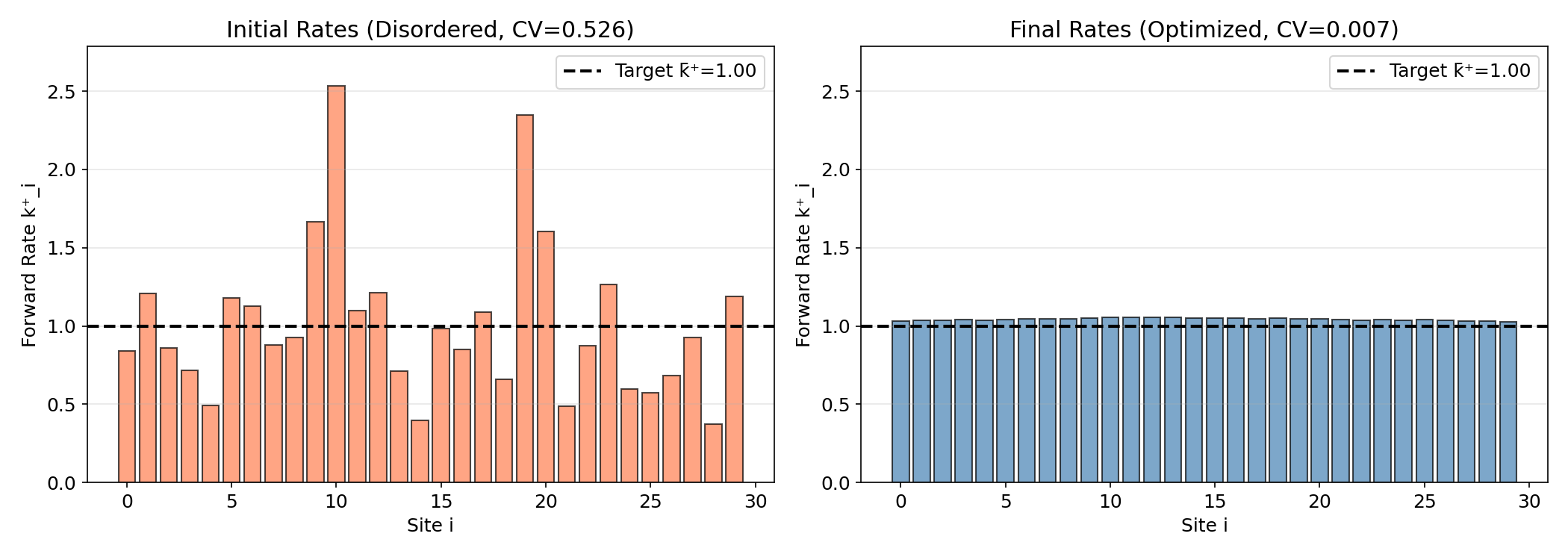}
\caption{\textbf{ASEP optimization details, $L=30$, $\rho=0.4$ ($N=12$).} Same format as Fig.~\ref{fig:asep-L10-d01}.}
\label{fig:asep-L30-d04}
\end{figure*}

\begin{figure*}[p]
\centering
\includegraphics[width=\textwidth]{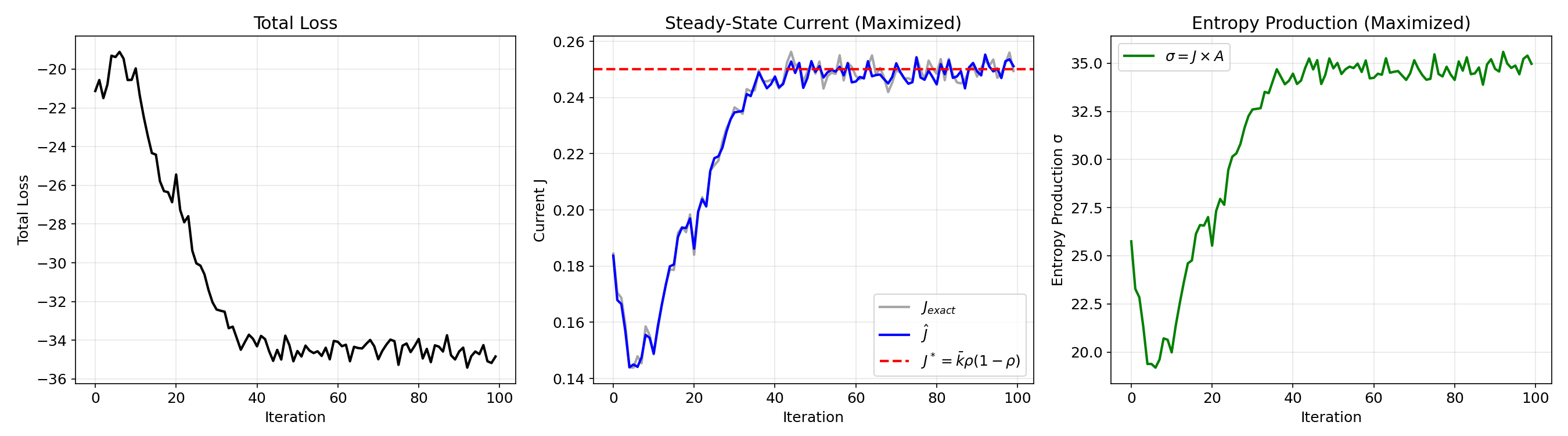}\\[0.3em]
\includegraphics[width=\textwidth]{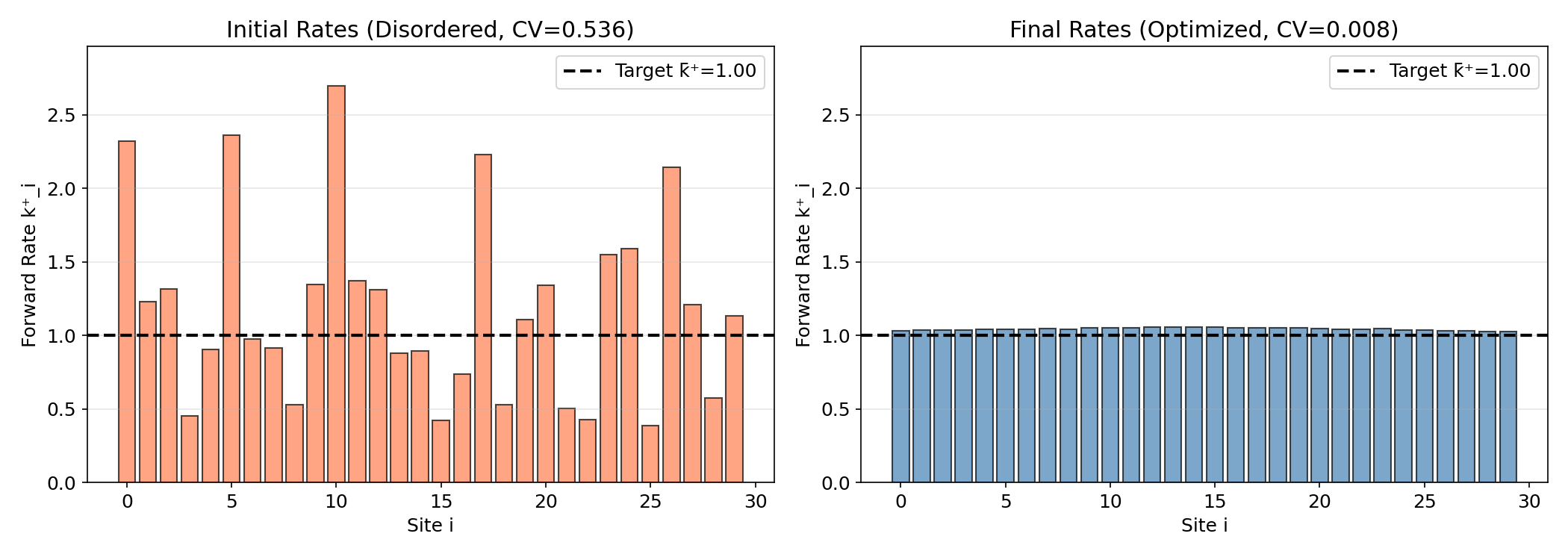}
\caption{\textbf{ASEP optimization details, $L=30$, $\rho=0.5$ ($N=15$).} Same format as Fig.~\ref{fig:asep-L10-d01}.}
\label{fig:asep-L30-d05}
\end{figure*}

\begin{figure*}[p]
\centering
\includegraphics[width=\textwidth]{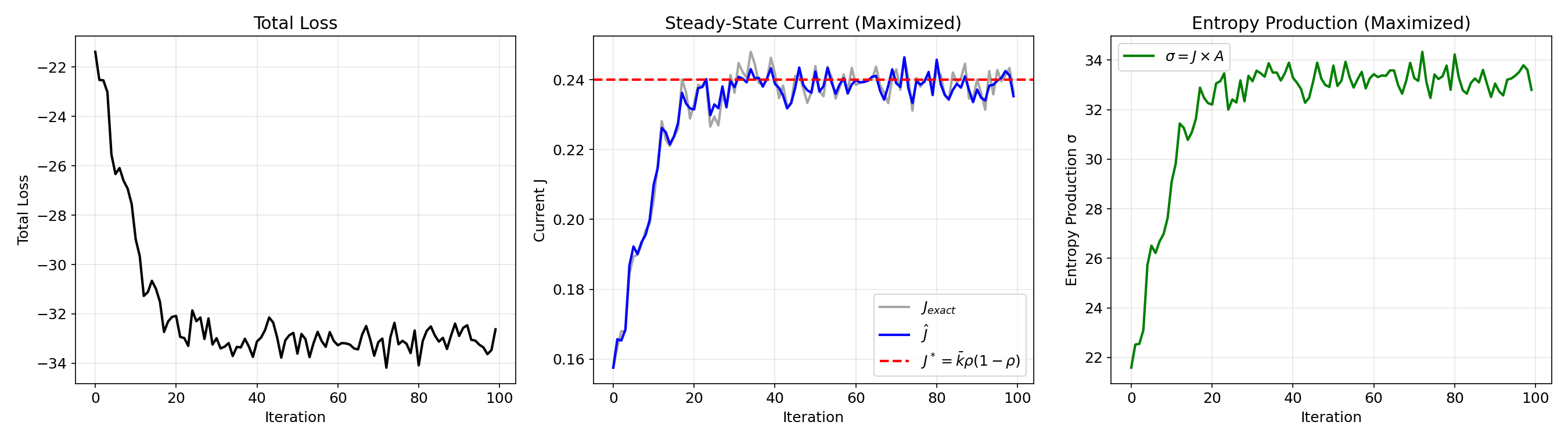}\\[0.3em]
\includegraphics[width=\textwidth]{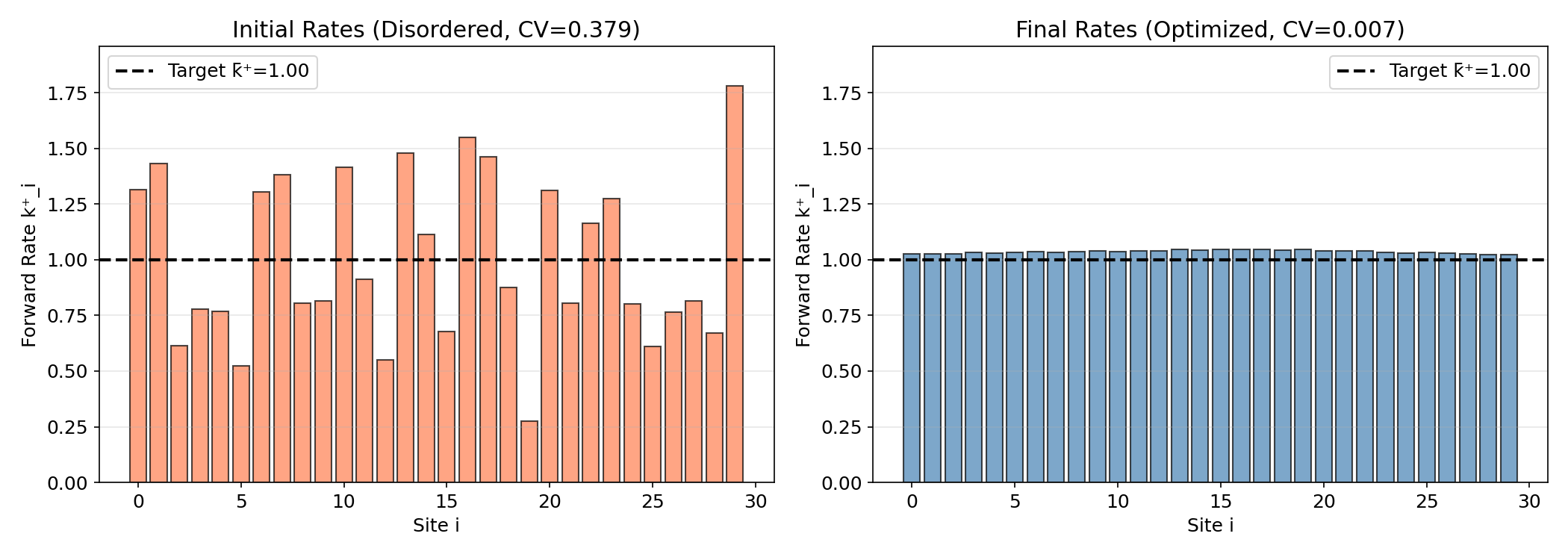}
\caption{\textbf{ASEP optimization details, $L=30$, $\rho=0.6$ ($N=18$).} Same format as Fig.~\ref{fig:asep-L10-d01}.}
\label{fig:asep-L30-d06}
\end{figure*}

\begin{figure*}[p]
\centering
\includegraphics[width=\textwidth]{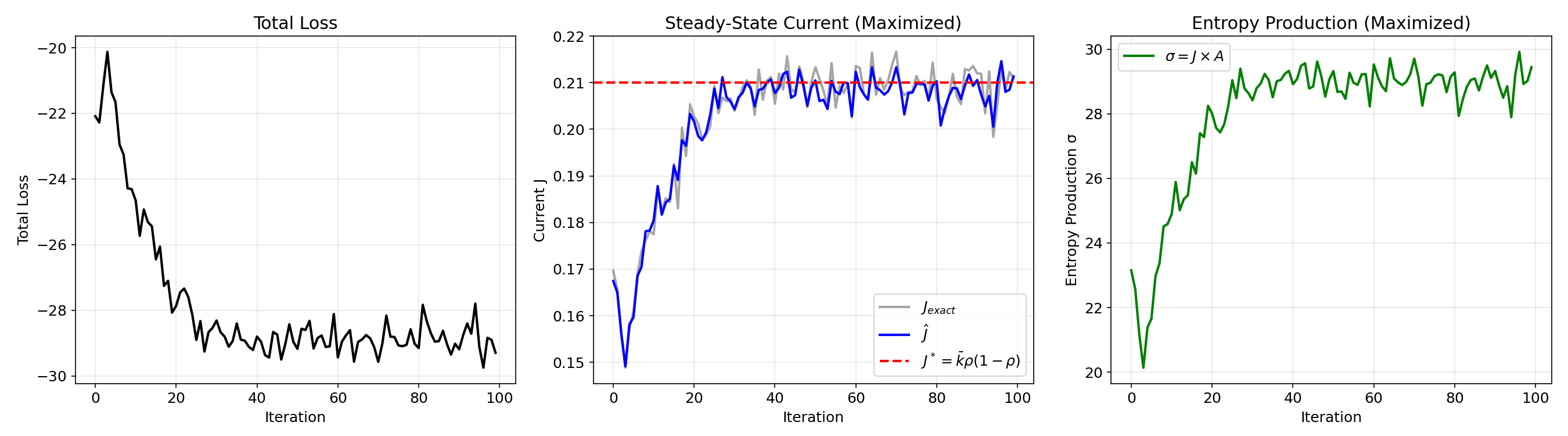}\\[0.3em]
\includegraphics[width=\textwidth]{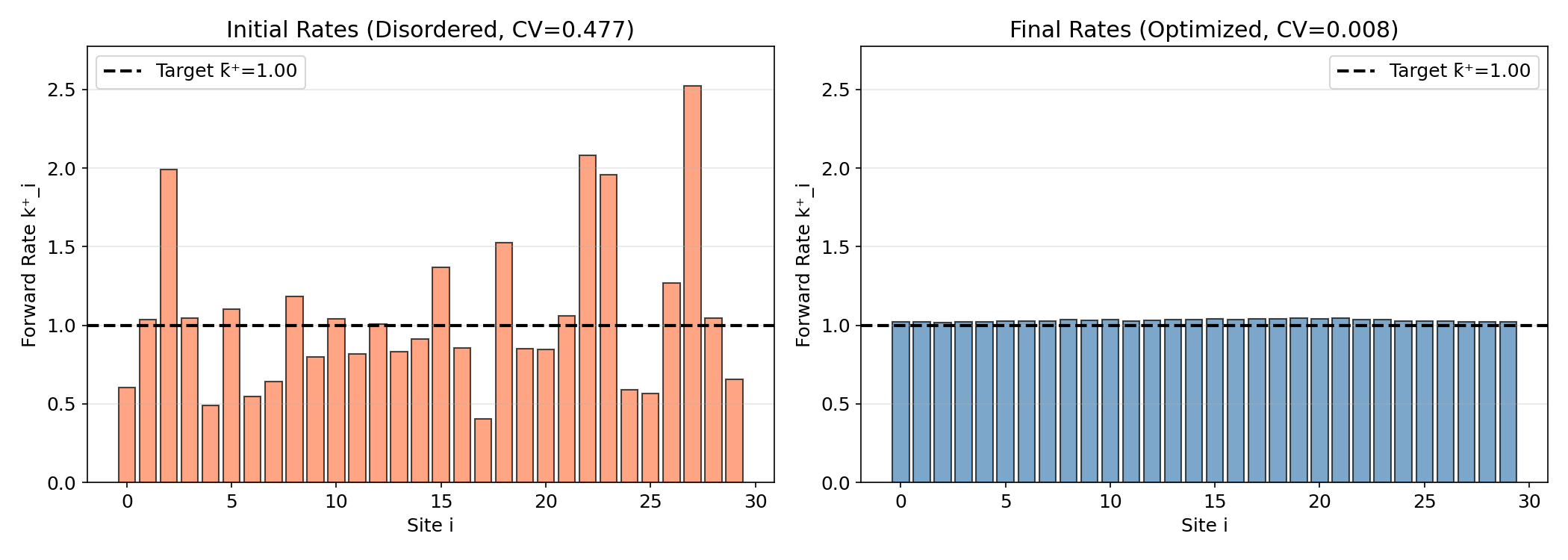}
\caption{\textbf{ASEP optimization details, $L=30$, $\rho=0.7$ ($N=21$).} Same format as Fig.~\ref{fig:asep-L10-d01}.}
\label{fig:asep-L30-d07}
\end{figure*}

\begin{figure*}[p]
\centering
\includegraphics[width=\textwidth]{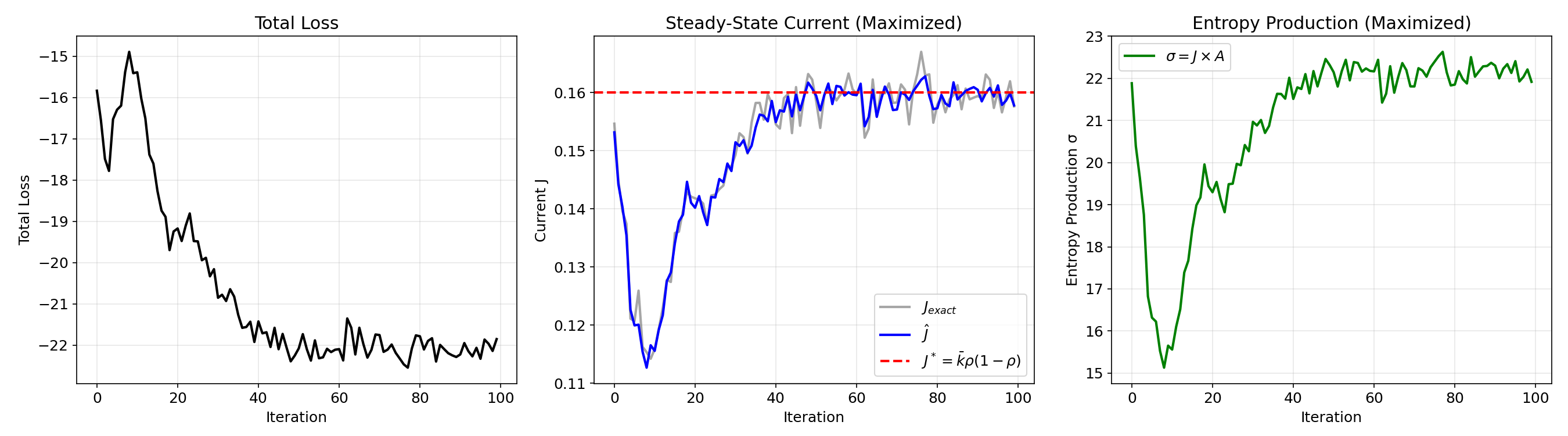}\\[0.3em]
\includegraphics[width=\textwidth]{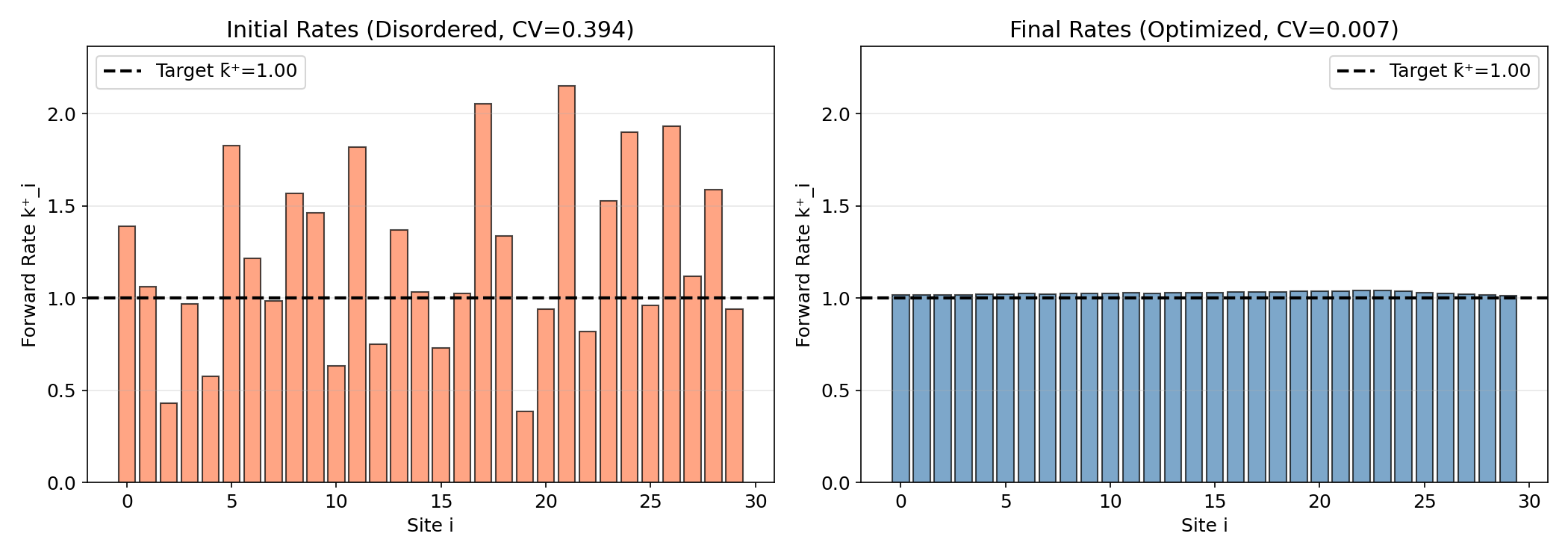}
\caption{\textbf{ASEP optimization details, $L=30$, $\rho=0.8$ ($N=24$).} Same format as Fig.~\ref{fig:asep-L10-d01}.}
\label{fig:asep-L30-d08}
\end{figure*}

\begin{figure*}[p]
\centering
\includegraphics[width=\textwidth]{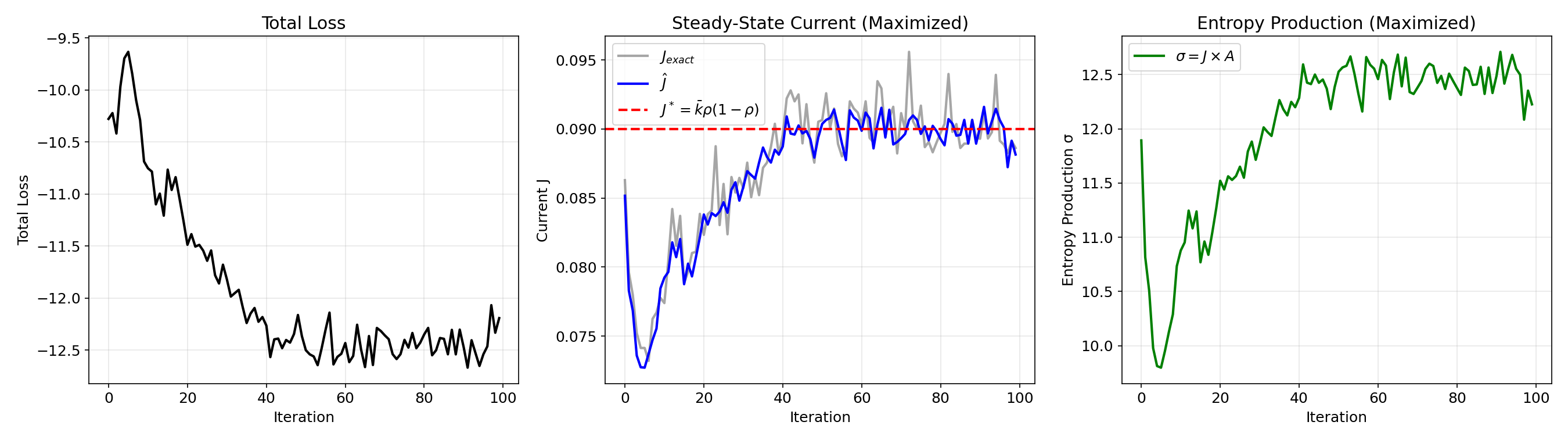}\\[0.3em]
\includegraphics[width=\textwidth]{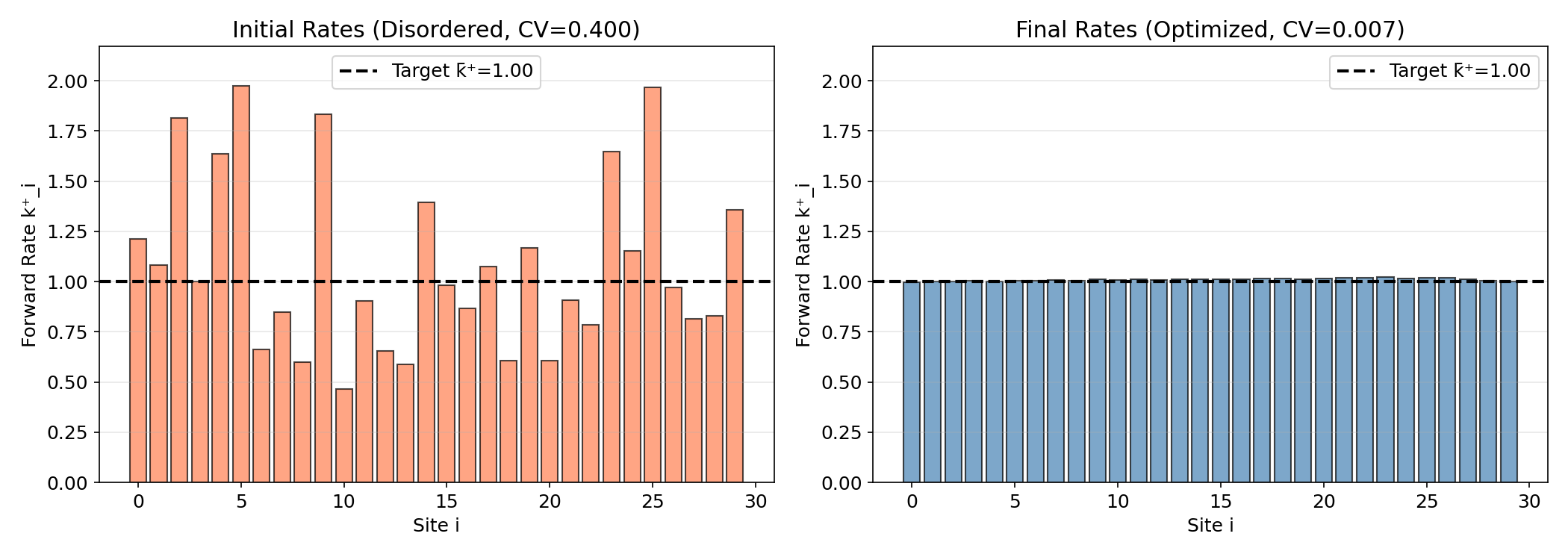}
\caption{\textbf{ASEP optimization details, $L=30$, $\rho=0.9$ ($N=27$).} Same format as Fig.~\ref{fig:asep-L10-d01}.}
\label{fig:asep-L30-d09}
\end{figure*}

\textbf{Additional Results.} Figs.~\ref{fig:asep-L10-d01}--\ref{fig:asep-L10-d09} show training traces and forward rate profiles (initial and optimized) for $L=10$ at all nine densities; Figs.~\ref{fig:asep-L30-d01}--\ref{fig:asep-L30-d09} show the corresponding results for $L=30$.


\subsection{L=10: Convergence to the Finite-Size Current Optimum}

The red dashed line in the training traces indicates the mean-field current $J_\text{MF} = (\bar{k}^+ - k^-)\rho(1-\rho)$. For $L=10$, the optimized current converges not to the mean-field prediction but to the finite-size corrected value $J^* = J_\text{MF} \cdot L/(L-1)$, as shown in the main text (Fig.~\ref{fig:asep}b).

\vspace{0.5cm}

\clearpage
\subsection{L=30: Approach to the Mean-Field Limit}

For $L=30$, the finite-size correction $L/(L-1) \approx 1.03$ is small, and the optimized current closely approaches the mean-field prediction across all densities (Fig.~\ref{fig:asep}b).

\clearpage
\section{Optimal Current-Dissipation Trade-Off in a Three-State Ring}
\label{sec:three-state-ring}

In addition to the ASEP problem presented in the main text, we apply our framework to a complementary inverse design problem in stochastic thermodynamics: characterizing the Pareto front of optimal trade-offs between steady-state current and entropy production in a driven system with fixed kinetic resources. This problem combines core ingredients of non-equilibrium thermodynamics (cycle currents, thermodynamic affinities, dissipation constraints) in a minimal setting that admits analytical solution, providing a rigorous benchmark.

\begin{figure*}[t]
    \includegraphics[width=\textwidth]{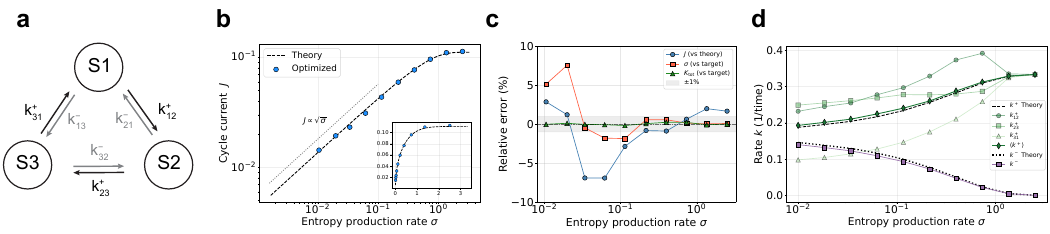}
    \caption{\textbf{Optimal current-dissipation trade-off in a three-state ring.} Optimization of cycle current $J$ under constraints on entropy production rate $\sigma$ and total kinetic budget $K_\text{tot}$. (a) Schematic of the three-state ring with forward ($k_{ij}^+$) and backward ($k_{ij}^-$) transition rates. (b) Optimized cycle current (blue circles) compared to the theoretical Pareto front (dashed line). Dotted line: asymptotic near-equilibrium scaling. Inset: Pareto front on linear axes. (c) Relative errors in $J$ (versus theory), $\sigma$, and $K_\text{tot}$ (versus target). Shaded band: $\pm 1\%$. (d) Optimized forward rates $k_{12}^+$, $k_{23}^+$, $k_{31}^+$ (light green symbols), their mean $\langle k^+ \rangle$ (green diamonds), and backward rate $k^-$ (squares), compared to theoretical predictions (dashed and dotted lines).}
\label{fig:loop}
\end{figure*}

We study a three-state ring~\cite{hill1989free} where a particle transitions cyclically among states $S_1 \to S_2 \to S_3 \to S_1$, with each edge supporting both forward and backward transitions (Fig.~\ref{fig:loop}a):
\begin{align*}
    S_1 &\xrightleftharpoons[k_{21}^-]{k_{12}^+} S_2, \qquad
    S_2 \xrightleftharpoons[k_{32}^-]{k_{23}^+} S_3, \qquad
    S_3 \xrightleftharpoons[k_{13}^-]{k_{31}^+} S_1
\end{align*}
Breaking detailed balance induces a net cycle current $J$ with entropy production rate $\sigma = J \mathcal{A}$, where $\mathcal{A} = \sum_i \log(k_i^+ / k_i^-)$ is the cycle affinity~\cite{schnakenberg1976network}. We impose a kinetic budget $K_\text{tot} = \sum_i (k_i^+ + k_i^-)$ bounding the system's dynamical activity~\cite{maes2020frenesy}.

The inverse design objective is to maximize $J$ at fixed $\sigma$ and $K_\text{tot}$. As in the ASEP problem, we estimate the current from time-averaged propensities along simulated trajectories:
$$
\hat{J} = \frac{1}{3T}\sum_n \left[\sum_i a_i^+(t_n) - \sum_i a_i^-(t_n)\right] \Delta t_n
$$
where $a_i^{\pm}(t_n)$ are the forward and backward propensities and $\Delta t_n$ is the waiting time at step $n$. This time-weighted propensity average is differentiable and provides gradients with respect to all rate parameters. The entropy production estimate follows as $\hat{\sigma} = \hat{J}\,\mathcal{A}$. Since thermodynamic behavior depends on rate ratios $k_i^+/k_i^-$, the same physical driving can arise from many absolute rate values. We constrain all backward rates to a common value (still learnable given the fixed total budget), establishing a reference scale that removes this degeneracy and makes the optimized forward rates directly comparable across edges. Constraints are enforced through logarithmic penalties, ensuring scale-invariant treatment across the range of target values:
$$
\mathcal{L} = -\hat{J} + \lambda_\sigma \left[\log\frac{\hat{\sigma}}{\sigma_\text{target}}\right]^2 + \lambda_K \left[\log\frac{K_\text{tot}}{K_\text{target}}\right]^2
$$
We optimize across target entropy production values spanning two orders of magnitude ($\sigma \in [0.01, 2.5]$). All optimizations start from the same random initial rates and are run to stable convergence using the modified Adam optimizer described in \textit{Methods}.

For the symmetric ring with fixed kinetic budget $K_\text{tot}$, the Pareto-optimal current at affinity $\mathcal{A}$ follows from steady-state analysis (Section~\ref{sec:pareto-theory}):
$$
J = \frac{K_\text{tot}}{9}\tanh\frac{\mathcal{A}}{6}, \qquad \sigma = J\mathcal{A}
$$
This defines a fundamental bound: no rate configuration with the given kinetic budget can achieve higher current at a given dissipation level.

The optimized cycle currents closely track the theoretical Pareto front across the full range of target dissipation values (Fig.~\ref{fig:loop}b). Agreement extends from the near-equilibrium regime ($\sigma \ll 1$), where $J \propto \sqrt{\sigma}$ (dotted line)~\cite{onsager1931reciprocal}, to the far-from-equilibrium regime ($\sigma \gtrsim 1$), where current saturates toward its maximum value $J_\text{max} = K_\text{tot}/9$. To assess constraint satisfaction, we evaluate relative errors in the optimized current (compared to the theoretical optimum), entropy production rate, and total kinetic budget (compared to their target values). The kinetic budget constraint is satisfied to within fractions of 1\% across all runs (Fig.~\ref{fig:loop}c, green triangles; Table~\ref{tab:pareto-results}), ensuring valid comparison with the theoretical bound. Relative errors in $J$ and $\sigma$ remain small throughout the moderate-to-high dissipation regime but increase at the lowest entropy production targets, where the near-equilibrium current becomes vanishingly small and stochastic fluctuations dominate.

The optimized common backward rate $k^-$ and mean forward rate $\langle k^+ \rangle$ closely follow theoretical predictions (Fig.~\ref{fig:loop}d). The individual forward rates, instead, display substantial spread near equilibrium before converging to the fully symmetric allocation at high dissipation (Fig.~\ref{fig:loop}d, light green symbols). This behavior is consistent with the close agreement between optimized currents and the Pareto bound: at small $\sigma$, the optimal rate configuration is only weakly determined at the level of individual edges. Many distinct allocations of the forward-rate budget yield indistinguishable $J(\sigma)$ once $\sigma$ and $K_{\text{tot}}$ are fixed, producing a broad family of near-degenerate solutions. Physically, although the smallest forward rate would nominally act as a bottleneck, transitions in the near-equilibrium regime are so infrequent that the delay from a single slow edge becomes negligible compared to the long waiting times throughout the cycle. Far from equilibrium, cycles complete rapidly enough that any rate asymmetry creates a persistent bottleneck, reducing the attainable current and driving the system toward symmetric rate allocation. Detailed optimization trajectories for each target $\sigma$ are provided in Figs.~\ref{fig:ring-sigma-0.01}--\ref{fig:ring-sigma-2.5}.

These results establish the framework as a practical tool for rational design of non-equilibrium systems under thermodynamic constraints, capable of tracking optimal solutions across different constraints.

\subsection{Derivation of the Pareto-Optimal Current Bound}
\label{sec:pareto-theory}

We derive here the theoretical upper bound on the steady-state current $J$ achievable in a three-state unicyclic network, subject to a fixed thermodynamic affinity $\mathcal{A}$ and a fixed total kinetic budget $K_\text{tot}$.

Consider a ring of $N=3$ states with periodic boundary conditions, where transitions from state $i$ to state $i+1$ (mod 3) occur with forward rate $k_i^+$ and backward rate $k_i^-$. The system is subject to two constraints. The first is a kinetic budget constraint on the total rate constant expenditure,
$$
K_\text{tot} = \sum_{i=1}^{3} (k_i^+ + k_i^-).
$$
The second is a thermodynamic constraint fixing the cycle affinity,
$$
\mathcal{A} = \ln \prod_{i=1}^{3} \frac{k_i^+}{k_i^-} = \sum_{i=1}^{3} \ln \frac{k_i^+}{k_i^-}.
$$
The affinity quantifies the thermodynamic driving force around the cycle and is related to the entropy production rate by $\sigma = J \mathcal{A}$.

In a three-state ring, current must pass through all three edges to complete a cycle. If rates are distributed unevenly, the edge with the smallest net forward bias acts as a bottleneck limiting the overall flux. Redistributing rates to equalize the edges removes this bottleneck without changing the total affinity. The optimal configuration is therefore symmetric:
$$
k_1^+ = k_2^+ = k_3^+ \equiv k_+, \qquad k_1^- = k_2^- = k_3^- \equiv k_-.
$$

In this symmetric regime, the constraints reduce to
$$
K_\text{tot} = 3(k_+ + k_-), \qquad \mathcal{A} = 3 \ln \frac{k_+}{k_-},
$$
which can be inverted to yield
$$
k_+ + k_- = \frac{K_\text{tot}}{3}, \qquad \frac{k_+}{k_-} = e^{\mathcal{A}/3}.
$$
By symmetry, the steady-state occupation probabilities are uniform, $\pi_i = 1/3$ for all $i$. The net current across any edge is therefore
$$
J = \pi_i k_+ - \pi_{i+1} k_- = \frac{1}{3}(k_+ - k_-).
$$

To express $J$ solely in terms of the constraint parameters, we rewrite the difference of rates as
$$
J = \frac{1}{3}(k_+ - k_-) = \frac{K_\text{tot}}{9} \cdot \frac{k_+ - k_-}{k_+ + k_-},
$$
where we have substituted the budget constraint in the equality. Dividing numerator and denominator by $k_-$ and introducing $\rho = k_+/k_- = e^{\mathcal{A}/3}$ gives
$$
J = \frac{K_\text{tot}}{9} \cdot \frac{\rho - 1}{\rho + 1} = \frac{K_\text{tot}}{9} \cdot \frac{e^{\mathcal{A}/3} - 1}{e^{\mathcal{A}/3} + 1}.
$$
Using the identity $(e^{2x} - 1)/(e^{2x} + 1) = \tanh(x)$ with $x = \mathcal{A}/6$, we obtain the Pareto-optimal bound,
$$
J^*(\mathcal{A}, K_\text{tot}) = \frac{K_\text{tot}}{9} \tanh\left(\frac{\mathcal{A}}{6}\right).
$$

This bound interpolates between two physical regimes. In the linear response regime where $\mathcal{A} \ll 1$, expanding $\tanh(x) \approx x$ gives the Onsager-like linear flux-force relation,
$$
J^* \approx \frac{K_\text{tot}}{54} \mathcal{A}.
$$
Using $\mathcal{A} = \sigma / J$ to eliminate the affinity in favor of entropy production, we obtain the square-root scaling of current with dissipation near equilibrium $J^* \propto \sqrt{\sigma}$. 

In the far-from-equilibrium limit $\mathcal{A} \to \infty$, the hyperbolic tangent saturates to unity and the current approaches a kinetically limited maximum,
$$
J^*_{\max} = \frac{K_\text{tot}}{9},
$$
independent of the driving force.

\subsection{Additional Details}

\textbf{Experiment Details.} We optimized the cycle current $J$ subject to constraints on entropy production rate $\sigma$ and total kinetic budget $K_\text{tot} = 1$. Target entropy production values were logarithmically spaced between $\sigma_\text{min} = 0.01$ and $\sigma_\text{max} = 2.5$, yielding 10 optimization targets. Constraints were enforced via penalty terms with coefficients $\lambda_\sigma = \lambda_{K_\text{tot}} = 100$. Rate constants were parameterized as $k = e^\phi$ and optimization was performed over the log-transformed variables $\phi$. Training ran for 1000 epochs using Adam with learning rate $\text{lr} = 0.05$ and modified hyperparameters ($\beta_1 = 0.8$, $\beta_2 = 0.9$). Gradient estimates were computed from batches of 32 independent trajectories, each simulated for $T_\text{sim} = 500$ time units.

\begin{figure*}[h!]
\centering
\includegraphics[width=\textwidth]{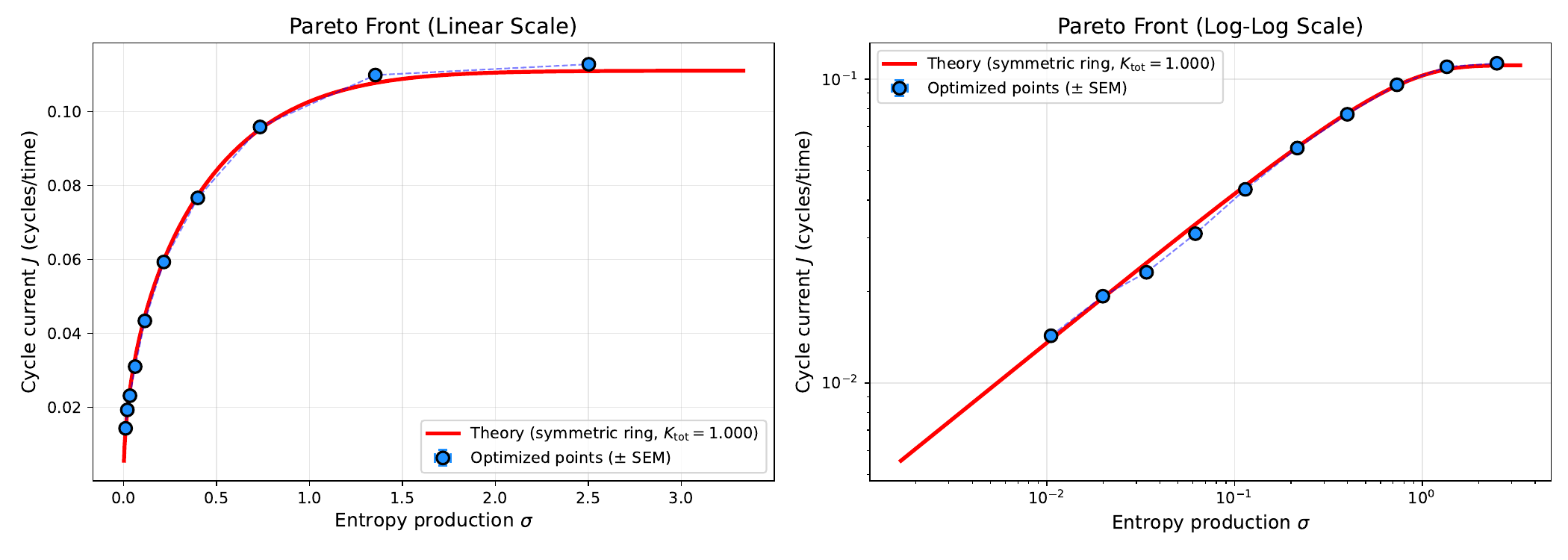}
\caption{\textbf{Pareto front for the three-state cycle model.} Optimized steady-state current $J$ as a function of entropy production rate $\sigma$ for fixed kinetic budget $K_\text{tot} = 1$. \textbf{(Left)} Linear scale. \textbf{(Right)} Log-log scale. Red curve: theoretical Pareto-optimal bound derived for symmetric rate allocation (Section~\ref{sec:pareto-theory}). Blue points: optimized values.}
\label{fig:pareto-front}
\end{figure*}

\begin{figure}[h!]
\centering
\includegraphics[width=0.8\textwidth]{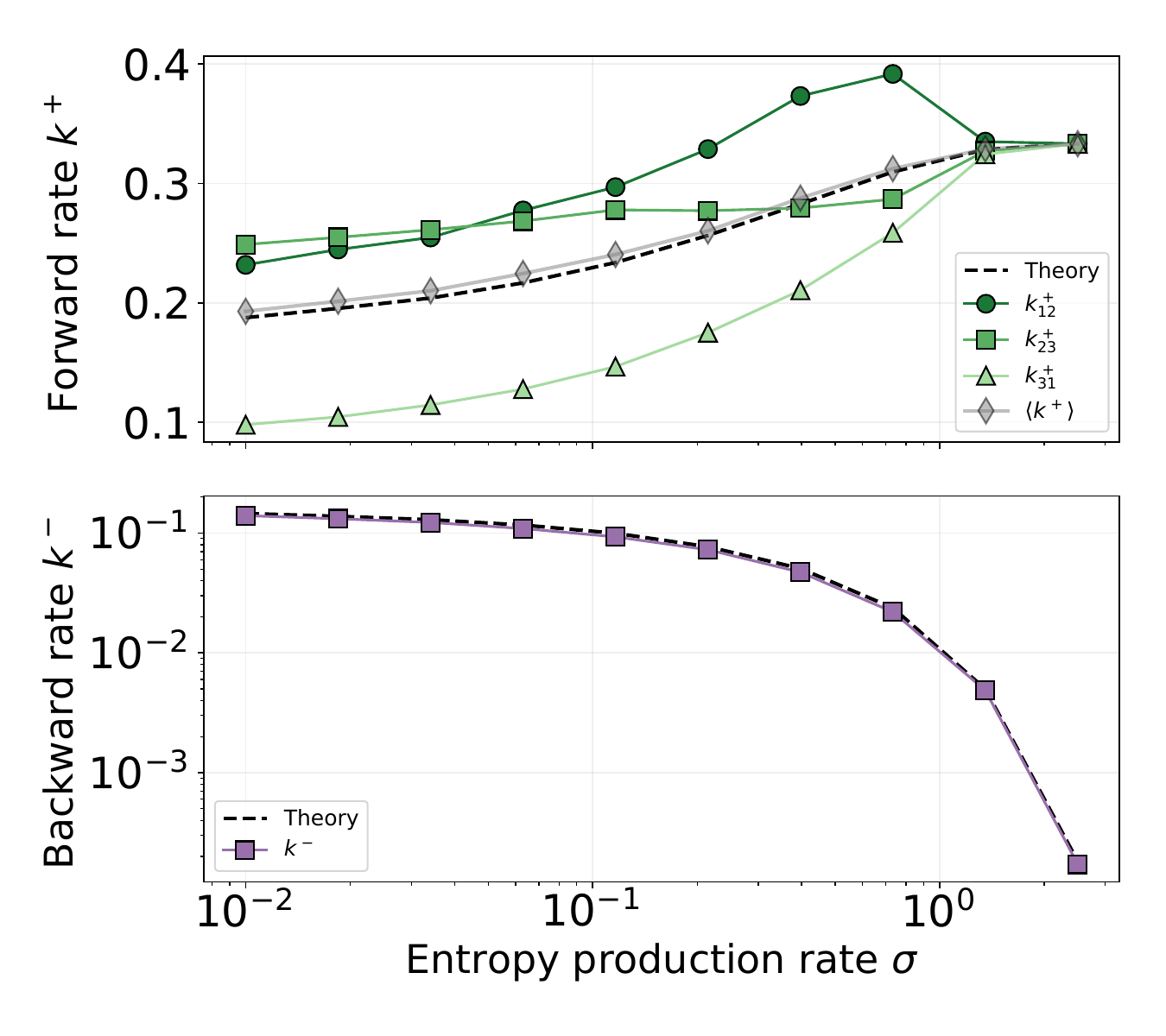}
\caption{\textbf{Optimized rate constants along the Pareto front.} \textbf{(Top)} Forward rates $k^+_{ij}$ for each edge of the three-state cycle. Individual rates (colored markers) show some heterogeneity, but the mean $\langle k^+ \rangle$ (gray diamonds) closely tracks the theoretical prediction for symmetric allocation (dashed line). \textbf{(Bottom)} Backward rates $k^-$ (purple squares) compared to theory.}
\label{fig:ring-rates}
\end{figure}

\begin{figure}[h!]
\centering
\begin{minipage}[t]{0.5\textwidth}
\centering
\includegraphics[width=\textwidth]{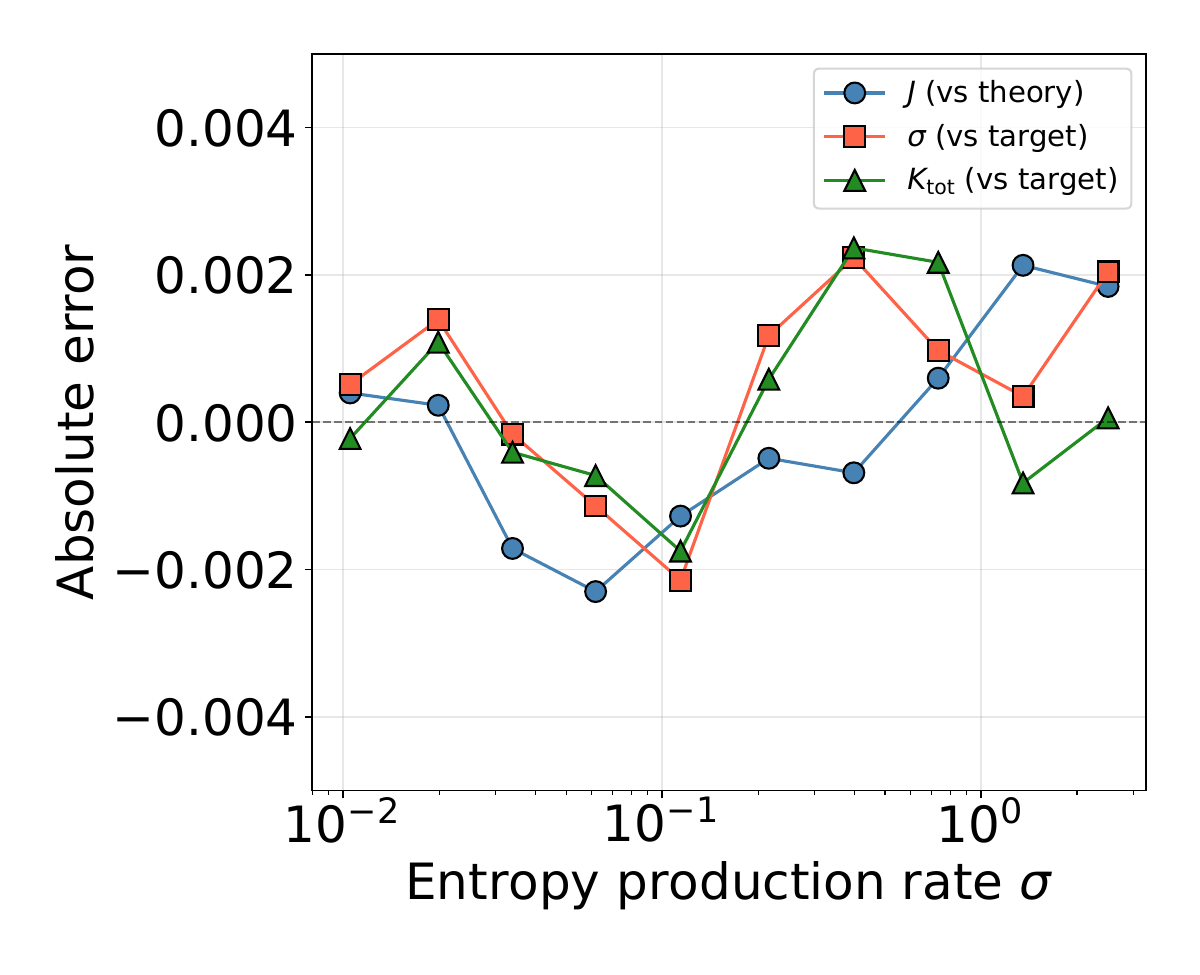}
\end{minipage}%
\hfill
\begin{minipage}[t]{0.5\textwidth}
\centering
\includegraphics[width=\textwidth]{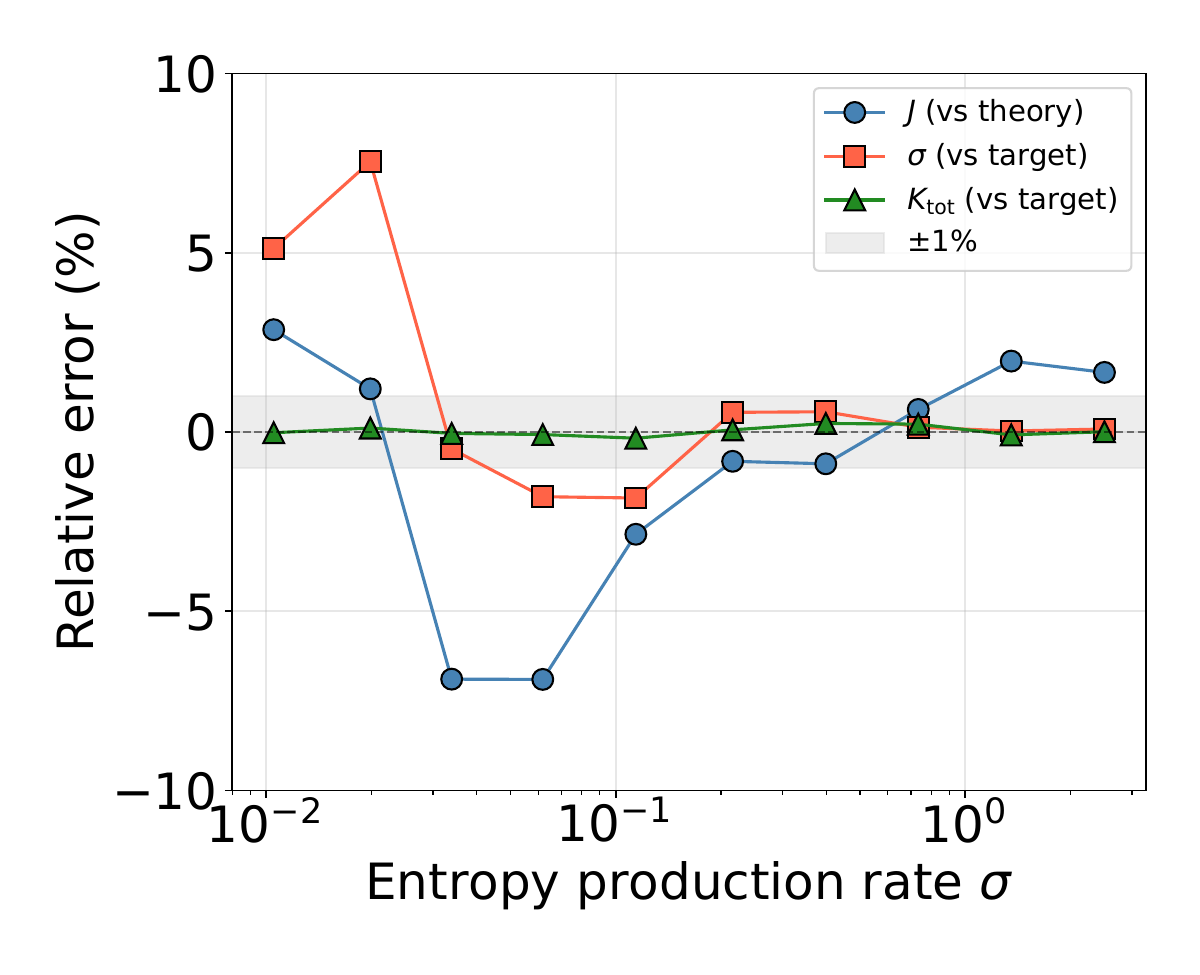}
\end{minipage}
\caption{\textbf{Optimization accuracy for the three-state cycle.} \textbf{(Left)} Absolute errors and \textbf{(Right)} relative errors for the optimized current $J$ (vs.\ theoretical bound), entropy production $\sigma$ (vs.\ target), and kinetic budget $K_\text{tot}$ (vs.\ target) as a function of entropy production rate.}
\label{fig:ring-errors}
\end{figure}

\begin{figure*}[p]
\centering
\includegraphics[width=\textwidth]{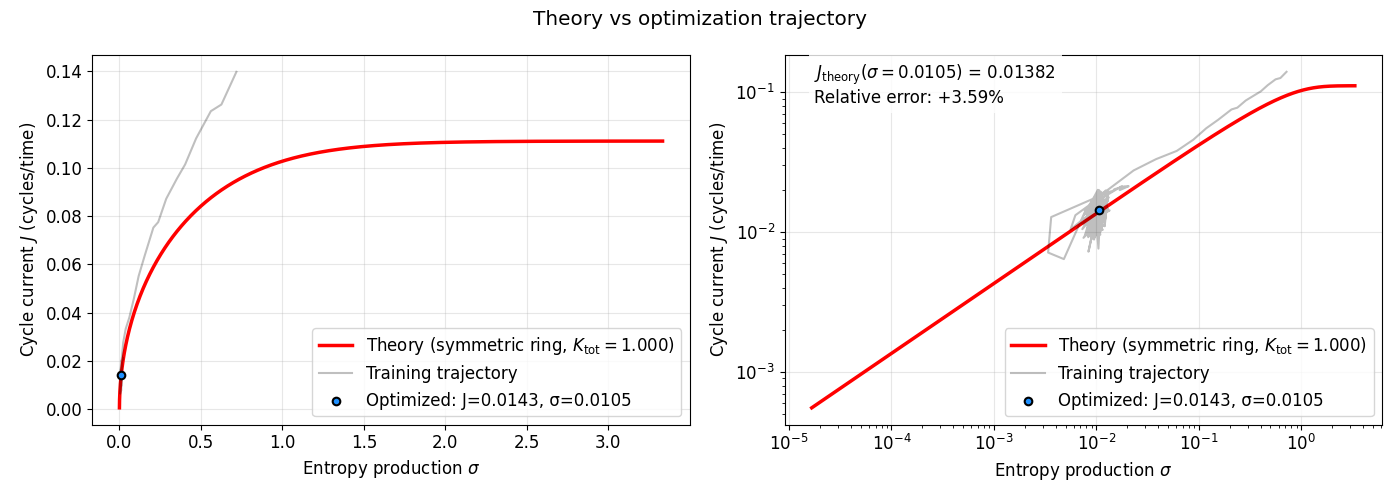}\\[0.3em]
\includegraphics[width=\textwidth]{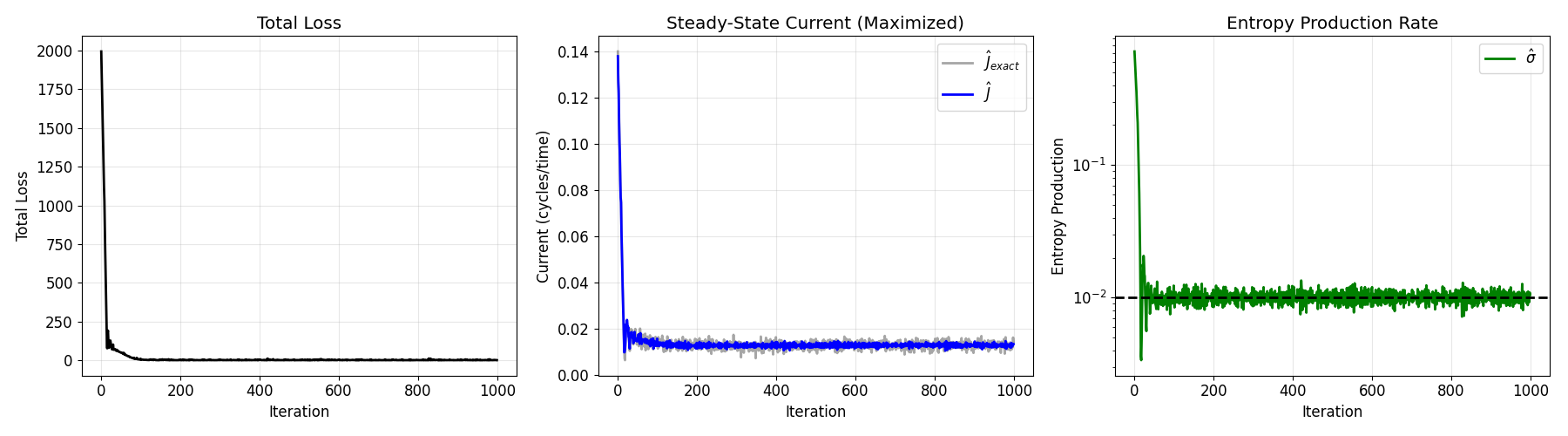}\\[0.3em]
\includegraphics[width=\textwidth]{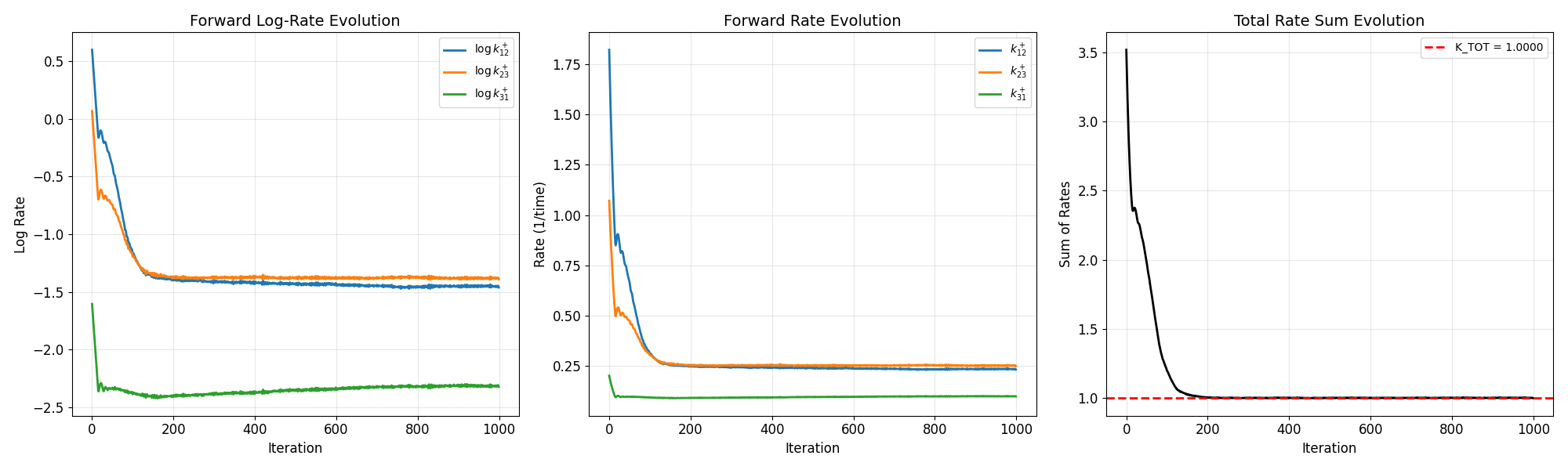}
\caption{\textbf{Optimization details for $\sigma = 0.01$.} \textbf{(Top)} Optimization trajectory in the current--dissipation plane; gray: training trajectory, blue: final point, red: theoretical Pareto front. \textbf{(Middle)} Training dynamics of total loss, steady-state current (gray: exact, blue: propensity-based estimate), and entropy production (dashed: target). \textbf{(Bottom)} Evolution of forward rates in log and linear scale, and total kinetic budget $K_\text{tot}$ (dashed red: target).}
\label{fig:ring-sigma-0.01}
\end{figure*}

\begin{figure*}[p]
\centering
\includegraphics[width=\textwidth]{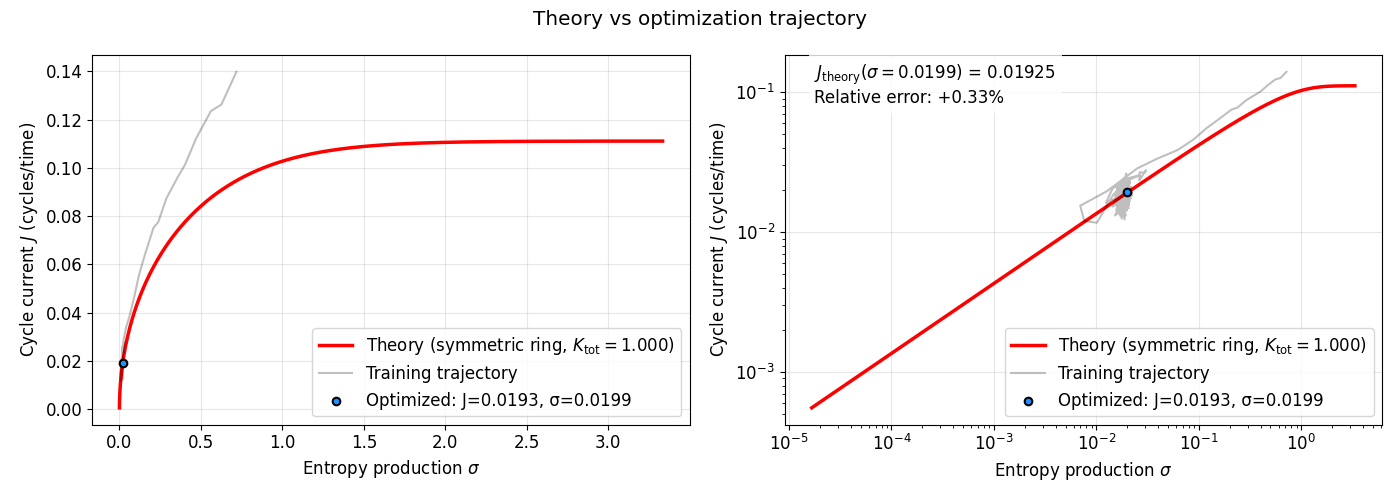}\\[0.3em]
\includegraphics[width=\textwidth]{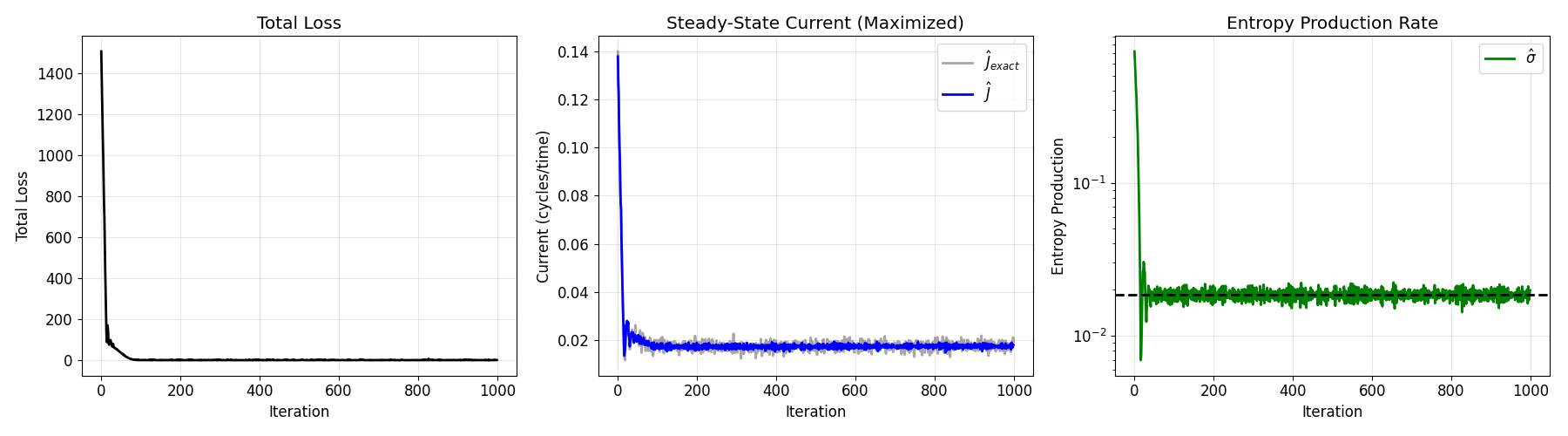}\\[0.3em]
\includegraphics[width=\textwidth]{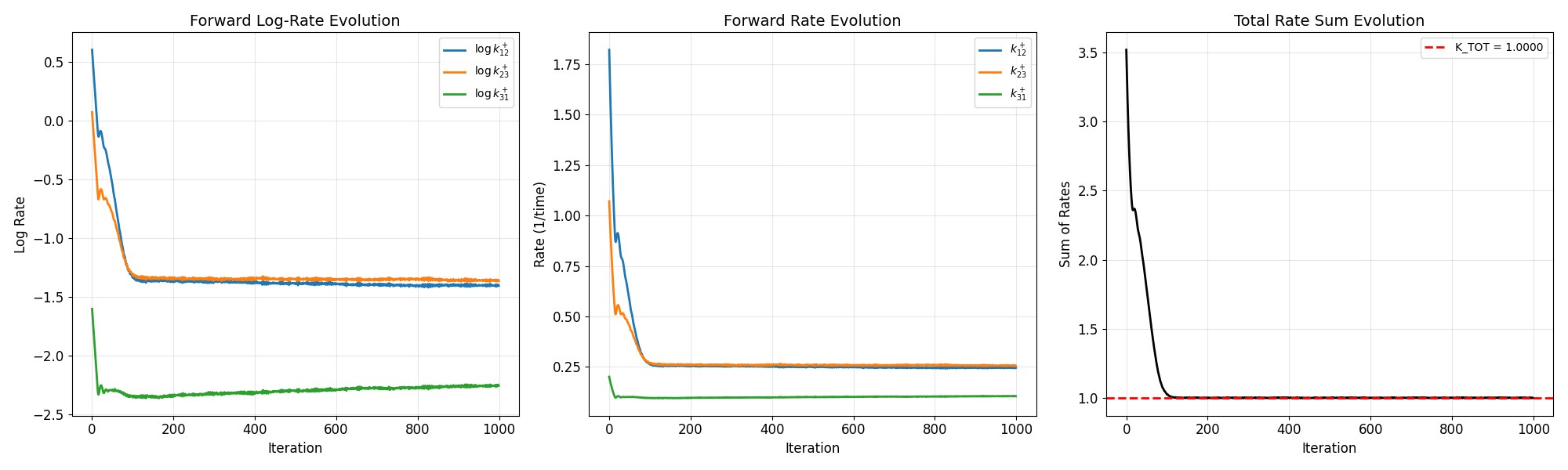}
\caption{\textbf{Optimization details for $\sigma = 0.018$.} Same format as Fig.~\ref{fig:ring-sigma-0.01}.}
\label{fig:ring-sigma-0.018}
\end{figure*}

\begin{figure*}[p]
\centering
\includegraphics[width=\textwidth]{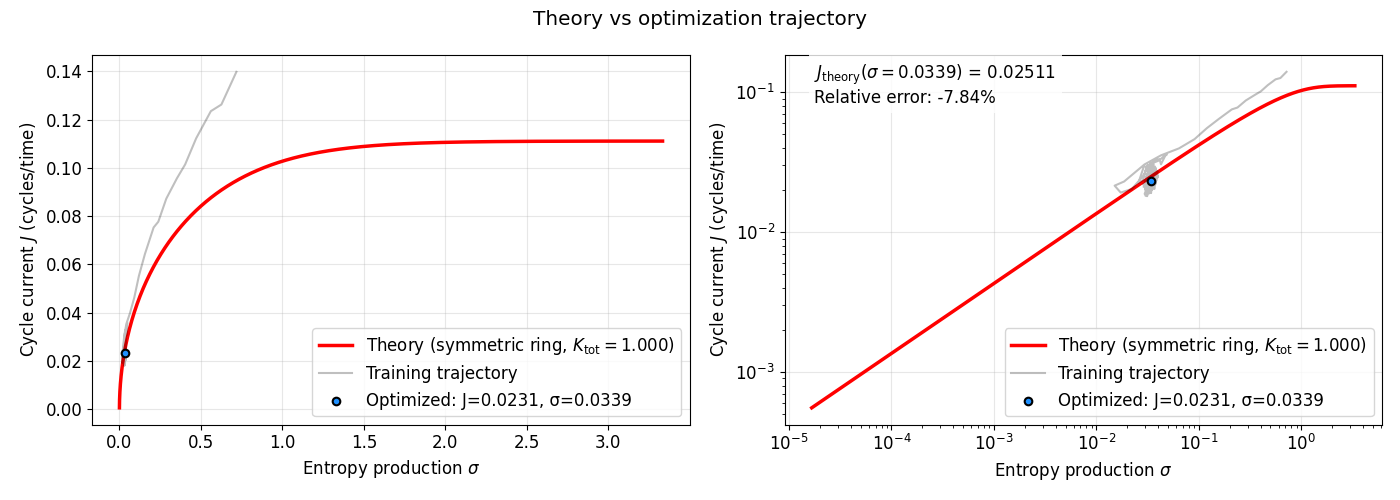}\\[0.3em]
\includegraphics[width=\textwidth]{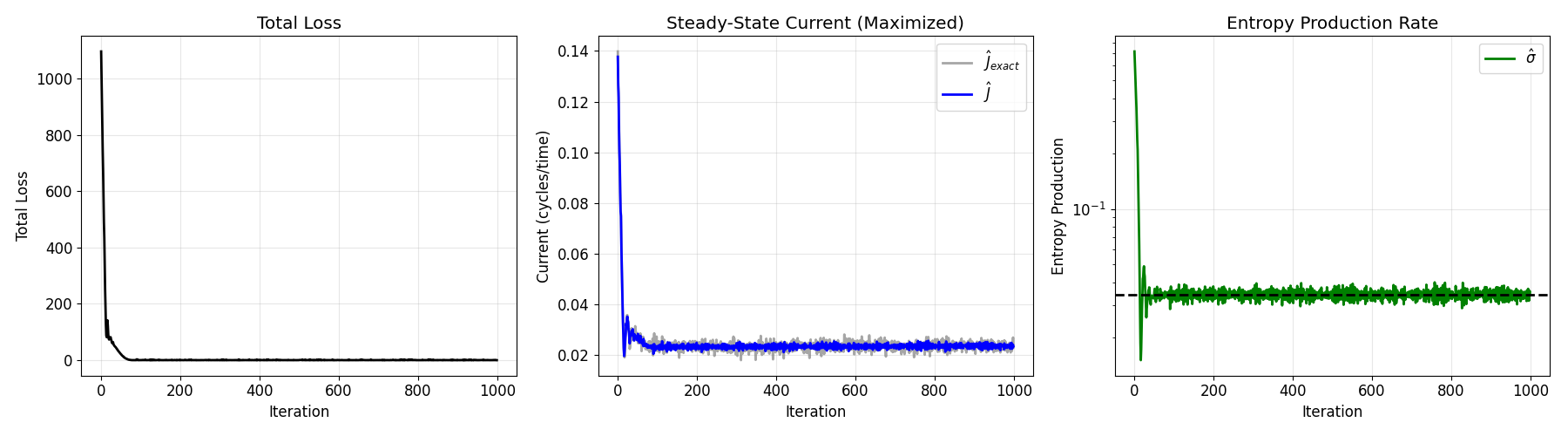}\\[0.3em]
\includegraphics[width=\textwidth]{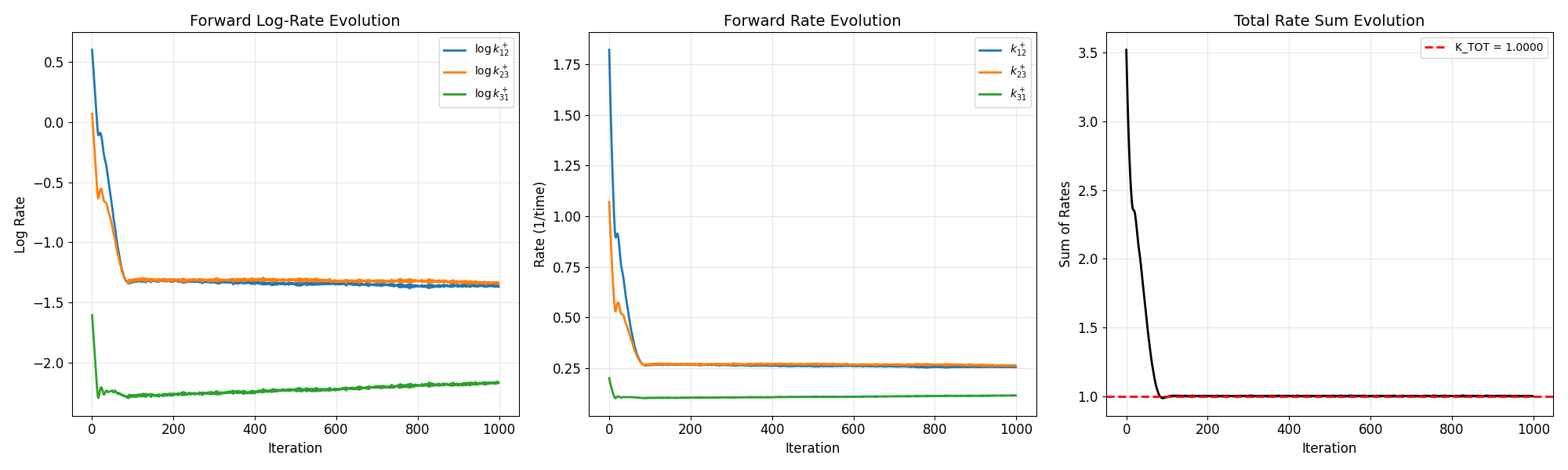}
\caption{\textbf{Optimization details for $\sigma = 0.034$.} Same format as Fig.~\ref{fig:ring-sigma-0.01}.}
\label{fig:ring-sigma-0.034}
\end{figure*}

\begin{figure*}[p]
\centering
\includegraphics[width=\textwidth]{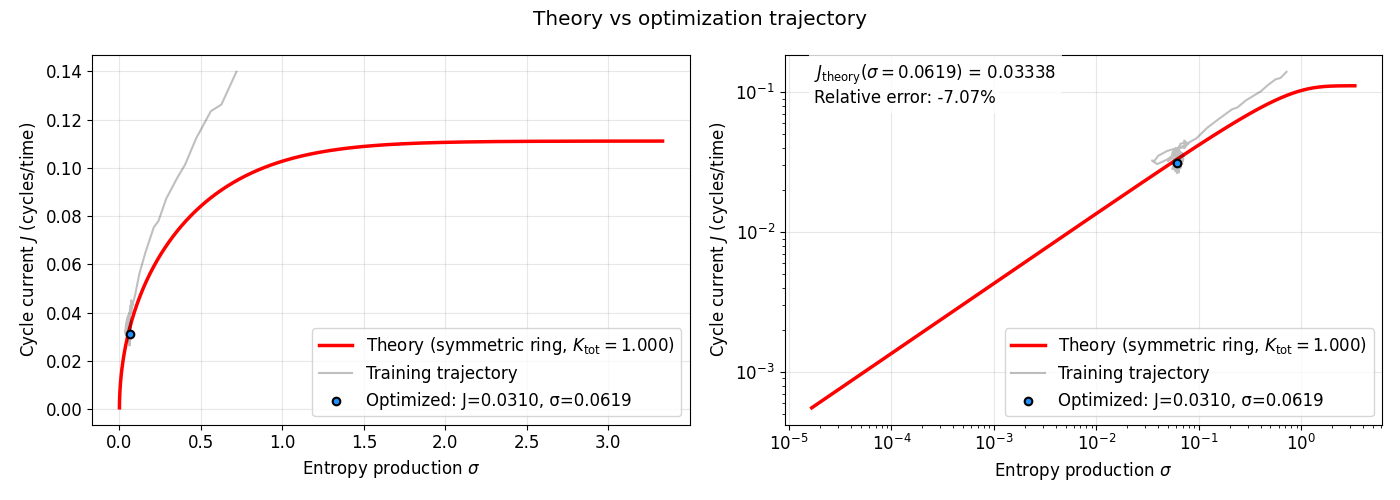}\\[0.3em]
\includegraphics[width=\textwidth]{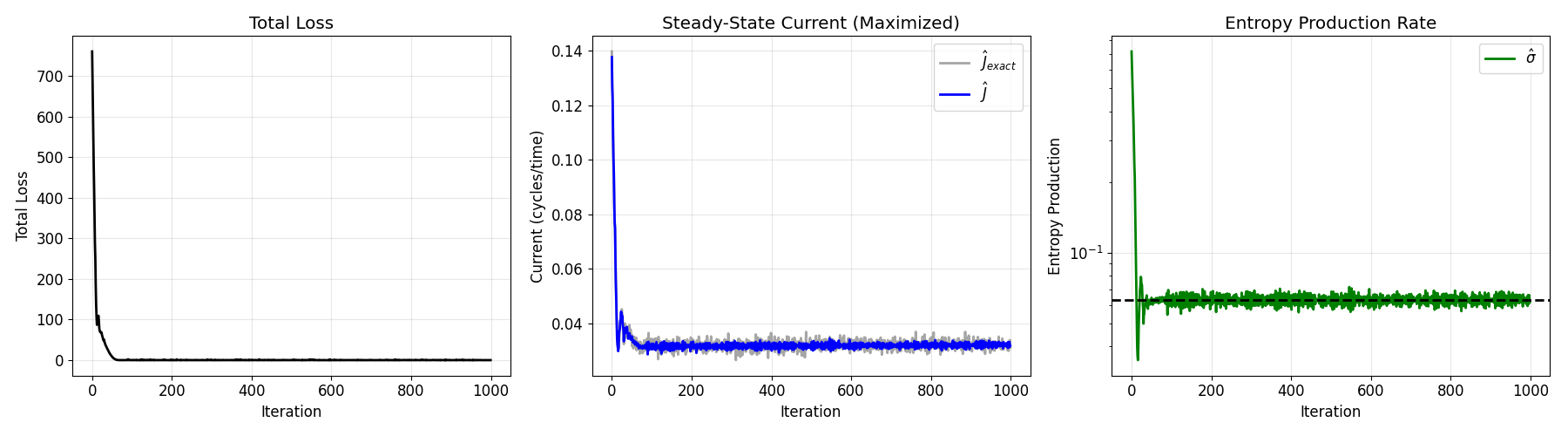}\\[0.3em]
\includegraphics[width=\textwidth]{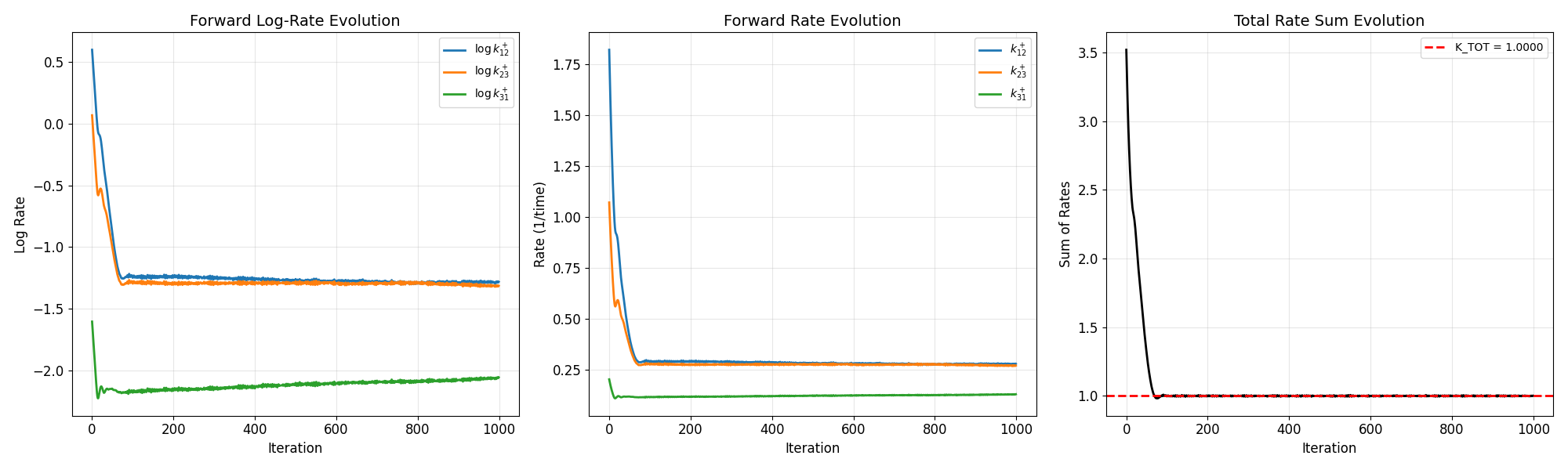}
\caption{\textbf{Optimization details for $\sigma = 0.063$.} Same format as Fig.~\ref{fig:ring-sigma-0.01}.}
\label{fig:ring-sigma-0.063}
\end{figure*}

\begin{figure*}[p]
\centering
\includegraphics[width=\textwidth]{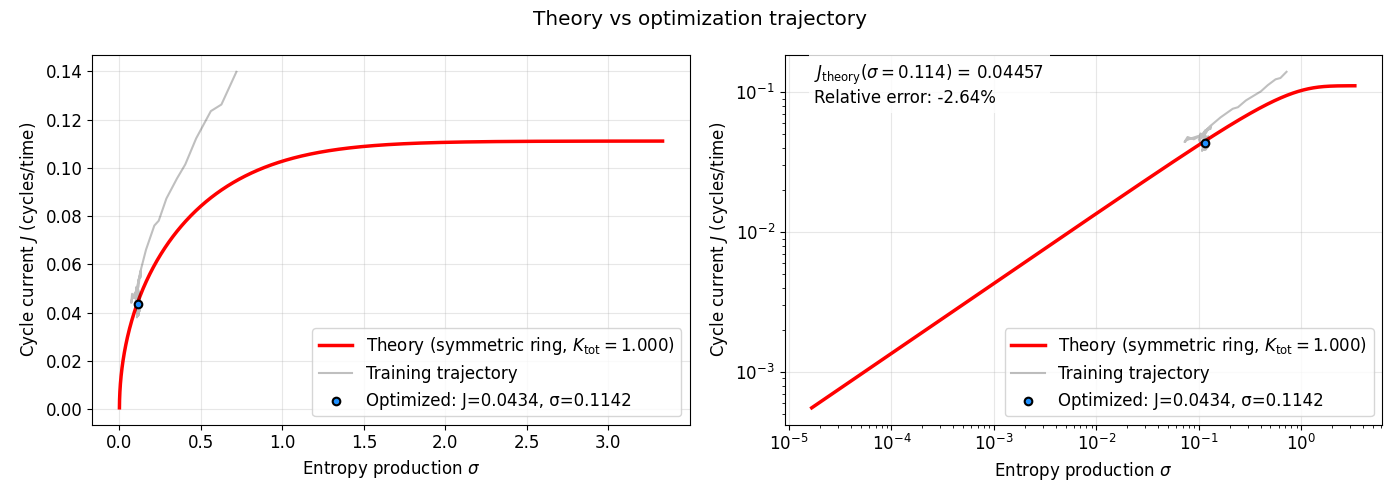}\\[0.3em]
\includegraphics[width=\textwidth]{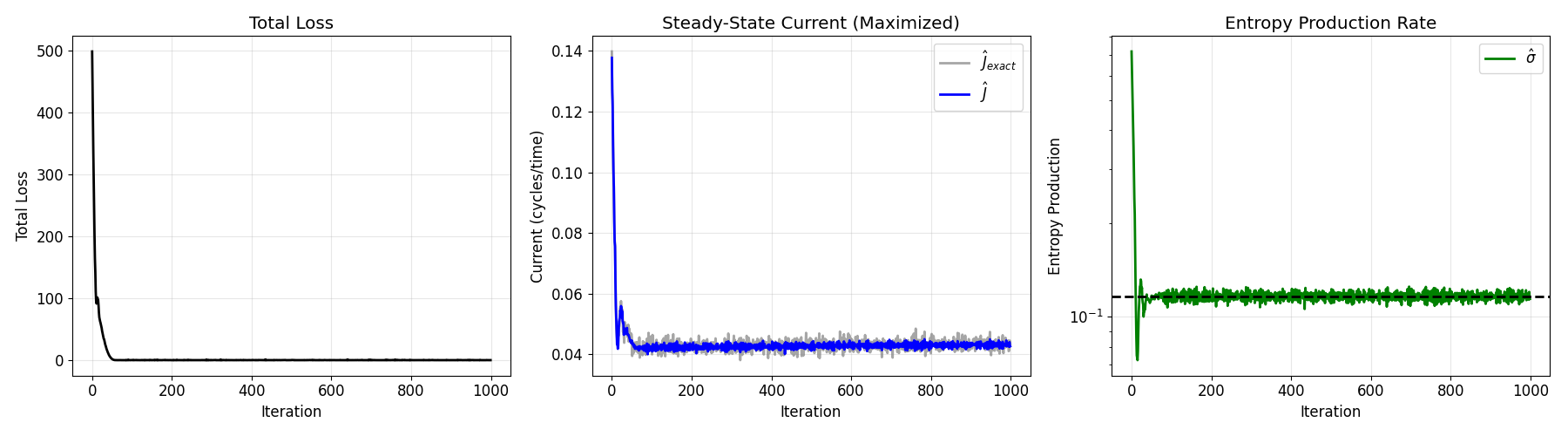}\\[0.3em]
\includegraphics[width=\textwidth]{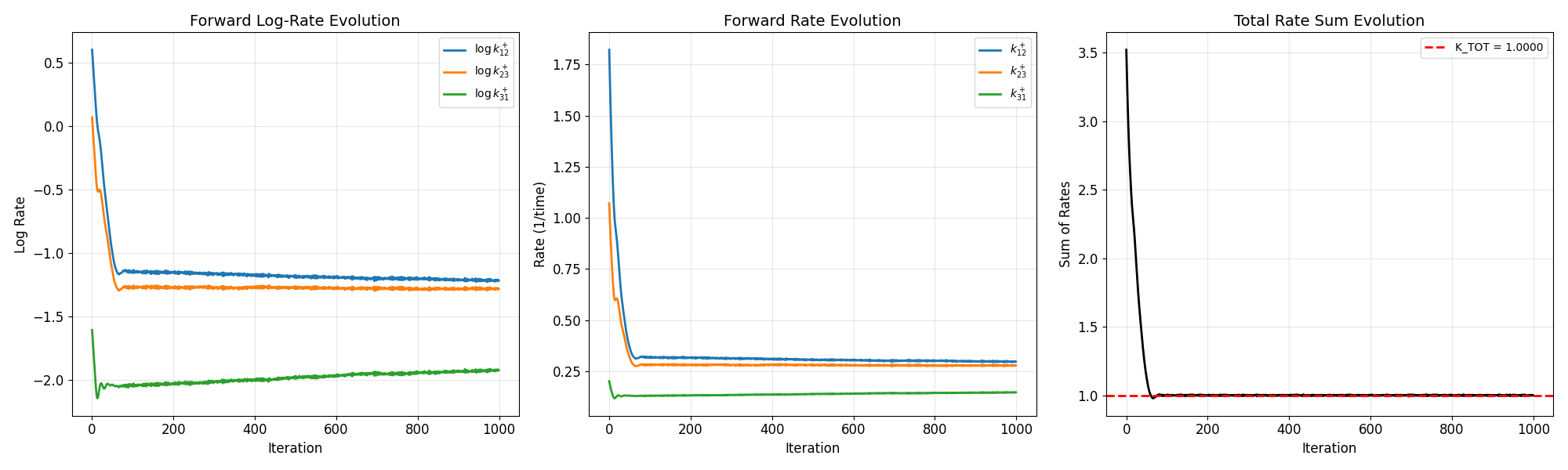}
\caption{\textbf{Optimization details for $\sigma = 0.12$.} Same format as Fig.~\ref{fig:ring-sigma-0.01}.}
\label{fig:ring-sigma-0.12}
\end{figure*}

\begin{figure*}[p]
\centering
\includegraphics[width=\textwidth]{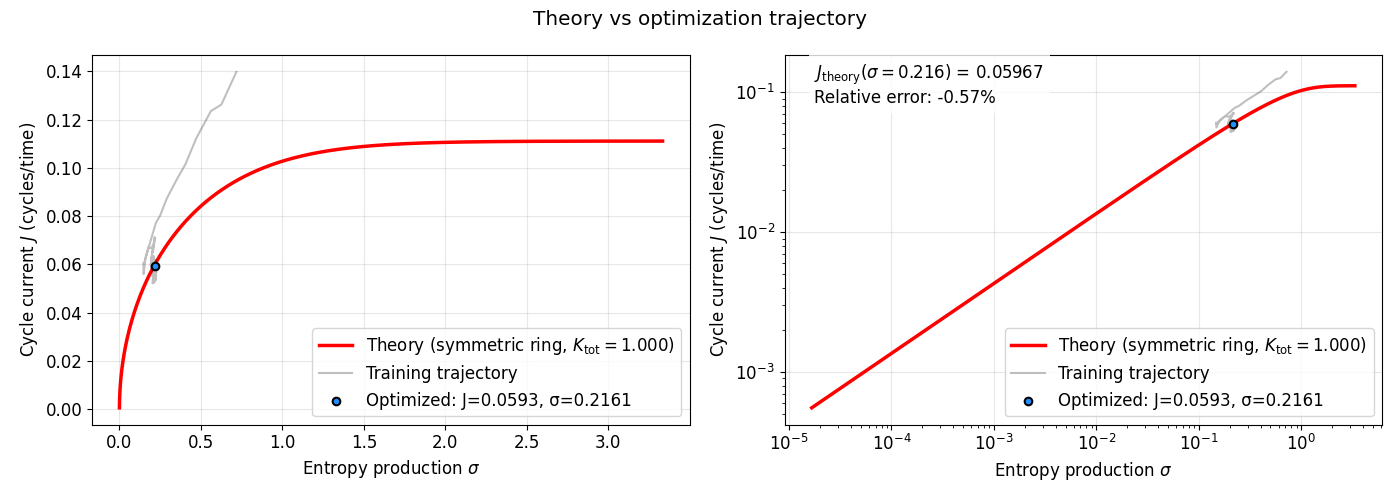}\\[0.3em]
\includegraphics[width=\textwidth]{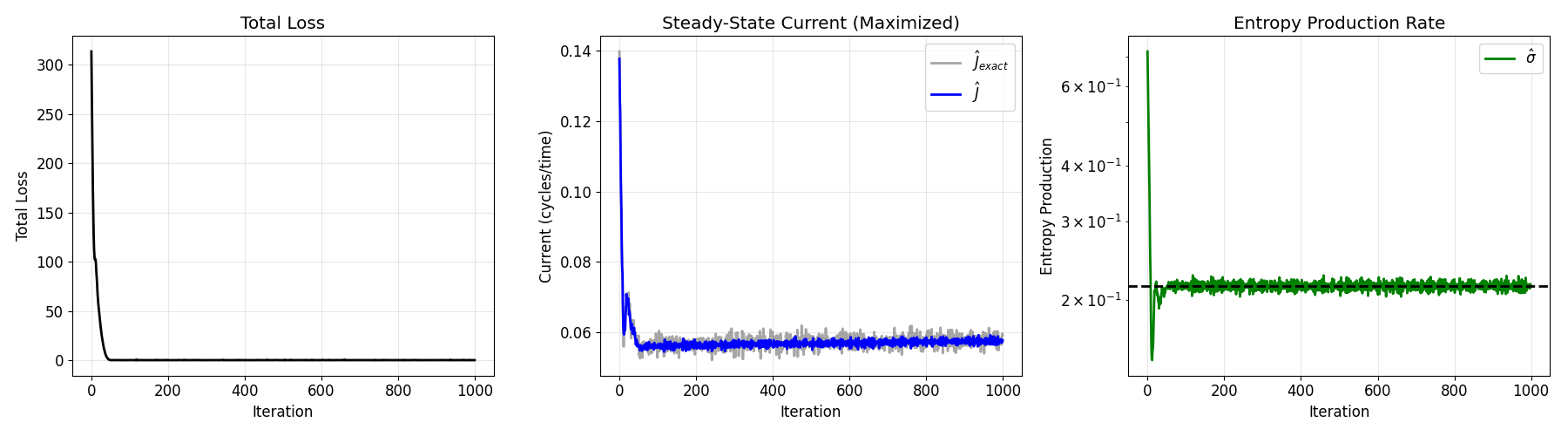}\\[0.3em]
\includegraphics[width=\textwidth]{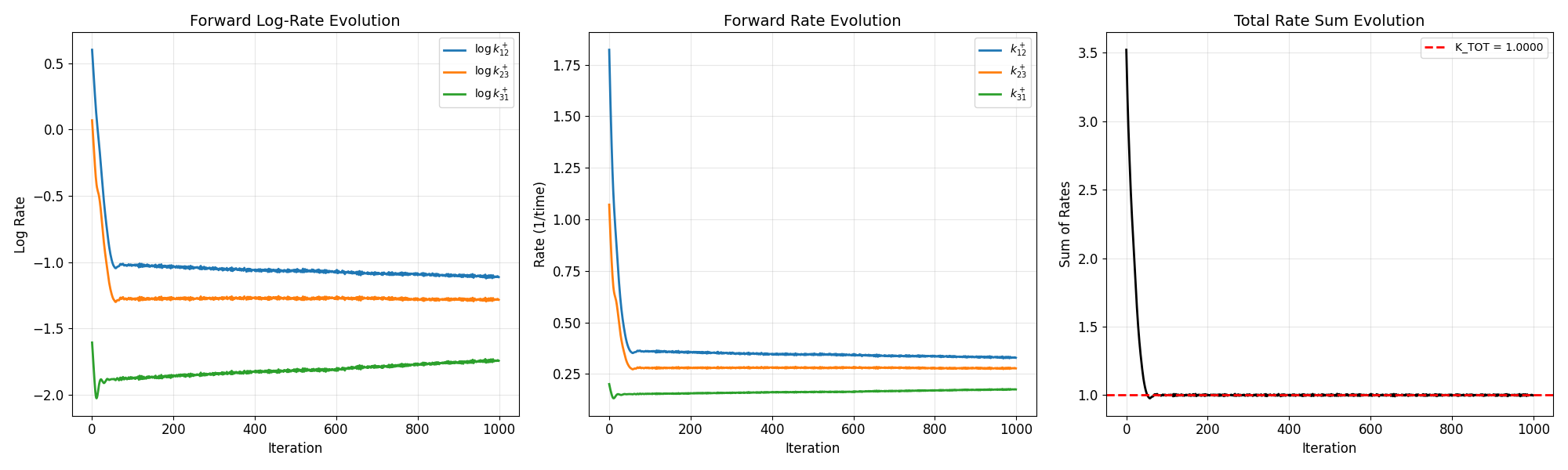}
\caption{\textbf{Optimization details for $\sigma = 0.21$.} Same format as Fig.~\ref{fig:ring-sigma-0.01}.}
\label{fig:ring-sigma-0.21}
\end{figure*}

\begin{figure*}[p]
\centering
\includegraphics[width=\textwidth]{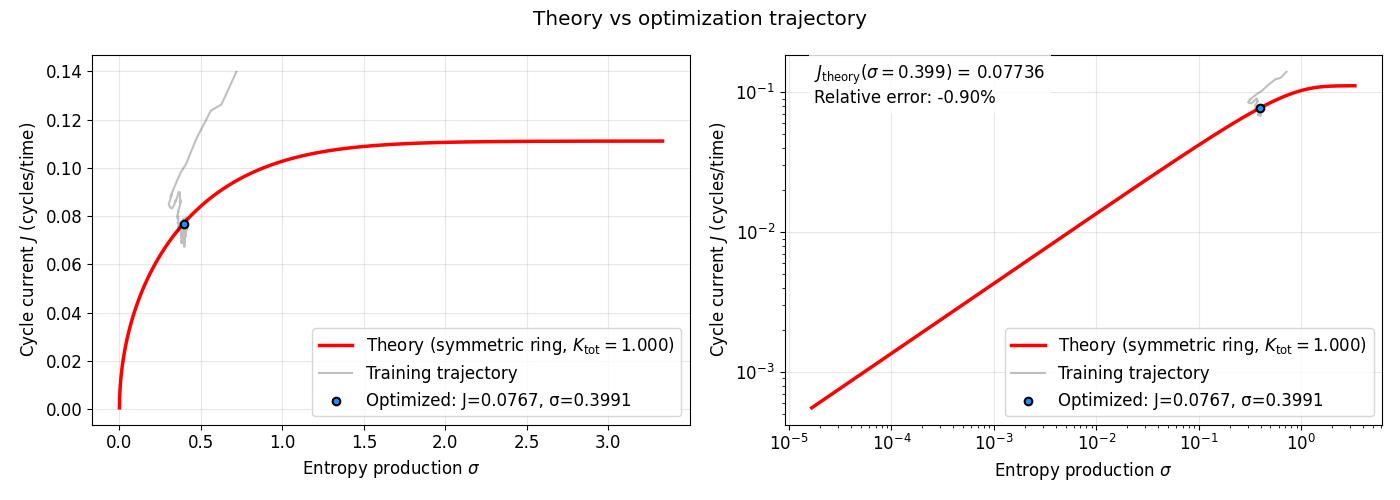}\\[0.3em]
\includegraphics[width=\textwidth]{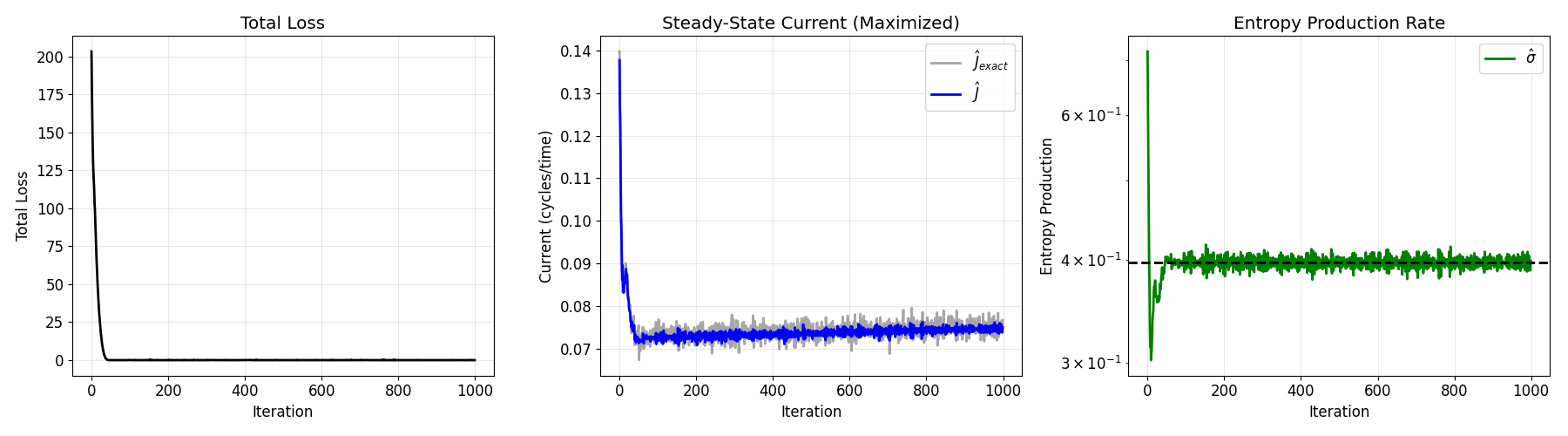}\\[0.3em]
\includegraphics[width=\textwidth]{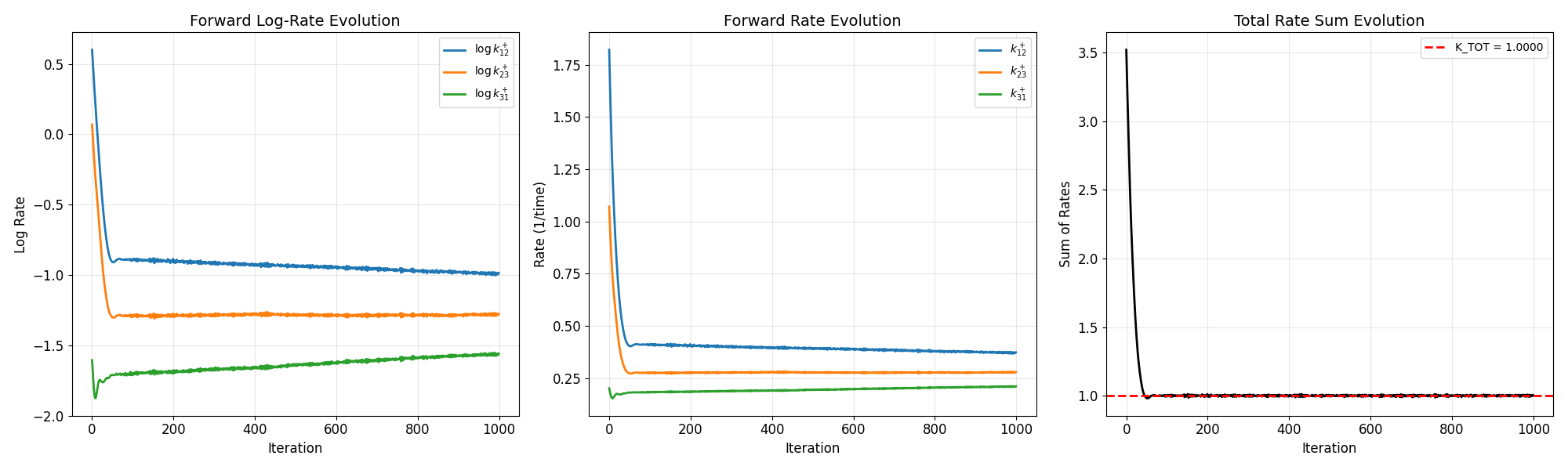}
\caption{\textbf{Optimization details for $\sigma = 0.40$.} Same format as Fig.~\ref{fig:ring-sigma-0.01}.}
\label{fig:ring-sigma-0.40}
\end{figure*}

\begin{figure*}[p]
\centering
\includegraphics[width=\textwidth]{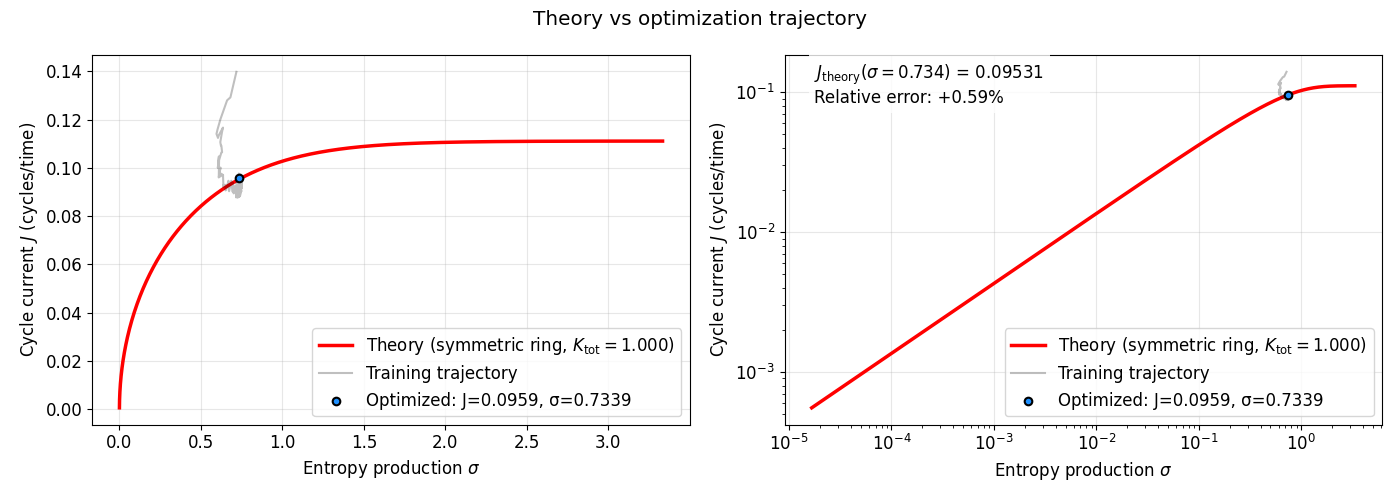}\\[0.3em]
\includegraphics[width=\textwidth]{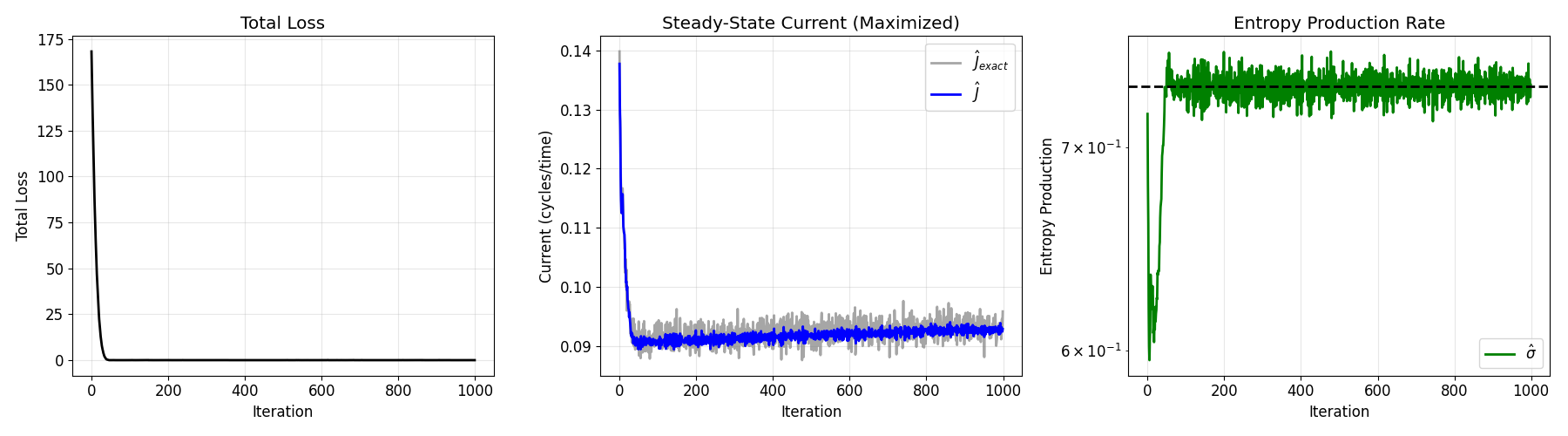}\\[0.3em]
\includegraphics[width=\textwidth]{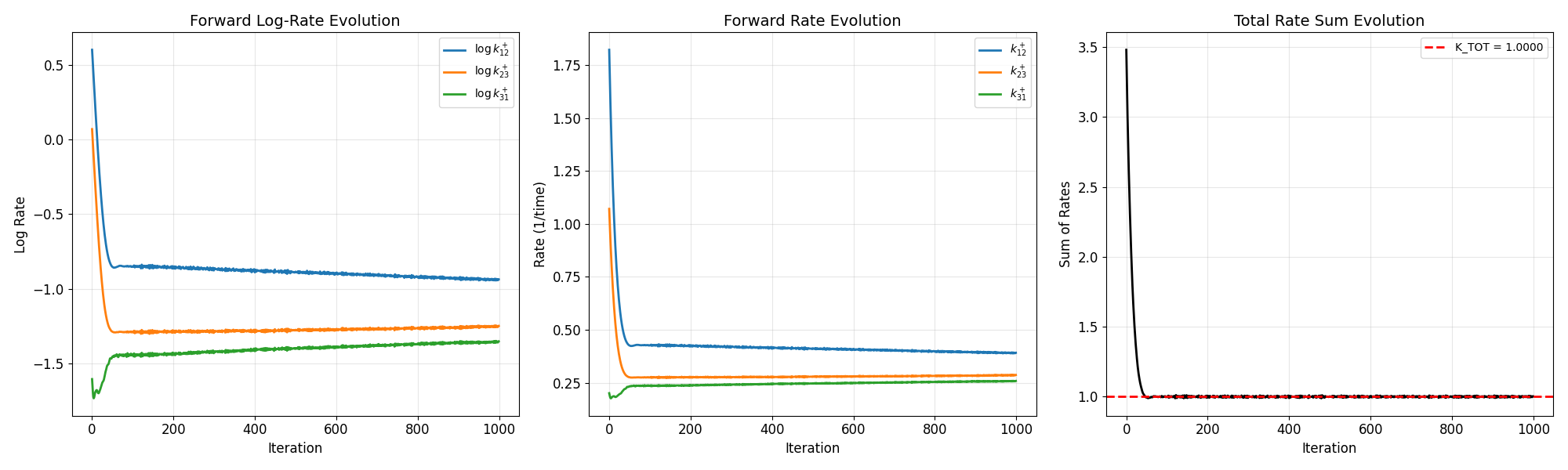}
\caption{\textbf{Optimization details for $\sigma = 0.73$.} Same format as Fig.~\ref{fig:ring-sigma-0.01}.}
\label{fig:ring-sigma-0.73}
\end{figure*}

\begin{figure*}[p]
\centering
\includegraphics[width=\textwidth]{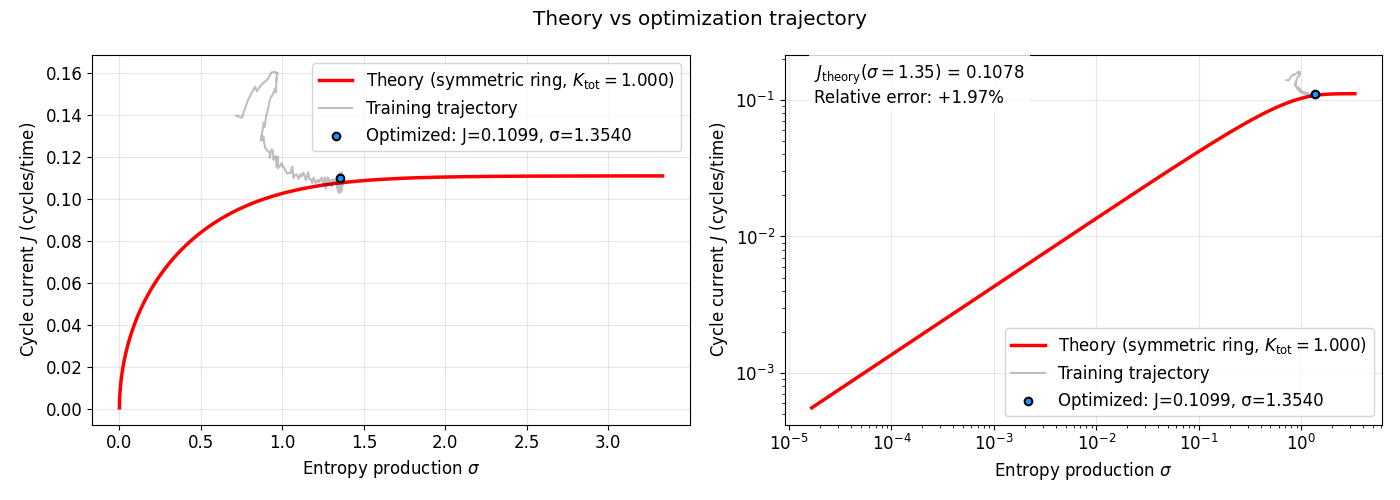}\\[0.3em]
\includegraphics[width=\textwidth]{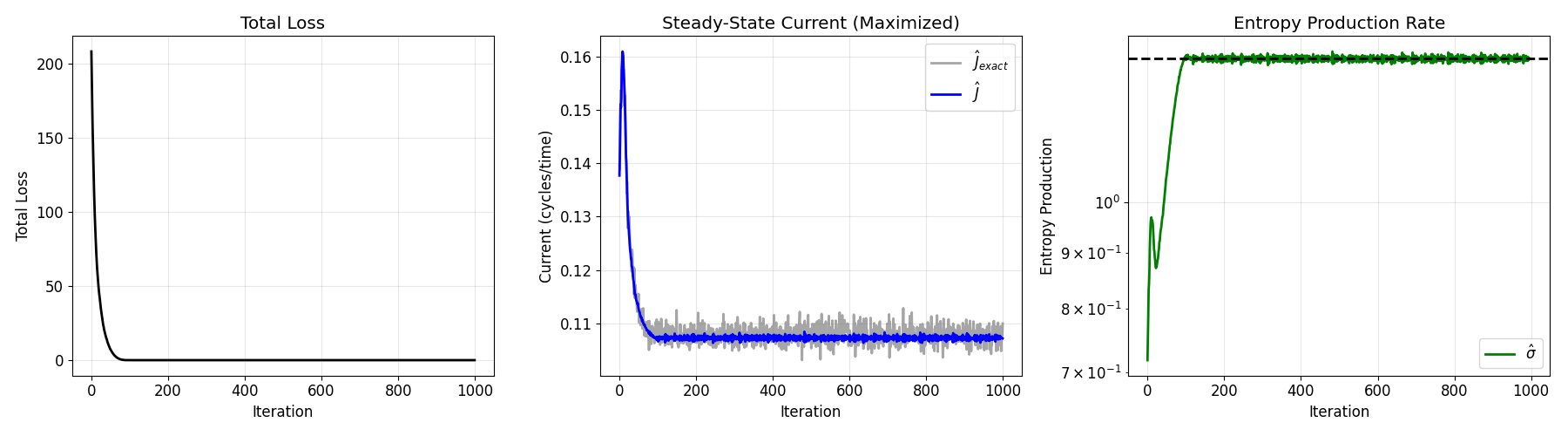}\\[0.3em]
\includegraphics[width=\textwidth]{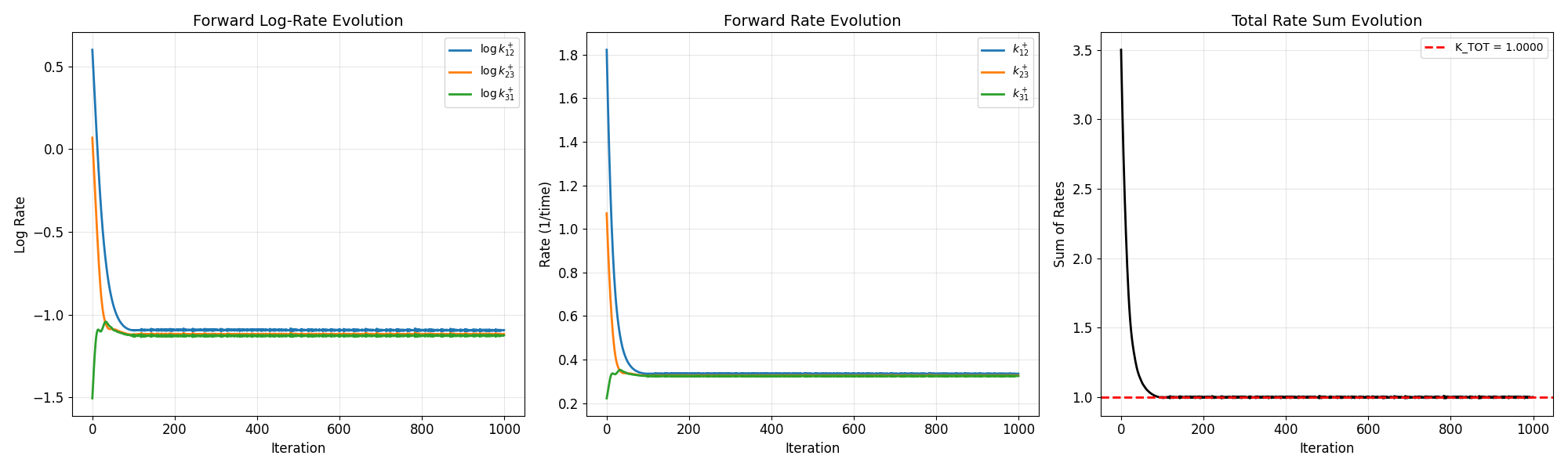}
\caption{\textbf{Optimization details for $\sigma = 1.35$.} Same format as Fig.~\ref{fig:ring-sigma-0.01}.}
\label{fig:ring-sigma-1.35}
\end{figure*}

\begin{figure*}[p]
\centering
\includegraphics[width=\textwidth]{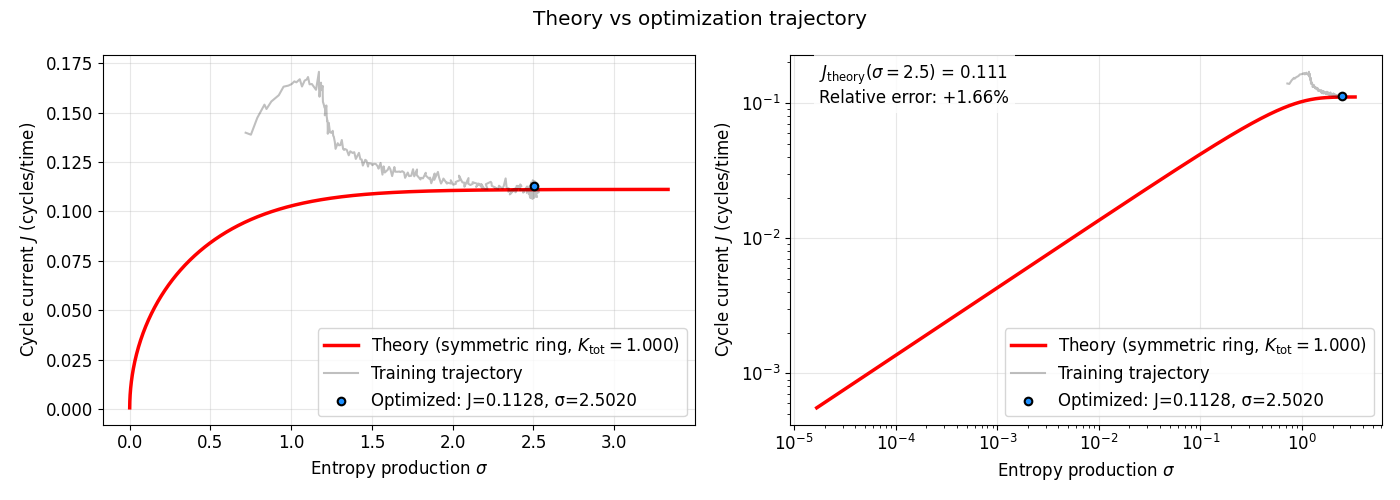}\\[0.3em]
\includegraphics[width=\textwidth]{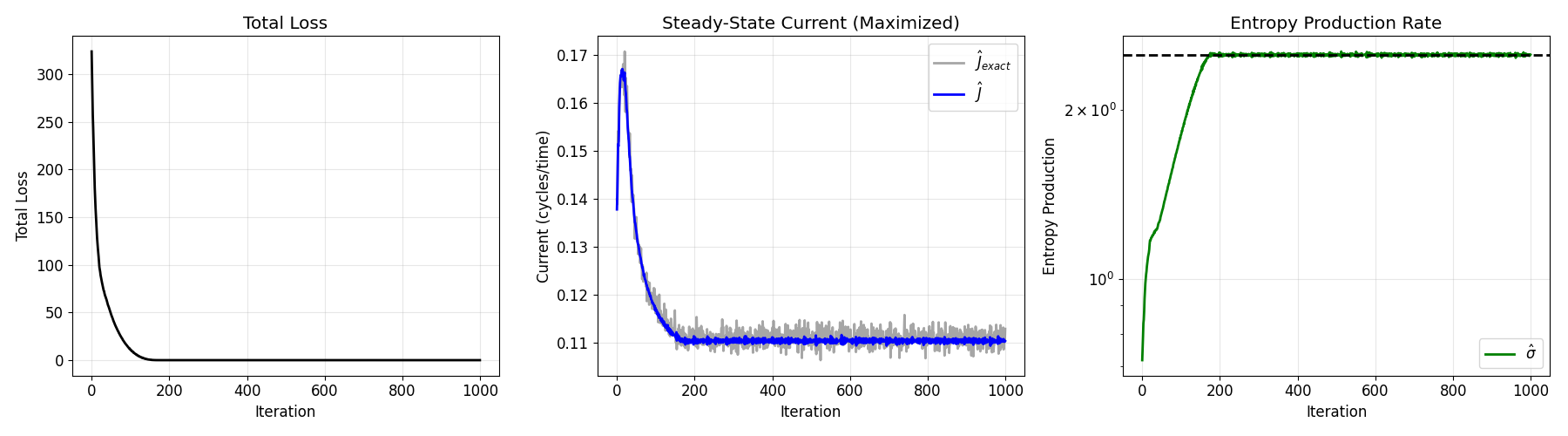}\\[0.3em]
\includegraphics[width=\textwidth]{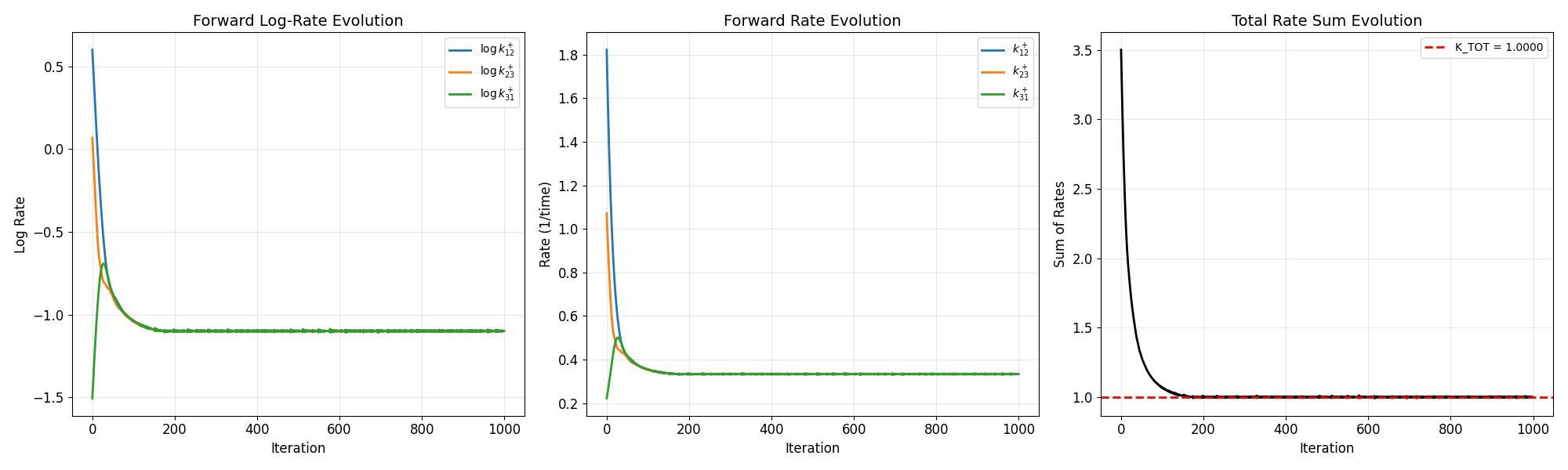}
\caption{\textbf{Optimization details for $\sigma = 2.5$.} Same format as Fig.~\ref{fig:ring-sigma-0.01}.}
\label{fig:ring-sigma-2.5}
\end{figure*}

\textbf{Additional Results.} Table~\ref{tab:pareto-results} reports the optimized current $J$, achieved entropy production $\sigma$, and total kinetic budget $K_\text{tot}$ for each target, along with relative errors compared to theoretical predictions. The kinetic budget constraint is satisfied to within 0.24\% across all runs. Relative errors in current range from $-6.9\%$ to $+2.9\%$, with larger deviations at low $\sigma$ where the absolute current is small. Table~\ref{tab:pareto-rates} reports the optimized rate constants. At high dissipation ($\sigma \gtrsim 1$), the forward rates converge to the symmetric allocation $k^+_{ij} \approx K_\text{tot}/9 \approx 0.333$. At low dissipation, individual forward rates show substantial heterogeneity while their mean remains close to theory. Fig.~\ref{fig:pareto-front} shows the Pareto front, Fig.~\ref{fig:ring-rates} the optimized rates, and Fig.~\ref{fig:ring-errors} the absolute and relative errors. Figs.~\ref{fig:ring-sigma-0.01}--\ref{fig:ring-sigma-2.5} show detailed optimization trajectories for each target $\sigma$.

\vspace{1cm}
\begin{table*}[h!]
\small
\begin{tabular*}{.8\textwidth}{@{\extracolsep{\fill}} c c c c c c c c}
\hline
$\sigma_\text{target}$ & $\sigma$ & $\Delta\sigma$ (\%) & $J$ & $J^*$ & $\Delta J$ (\%) & $K_\text{tot}$ & $\Delta K_\text{tot}$ (\%) \\
\hline
0.010 & 0.0105 & $+5.12$ & 0.0143 & 0.0139 & $+2.85$ & 0.9998 & $-0.02$ \\
0.018 & 0.0199 & $+7.54$ & 0.0193 & 0.0191 & $+1.20$ & 1.0011 & $+0.11$ \\
0.034 & 0.0339 & $-0.47$ & 0.0231 & 0.0249 & $-6.90$ & 0.9996 & $-0.04$ \\
0.063 & 0.0619 & $-1.81$ & 0.0310 & 0.0333 & $-6.90$ & 0.9993 & $-0.07$ \\
0.116 & 0.1142 & $-1.84$ & 0.0434 & 0.0447 & $-2.86$ & 0.9982 & $-0.18$ \\
0.215 & 0.2161 & $+0.55$ & 0.0593 & 0.0598 & $-0.82$ & 1.0006 & $+0.06$ \\
0.397 & 0.3991 & $+0.56$ & 0.0767 & 0.0774 & $-0.89$ & 1.0024 & $+0.24$ \\
0.733 & 0.7339 & $+0.13$ & 0.0959 & 0.0953 & $+0.63$ & 1.0022 & $+0.22$ \\
1.354 & 1.3540 & $+0.03$ & 0.1099 & 0.1078 & $+1.98$ & 0.9992 & $-0.08$ \\
2.500 & 2.5020 & $+0.08$ & 0.1128 & 0.1110 & $+1.66$ & 1.0001 & $+0.01$ \\
\hline
\end{tabular*}
\centering
\caption{\textbf{Pareto front optimization results.} For each target entropy production $\sigma_\text{target}$: achieved entropy production $\sigma$, optimized current $J$, theoretical current $J^*$, achieved kinetic budget $K_\text{tot}$, and relative errors. Theory assumes optimal symmetric allocation.}
\label{tab:pareto-results}
\end{table*}

\vspace{1cm}
\begin{table*}[h!]
\small
\begin{tabular*}{.7\textwidth}{@{\extracolsep{\fill}} c c c c c c c}
\hline
$\sigma_\text{target}$ & $k^+_{12}$ & $k^+_{23}$ & $k^+_{31}$ & $k^-$ & $k^+_\text{theory}$ & $k^-_\text{theory}$ \\
\hline
0.010 & 0.232 & 0.249 & 0.098 & 0.140 & 0.188 & 0.146 \\
0.018 & 0.245 & 0.255 & 0.104 & 0.132 & 0.195 & 0.138 \\
0.034 & 0.255 & 0.261 & 0.114 & 0.123 & 0.204 & 0.129 \\
0.063 & 0.277 & 0.268 & 0.128 & 0.109 & 0.217 & 0.117 \\
0.116 & 0.297 & 0.278 & 0.147 & 0.093 & 0.234 & 0.100 \\
0.215 & 0.329 & 0.277 & 0.175 & 0.073 & 0.256 & 0.077 \\
0.397 & 0.373 & 0.279 & 0.210 & 0.047 & 0.283 & 0.051 \\
0.733 & 0.392 & 0.287 & 0.259 & 0.022 & 0.310 & 0.024 \\
1.354 & 0.335 & 0.327 & 0.324 & 0.005 & 0.328 & 0.005 \\
2.500 & 0.333 & 0.333 & 0.333 & 0.0002 & 0.333 & 0.0002 \\
\hline
\end{tabular*}
\centering
\caption{\textbf{Optimized rate constants.} Forward rates $k^+_{12}$, $k^+_{23}$, $k^+_{31}$ and common backward rate $k^-$ for each target entropy production. Theory columns show predictions for optimal symmetric allocation.}
\label{tab:pareto-rates}
\end{table*}

\end{document}